\documentclass[fleqn,usenatbib]{mnras}

\usepackage{newtxtext,newtxmath}

\usepackage[T1]{fontenc}

\DeclareRobustCommand{\VAN}[3]{#2}
\let\VANthebibliography\thebibliography
\def\thebibliography{\DeclareRobustCommand{\VAN}[3]{##3}\VANthebibliography}

\usepackage{graphicx}	
\usepackage{amsmath}	
\usepackage[dvipsnames]{xcolor}
\usepackage{tabularx}
\usepackage{subcaption}
\usepackage{stfloats} 
\usepackage[nameinlink,capitalize]{cleveref}
\usepackage{enumitem}
\usepackage{siunitx} 
\newcommand{\ifm}[1]{\relax\ifmmode#1\else$\mathsurround=0pt #1$\fi}

\newcommand{\parfrac}[2]{\left(\frac{#1}{#2}\right)}

\newcommand{\software}{\textsc}
\newcommand{\run}{\texttt}

\DeclareSIUnit{\Msun}{{\rm M}_{\sun}}
\DeclareSIUnit{\Lsun}{{\rm L}_{\sun}}
\DeclareSIUnit{\kms}{{\rm km}\,{\rm s}^{-1}}
\DeclareSIUnit{\Mpc}{{\rm Mpc}}
\DeclareSIUnit{\kpc}{{\rm kpc}}
\DeclareSIUnit{\pc}{{\rm pc}}
\DeclareSIUnit{\Gyr}{{\rm Gyr}}
\DeclareSIUnit{\Myr}{{\rm Myr}}
\DeclareSIUnit{\yr}{{\rm yr}}
\DeclareSIUnit{\K}{{\rm K}}
\DeclareSIUnit{\Mpcc}{{\rm Mpc}^{-3}}
\DeclareSIUnit{\cmc}{{\rm cm}^{-3}}
\DeclareSIUnit{\cms}{{\rm cm}^{-2}}
\DeclareSIUnit{\ergs}{{\rm erg}\,{\rm s}^{-1}}
\DeclareSIUnit{\sy}{M_{\sun}\,{\rm yr}^{-1}}
\DeclareSIUnit{\skpc}{M_{\sun}\,{\rm kpc}^{-1}}

\DeclareSIUnit{\hmpc}{\,\ifm{h^{-1}}{\rm Mpc}}
\DeclareSIUnit{\hkpc}{\,\ifm{h^{-1}}{\rm kpc}}

\newcommand{\kb}{k_{\rm B}}

\newcommand{\dotm}[1]{\dot{m}_{\rm #1}}

\newcommand{\Ax}[1]{A_{\rm #1}}
\newcommand{\Sx}[1]{S_{\rm #1}}
\newcommand{\Mx}[1]{M_{\rm #1}}
\newcommand{\mx}[1]{m_{\rm #1}}
\newcommand{\tx}[1]{t_{\rm #1}}
\newcommand{\Rx}[1]{R_{\rm #1}}
\newcommand{\rx}[1]{r_{\rm #1}}
\newcommand{\Tx}[1]{T_{\rm #1}}
\newcommand{\cx}[1]{c_{\rm #1}}
\newcommand{\Px}[1]{P_{\rm #1}}
\newcommand{\px}[1]{p_{\rm #1}}

\newcommand{\vx}[1]{v_{\rm #1}}
\newcommand{\rhox}[1]{\rho_{\rm #1}}
\newcommand{\Nx}[1]{N_{\rm #1}}
\newcommand{\nx}[1]{n_{\rm #1}}
\newcommand{\fx}[1]{f_{\rm #1}}
\newcommand{\chix}[1]{\chi_{\rm #1}}
\newcommand{\Thetax}[1]{\Theta_{\rm #1}}
\newcommand{\thetax}[1]{\theta_{\rm #1}}
\newcommand{\Max}[1]{\mathcal{M}_{\rm #1}} 
\newcommand{\Lambdax}[1]{\Lambda_{\rm #1}}
\newcommand{\lambdax}[1]{\lambda_{\rm #1}}
\newcommand{\mux}[1]{\mu_{\rm #1}}
\newcommand{\zx}[1]{z_{\rm #1}}
\newcommand{\Zx}[1]{Z_{\rm #1}}

\newcommand{\lsh}{\ell_{\rm shatter}}

\title[Stream Penetration]{Cold Stream Penetration of Virial Shocks: Fragmentation, Coagulation, and Disruption in the Hot Circumgalactic Medium}

\author[Yao et al.]{
Zhiyuan Yao$^{1}$\thanks{E-mail: zhiyuan.yao@mail.huji.ac.il},
Nir Mandelker$^{1,2}$,
S. Peng Oh$^{3}$
\\
$^{1}$Racah Institute of Physics, The Hebrew University, Jerusalem 91904, Israel\\
$^{2}$University of Washington, Department of Astronomy, Seattle, WA 98195, USA\\
$^{3}$Department of Physics, University of California, Santa Barbara, CA 93106, USA\\
}
\date{Accepted XXX. Received YYY; in original form ZZZ}

\pubyear{\the\year}

\begin{document}
\label{firstpage}
\pagerange{\pageref{firstpage}--\pageref{lastpage}}
\maketitle

\begin{abstract}
Cold streams penetrating virial shocks of massive halos along cosmic web filaments are expected to fuel galaxy growth at high redshift, yet the physical processes governing their penetration remain uncertain. We investigate cylindrical cold streams penetrating a hot circumgalactic medium (CGM) using idealized three-dimensional simulations. We systematically vary the stream radius, Mach number, and initial stream-to-CGM pressure contrast across three density contrasts, while controlling stream properties after pressure equilibrium is re-established. We identify three evolutionary regimes: coagulation, fragmentation, and disruption, plus a borderline regime in which the stream core marginally survives while detached fragments are disrupted. At modest pressure contrast, survival depends primarily on the competition between velocity shear and radiative cooling. Increasing pressure contrast produces a transient response during pressure restoration, temporarily enhancing or suppressing cold-gas mass and cold-hot interfacial area before evolution converges to a shear-dominated state. At larger pressure contrasts, the oblique shock steepens into a bow shock, and the final outcome is determined by the ratio of post-shock cooling time to virial crossing time. In all survival cases, post-equilibration evolution is well described by turbulent radiative entrainment at the stream-CGM interface: cold-gas mass flux increases while mean streamwise momentum flux remains approximately conserved. Applying these results to a galaxy evolution framework, we find that cold streams are expected to penetrate virial shocks around massive halos at $z>2$, though they may fragment into multiphase structures within the hot CGM at $\Mx{v}\gtrsim\SI{2e12}{\Msun}$. Toward lower redshift, bow-shock heating and inefficient post-shock cooling can suppress cold-stream penetration.
\end{abstract}



\section{Introduction}
\label{sec:intro}

\begin{figure*}
    \centering	\includegraphics[width=\textwidth]{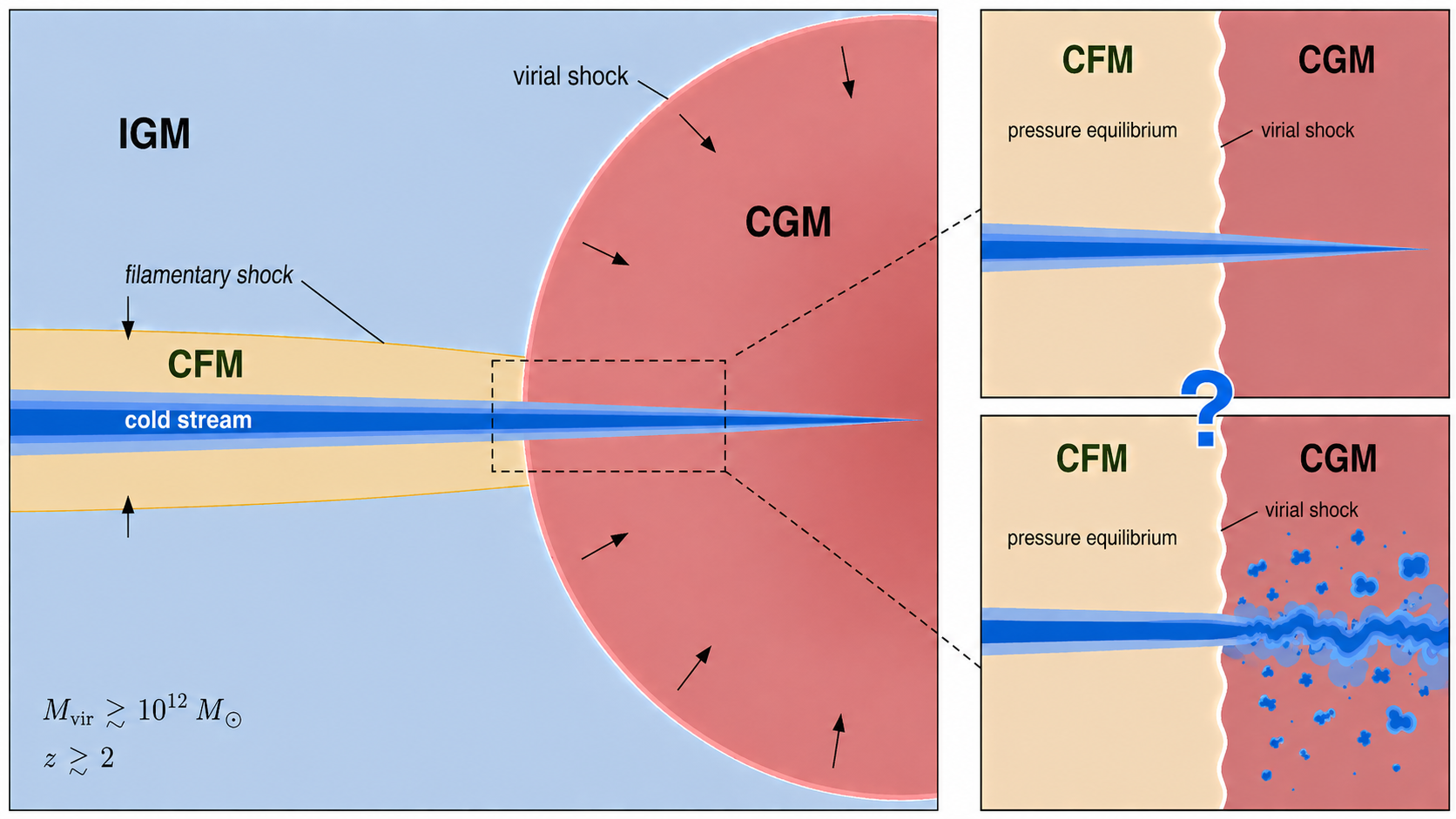}
   \caption{A schematic illustration highlighting the fate of cold gas as a cold stream penetrates the virial shock. In this example, radiative cooling allows the stream to survive its interaction with the hot CGM. An open question, however, is whether such streams can seed the population of cold clumps observed in the CGM. The strong pressure contrast between the hot halo gas and the incoming, unshocked stream, arising from different virialized properties between galactic halos and intergalactic filaments, may drive fragmentation and promote the formation of cold structures.} 
   \label{fig:cartoon}
\end{figure*}

A central puzzle in galaxy formation is how galaxies at high redshift sustain their rapid growth despite the presence of virial shocks and energetic feedback. Observations indicate that the circumgalactic medium (CGM) around star-forming galaxies at $z\sim2$--$3$ contains a substantial reservoir of cool gas, with large covering fractions inferred from absorption-line studies at $\sim 0.1$--$1\,R_{\rm v}$ \citep[e.g.][]{Steidel.etal.10,Rudie.etal.12,Prochaska.etal.13}. Complementary evidence from emission-line observations comes from the widespread detection of extended Ly$\alpha$ halos around galaxies at $z\gtrsim2$, indicating that cool, radiatively emitting gas occupies large volumes around these galaxies \citep[e.g.][]{Cantalupo.etal.14,Wisotzki.etal.16,Leclercq.etal.17,Wisotzki.etal.18}. Together, these data support a picture in which cool ($T\lesssim 10^{4.5} {\rm K}$) CGM phases are common at high redshift, and they motivate theoretical efforts to understand how this gas can survive and remain spatially extended while embedded in a hotter halo atmosphere \citep[for reviews, see][]{Tumlinson.etal.17,Faucher-Giguere.Oh.23,Gronke.Schneider.26}.

\smallskip
Several mechanisms have been proposed to explain this abundant cool gas, but their relative contributions remain uncertain. Galactic winds driven by stellar feedback can carry cool interstellar gas into the halo \citep{Schneider.etal.20,Fielding.Bryan.22}, while radiative cooling of initially hot outflows can form additional cool material \citep{Thompson.etal.16}. In halos containing a hot atmosphere, local thermal instability offers another route to cool-gas formation through condensation, regulated by the competition between radiative cooling and buoyancy in the stratified gas \citep{McCourt.etal.12,Sharma.etal.12,Voit.etal.17}. These channels can coexist.

\smallskip
In the standard framework of galaxy formation, a significant fraction of this cool CGM is supplied by filamentary ``cold streams'' that penetrate hot halos and deliver fresh fuel to galaxies \citep[e.g.][]{Birnboim.Dekel.03,Keres.etal.05,Dekel.etal.09}. A longstanding concern for this scenario is that cold streams are subject to strong hydrodynamic instabilities and turbulent mixing with the ambient hot medium, which could in principle disrupt the stream before it reaches the central galaxy. In particular, velocity shear at the stream--CGM interface excites the Kelvin--Helmholtz instability (KHI), whose nonlinear evolution depends sensitively on the flow regime, geometry, and the efficiency of radiative cooling\footnote{As well as other processes such as self-gravity \citep{Aung.etal.19}, magnetic fields, and thermal conduction (see discussion in \citealt{Mandelker.etal.20a}).} in the turbulent mixing layer \citep[e.g.][]{Mandelker.etal.16,Mandelker.etal.19,Mandelker.etal.20a,Padnos.etal.18}.

\smallskip
The survival of radiatively cooling cold streams in the presence of shear has been investigated in recent work \citep{Mandelker.etal.20a,Mandelker.etal.20b,Aung.etal.24}. Building on earlier studies of radiatively cooling clouds embedded in hot winds \citep[e.g.][]{Gronke.Oh.18,Gronke.Oh.20a,Banda-Barragan.etal.21,Antipov.etal.25}, these works show that efficient cooling within the turbulent mixing layer can prevent stream disruption by Kelvin--Helmholtz instabilities, and may even promote net condensation of hot CGM gas onto the cold phase, provided that the stream radius exceeds a critical threshold. As a result, many cold streams are expected to survive under the physical conditions characteristic of high-redshift halos. The remaining challenge, however, is not merely the survival of the stream core, but the redistribution of cold gas from narrow, coherent streams into the extended CGM volume implied by observations, as illustrated schematically in \cref{fig:cartoon}.

\smallskip
Recently, \citet{Yao.etal.25} proposed a distinct channel for generating small-scale cold structure through \emph{pressure-driven} fragmentation. In that picture, a cold cloud or stream that is initially underpressurized relative to its surroundings undergoes an implosion followed by a rebound, during which impulsive acceleration of the interface can excite the Richtmyer--Meshkov instability (RMI) and promote fragmentation \citep[see also][]{McCourt.etal.18,Gronke.Oh.20b}. This mechanism is particularly relevant in environments where cold gas is subjected to strong pressure contrasts, such as when accreting material encounters the virial shock commonly present near the virial radius of massive halos.

\smallskip
However, an important caveat of the analysis in \citet{Yao.etal.25} is that it considered pressure-driven evolution in the absence of velocity shear. Moreover, in their fiducial setup the effective pressure contrast arose from thermal instability rather than being imposed by a supersonic shock. In realistic cold-stream penetration, pressure contrasts and velocity shear coexist. A stream entering the halo is expected to be both underpressurized relative to the hot CGM and moving at substantial speed. Thus, the same event that produces a pressure jump also establishes strong velocity shear, promoting shear-driven mixing that can influence the survival of fragments.

\smallskip
In this work, we isolate and quantify the role of the pressure contrast experienced by a cold stream as it penetrates a galactic halo, while explicitly accounting for velocity shear. Using idealized three-dimensional simulations of a cylindrical cold stream interacting with a hot CGM, we systematically vary the final stream size in the CGM, \ifm{\rx{s,f}}, the final density contrast between the cold stream and the hot CGM, \ifm{\chix{f}\equiv\rhox{s,f}/\rhox{h}} with $\rho_{\rm s,f}$ and $\rho_{\rm h}$ denoting, respectively, the densities of the cold stream and the hot CGM when they are in pressure equilibrium, the initial pressure contrast between the CGM and cold stream before it enters the halo, \ifm{\mathcal{P} \equiv P_{\rm h}/P_{\rm s,i}}, and the stream Mach number relative to the ambient CGM, \ifm{\Max{s}\equiv\vx{s,i}/\cx{h}} where $v_{\rm s,i}$ is the stream velocity and $c_{\rm h}$ is the sound speed in the hot CGM, while holding other post-equilibration stream properties fixed. Across the region of parameter space relevant for high-redshift halos, we identify three distinct regimes that govern the evolution of the cold-stream mass and morphology. In the fragmentation regime, clumps forming around the cold stream are long-lived; in the coagulation regime they are short-lived and quickly merge back onto the cold stream core; and in the disruption regime clumps are destroyed shortly after the stream fragments. We find that, at moderate values, the pressure contrast primarily acts as an amplifier of the processes driving the evolution of the cold-gas mass and the cold--hot interface during the transient restoring phase preceding pressure equilibrium. After equilibrium is established, the evolution closely resembles that of a shear-dominated, radiatively cooled flow. At sufficiently large pressure contrasts, however, the oblique shock driven into the stream steepens into a bow shock, and survival is instead governed by whether the shock-heated gas can cool within the halo-crossing time. We discuss the implications of this behavior for the prevalence and structure of cool CGM gas in massive halos at high redshifts.

\smallskip
The rest of this paper is structured as follows. In \cref{sec:theory}, we develop a theoretical framework that defines three evolutionary regimes based on two physically motivated criteria. In \cref{sec:methods}, we describe the numerical methods and simulation setup. The results are presented in \cref{sec:size,sec:pres,sec:shear}, where we systematically explore the dependence on stream size, pressure contrast, and Mach number at three fixed density contrasts, varying one parameter at a time while holding the others constant. In \cref{sec:discuss}, we discuss the physical implications of our findings and present caveats to our analysis. Finally, we summarize our results in \cref{sec:summary}.

\section{Theoretical framework}
\label{sec:theory}
In this section we review the theoretical framework relevant to cold-stream penetration in the presence of radiative cooling. Two physical processes are central. The first is the velocity shear between the cold stream and the surrounding CGM, which triggers KHI and governs whether the stream survives or is disrupted. The second is the pressure contrast experienced by the stream as it penetrates the halo, which can excite RMI and can thereby determine the morphology of the stream, provided it survives. We discuss these two criteria in detail below.

\subsection{The survival criterion}
Velocity shear between the cold stream and the hot CGM drives KHI \citep{Mandelker.etal.16,Padnos.etal.18,Mandelker.etal.19}. The evolution of KHI depends heavily on the flow regime and geometry. In subsonic and transonic flows, surface modes dominate, manifesting as a growing shear layer characterized by the classic ``cat's-eye'' vortex morphology. In the supersonic regime, the behavior diverges based on dimensionality. While surface modes stabilize in 2D slab geometries, leaving body modes as the dominant driver of instability, high-azimuthal-wavenumber (\ifm{m}) surface modes remain unstable in 3D cylinders. Consequently, the dominant mode in 3D depends on the initial width of the interface transition region; a sufficiently wide interface suppresses these high-\ifm{m} surface modes. For the Mach numbers considered here, surface modes are expected to dominate at \ifm{\Max{s}<1}; at \ifm{\Max{s}=1} the flow lies near the threshold for exciting unstable body modes, and at \ifm{\Max{s}=2} body modes may contribute substantially.

\smallskip
However, the stream's fate is not governed by non-radiative hydrodynamics alone. \citet{Mandelker.etal.20a} investigated KHI in the presence of radiative cooling and a UV background, identifying the ratio of the cooling timescale of the mixed gas, $t_\mathrm{cool,mix}$, to the shear disruption timescale, $t_\mathrm{shear}$, as the decisive parameter. This ratio defines a critical stream radius:
\begin{equation} 
    \label{eq:Rshear}
    \rx{crit,surv}\simeq\SI{0.3}{\kpc}\,\alpha_{0.1}\chix{f,100}^{3/2}\Max{s}\frac{\Tx{s,4}}{\nx{s,f,0.01}\Lambdax{mix,-22.5}},
\end{equation}
where \ifm{\alpha_{0.1}=\alpha/0.1}, \ifm{\chix{f,100}=\chix{f}/100}, \ifm{\Tx{s,4}=\Tx{s}/\SI{1e4}{K}}, \ifm{\nx{s,f,0.01}=\nx{s,f}/\SI{0.01}{cm^{-3}}}, \ifm{\Lambdax{mix,-22.5}=\Lambdax{mix}/10^{-22.5}\,\SI{}{erg\,s^{-1}\,cm^3}}. $T_{\rm s}$ is the stream temperature, $n_{\rm s,f}$ is its hydrogen number density after pressure equilibrium is re-established, and $\Lambda_{\rm mix}$ is the cooling function in the mixing layer, assumed to have characteristic density and temperature equal to the geometric mean of the stream and the CGM, $\rho_{\rm mix}=(\rho_{\rm s,f}\rho_{\rm h})^{1/2}$ and $T_{\rm mix}=(T_{\rm s}T_{\rm h})^{1/2}$.  The dimensionless shear-layer growth coefficient is \ifm{\alpha\simeq0.21\times\left[0.8\,\mathrm{exp}(-3\Max{tot}^2)+0.2\right]}, with \ifm{\Max{tot}=\vx{s,i}/(\cx{s}+\cx{h})} \citep{Dimotakis.91}. If the final stream radius after reaching pressure equilibrium, $\rx{s,f}$, is smaller than the critical survival radius, $\rx{crit,surv}$, the stream is gradually disrupted by Kelvin--Helmholtz instability and loses mass to the CGM. Conversely, if $\rx{s,f}>\rx{crit,surv}$, radiative cooling of the entrained background gas outpaces shear-driven disruption, resulting in net mass growth of the cold stream.

\subsection{The fragmentation criterion}
An implicit assumption underlying the survival criterion derived above is pressure equilibrium between the CGM and the cold stream. In practice, this condition is not always satisfied during infall. In massive halos, a stable virial shock is expected for \ifm{M_{\rm v}\gtrsim M_{\rm shock}\sim\SI{6e11}{\Msun}} \citep{Birnboim.Dekel.03,Keres.etal.05,Dekel.Birnboim.06,vandevoort.etal.11,Fielding.etal.17,Stern.etal.20}, implying a sudden increase in external pressure as gas enters the halo. However, the cold streams that form the central spine of cosmic filaments at \ifm{z\gtrsim2} \citep[e.g.][]{Ramsoy.etal.21,Lu.etal.24} are typically not shock-heated at the virial radius. Owing to their high densities, short cooling times, and large inflow velocities, these streams can efficiently penetrate the virial shock \citep{Dekel.etal.09}. Nevertheless, upon entering the halo, the streams encounter a sudden increase in confining pressure from the hot CGM. This pressure contrast drives a cylindrical shock that propagates inward and subsequently reflects off the stream axis. The interaction between this reflected shock and the contact discontinuity at the stream interface induces the RMI \citep{Richtmyer.60,Meshkov.69,Zhou.17a,Zhou.17b} through baroclinic vorticity generation, producing mushroom-like structures that fragment into small cloudlets. \citet{Yao.etal.25} investigated the effects of RMI on such streams, accounting for radiative cooling and a UV background in the absence of shear. They identified a competition between this fragmentation and the subsequent coagulation of cloudlets due to entrainment from the surrounding hot gas. Their analysis yields a critical condition for sustained fragmentation, which can be written as a critical stream radius once the stream regains pressure equilibrium:
\begin{equation}
    \label{eq:Rcoag} 
    \rx{crit,frag}\simeq\SI{5}{\kpc}\,\chix{f,100}^4\ell_\mathrm{shatter,200}, 
\end{equation}
where \ifm{\ell_\mathrm{shatter,200}=\lsh/\SI{200}{pc}} and \ifm{\lsh=\min(\cx{s}\tx{cool})\simeq\SI{200}{pc}} for \ifm{Z=0.03\,Z_{\sun}} under a \ifm{z=2} UV background (UVB). If $\rx{s,f}<\rx{crit,frag}$, the stream remains fragmented. Conversely, if \ifm{\rx{s,f}>\rx{crit,frag}}, cooling and entrainment of the surrounding hot gas overwhelm fragment dispersion, leading to coagulation within a few sound-crossing times.

\begin{figure}
    \centering	\includegraphics[width=\columnwidth]{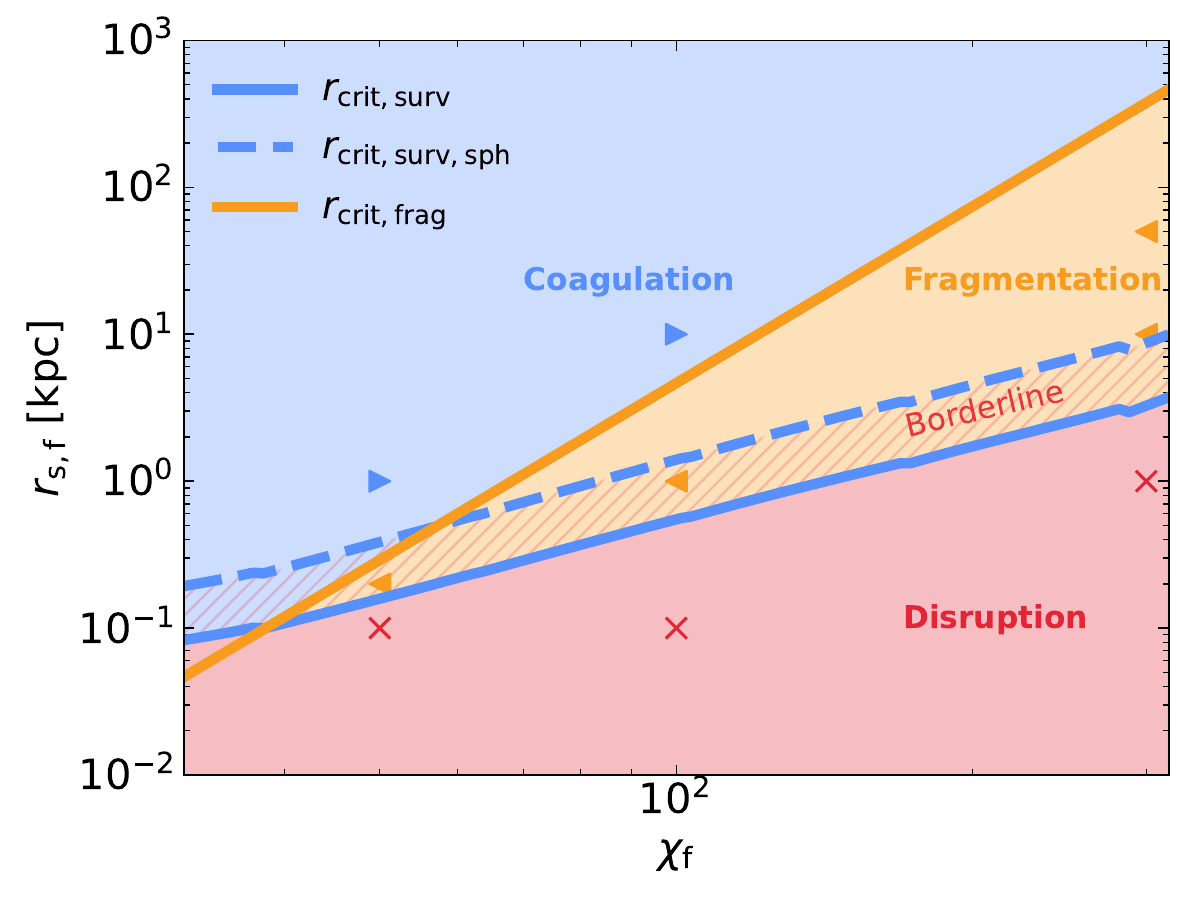}
   \caption{Regimes of stream evolution in the CGM subject to shear and a pressure contrast. We show results in the $\rx{s,f}$--$\chix{f}$ plane for a fixed stream Mach number $\Max{s}=1$, metallicity \ifm{Z=0.03\,Z_{\sun}}, and final stream hydrogen number density (after pressure equilibrium is re-established) \ifm{\nx{s,f}=\SI{0.01}{cm}^{-3}}. The blue and orange curves denote the critical radii for survival and fragmentation, given by \cref{eq:Rshear} and \cref{eq:Rcoag}, respectively. Together, these criteria partition the parameter space into three regimes: coagulation, fragmentation, and disruption, indicated by the blue, orange, and red shaded regions. The blue dashed line denotes the spherical survival criterion, \cref{eq:Rsurv_sph}, from \citet{Abruzzo.etal.23}, evaluated under the same assumptions. The red hatched region denotes the parameter space between the cylindrical and spherical survival criteria. In this borderline regime, newly formed fragments are rapidly disrupted while the stream core remains intact, making the distinction between coagulation and fragmentation less well defined. The nine simulations in the stream-size subset of \cref{tab:sim_table} are marked by blue right-pointing triangles (coagulation), orange left-pointing triangles (fragmentation or borderline), and red crosses (disruption).} 
   \label{fig:diag_size_chi}
\end{figure}

\subsection{Combining the survival and fragmentation criteria}

Cold-stream penetration of the virial shock is governed jointly by velocity shear and pressure contrast. In \cref{fig:diag_size_chi}, we present the two associated critical radii in the $\rx{s,f}$--$\chix{f}$ plane, adopting a metallicity of \ifm{0.03\,Z_{\sun}}, a stream density of \ifm{\nx{s,f}=\SI{0.01}{cm}^{-3}}, and a Mach number of \ifm{\Max{s}=1}. The diagram is insensitive to the exact pressure contrast \ifm{\mathcal{P}}, provided it is sufficiently large to trigger the RMI but below the bow-shock threshold discussed in \cref{sec:pres}. The resulting critical curves divide the parameter space into three regimes: coagulation, fragmentation, and disruption, indicated by blue, orange, and red regions, respectively. Both the coagulation and fragmentation regimes lie within the broader survival domain of the cold stream, as defined by the shear-driven survival criterion. For reference, we also plot the spherical survival criterion derived by \citet{Abruzzo.etal.23},
\begin{equation}
\label{eq:Rsurv_sph}
    \rx{crit,surv,sph}\simeq\SI{0.43}{\kpc}\,\chix{f,100}^{3/2}\Max{s}\frac{\Tx{s,4}}{\nx{s,f,0.01}\Lambdax{mix,-22.5}},
\end{equation}
which roughly lies a factor of two above the cylindrical stream criterion. This offset provides a useful benchmark for assessing the survivability of individual clumps produced by fragmentation. In particular, it defines a region of parameter space between the two survival criteria where the cold stream core can persist as a coherent structure, while individual fragments are unstable. We designate this intermediate region as a ``borderline'' regime. In this regime, fragmentation may occur transiently, but the resulting clumps are rapidly disrupted. As a consequence, the physical distinction between the coagulation and fragmentation outcomes becomes ambiguous.

\smallskip
This highlights the importance of explicitly comparing the timescales associated with the relevant physical processes. The two criteria introduced above are derived assuming sufficiently long evolution times. In practice, however, the outcome can differ if the available time is limited, particularly at high Mach numbers where the virial crossing time is short. 

\smallskip
For a spherical cloud, the disruption timescale is a few times the cloud-crushing time, \ifm{\tx{cc}=\chix{f}^{1/2}\rx{s}/\vx{s,i}}. For a cylindrical stream, the disruption timescale is expected to be somewhat longer because its volume-to-area ratio exceeds that of a sphere. Nevertheless, once turbulent mixing develops, the stream first breaks into segments and then disrupts in a manner broadly analogous to spherical clouds, as demonstrated below. It is therefore reasonable to adopt the cloud-crushing timescale as the characteristic disruption timescale for streams as well. The fragmentation and coagulation processes, by contrast, are governed by the sound-crossing time of the cold stream, \ifm{\tx{sc}=\rx{s}/\cx{s}=\Max{s}\tx{cc}}, where \ifm{\Max{s}=\vx{s,i}/\cx{h}} and \ifm{\cx{h}=\chix{f}^{1/2}\cx{s}}. The entrainment timescale of the cold stream is further regulated by radiative cooling and can be expressed as \ifm{\tx{ent}=\mx{s}/\dotm{s}}, where \ifm{\dotm{s}=\rhox{h}\vx{mix}\Ax{s}} and \ifm{\vx{mix}\sim0.4\cx{s}(\rx{s}/\lsh)^{1/4}} \citep[e.g.][]{Gronke.Oh.20a,Yao.etal.25}. For cylindrical geometry, this yields \ifm{\tx{ent}\sim\chix{f}(\rx{s}/\lsh)^{-1/4}\tx{sc}}. 

\smallskip
To assess whether these processes can fully develop before the stream reaches the central galaxy, all relevant timescales should be compared to the virial crossing time, \ifm{\tx{v}=\Rx{v}/\vx{s,i}}, where \ifm{\Rx{v}} is the virial radius of the halo. This yields
\begin{equation}
\label{eq:tcc2tv}
    \frac{\tx{cc}}{\tx{v}}=\frac{\chix{f}^{1/2}\rx{s}}{\Rx{v}},
\end{equation}
\begin{equation}
\label{eq:tsc2tv}
    \frac{\tx{sc}}{\tx{v}}=\frac{\Max{s}\chix{f}^{1/2}\rx{s}}{\Rx{v}},
\end{equation}
and 
\begin{equation}
\label{eq:tent2tv}
    \frac{\tx{ent}}{\tx{v}}=\frac{\Max{s}\chix{f}^{3/2}\rx{s}^{3/4}\lsh^{1/4}}{\Rx{v}}.
\end{equation}
At sufficiently large density contrasts \ifm{\chix{f}}, these three processes---disruption, entrainment, and fragmentation---can become inefficient, allowing the cold stream to remain largely intact until it feeds the central galaxy. Moreover, a large stream Mach number \ifm{\Max{s}} further delays the dynamical response to the pressure contrast, so that the cold stream could remain underpressurized throughout its infall and reach the central galaxy before pressure equilibrium is established. In this regime, the distinction between coagulation and fragmentation loses physical significance. The halo potential may further increase the departure from pressure equilibrium by compressing the infalling stream and increasing the ambient CGM pressure along its trajectory. We will discuss these timescales in a galaxy evolution framework in \cref{sec:discuss}.

\begin{table*}
   \centering
   \caption{Summary of the simulations analyzed in this work. From left to right, we list the run name, constructed from the final cold-stream radius, final density contrast, initial pressure contrast, and stream Mach number; the initial stream radius at the injection boundary, \ifm{\rx{s,i}}, in kpc; the final stream radius after reaching pressure equilibrium with the background, \ifm{\rx{s,f}}, in kpc; the density contrast between the injected stream and the background, \ifm{\chix{i}\equiv \rhox{s,i}/\rhox{h}}; the final density contrast after the stream reaches pressure equilibrium with the background gas, \ifm{\chix{f}\equiv \rhox{s,f}/\rhox{h}}; the ratio of the final to injected stream density, \ifm{\eta\equiv \rhox{s,f}/\rhox{s,i}=\chix{f}/\chix{i}}; the pressure contrast between the hot background and the injected cold stream, \ifm{\mathcal{P}\equiv \Px{h}/\Px{s,i}}; the Mach number of the injected stream relative to the background, \ifm{\Max{s}\equiv \vx{s,i}/\cx{h}}; the resulting evolutionary regime; and the section where the simulation is discussed. The labels ``C'', ``F'', ``B'', and ``D'' denote coagulation, fragmentation, borderline, and disruption, respectively. Depending on which parameters are varied, we group the simulations into three subsets: stream size, pressure contrast, and Mach number. Within each subset, we further vary the final density contrast, \ifm{\chix{f}=50,100,300}, to more fully sample the parameter space. Note that the fiducial run with \ifm{\rx{s,f}=\SI{1}{kpc}}, \ifm{\mathcal{P}=7}, and \ifm{\Max{s}=1} at each \ifm{\chix{f}} is repeated across subsets.
   }
   \begin{tabular*}{0.95\textwidth}{@{\extracolsep{\fill}}cccccccccc}
       \hline
       Name & \ifm{\rx{s,i}} [kpc]& \ifm{\rx{s,f}} [kpc] & \ifm{\chix{i}} & \ifm{\chix{f}} & \ifm{\eta} & \ifm{\mathcal{P}} & \ifm{\Max{s}} & Regime & Section \\
       \hline
       \multicolumn{10}{c}{\textbf{Stream Sizes}} \\
       \run{R01D50P7M1} & 0.3  & 0.1 & 5  &50  & 10 &  7.2 &  1 & D & \ref{sec:size}\\
       \run{R02D50P7M1} & 0.6  & 0.2 & 5  &50  & 10 &  7.2 &  1 & B & \ref{sec:size}, \ref{sec:discuss}\\
       \run{R1D50P7M1} & 3.2  & 1   & 5  &50  & 10 &  7.2 &  1 & C & \ref{sec:size}, \ref{sec:discuss}\\
       \run{R01D100P7M1} & 0.3  & 0.1 & 10 &100 & 10 &  7.2 &  1 & D & \ref{sec:size}\\
       \run{R1D100P7M1} & 3.2  & 1   & 10 &100 & 10 &  7.2 &  1 & B & \ref{sec:size}, \ref{sec:discuss}\\
       \run{R10D100P7M1} & 31.6 & 10  & 10 &100 & 10 &  7.2 &  1 & C & \ref{sec:size}, \ref{sec:discuss}\\     
       \run{R1D300P7M1} & 3.2  & 1   & 30 &300 & 10 &  7.2 &  1 & D & \ref{sec:size}\\
       \run{R10D300P7M1} & 31.6 & 10  & 30 &300 & 10 &  7.2 &  1 & B & \ref{sec:size}, \ref{sec:discuss}\\
       \run{R50D300P7M1} & 158.1& 50  & 30 &300 & 10 &  7.2 &  1 & F & \ref{sec:size}, \ref{sec:discuss}\\ 
       \multicolumn{10}{c}{\textbf{Pressure Contrasts}} \\
       \run{R1D50P1M1} & 1 & 1   & 50   &50  & 1 &  1   &  1 & C  & \ref{sec:pres}, \ref{sec:discuss}\\
       \run{R1D50P7M1} & 3.2  & 1   & 5  &50  & 10 &  7.2 &  1 & C & \ref{sec:pres}, \ref{sec:discuss}\\
       \run{R1D50P18M1} & 5.5 & 1   & 1.7 &50  & 30 &  18.4&  1 & C & \ref{sec:pres}, \ref{sec:discuss} \\
       \run{R1D100P1M1} & 1   & 1   & 100  &100 & 1 &  1   &  1 & B & \ref{sec:pres}, \ref{sec:discuss}\\
       \run{R1D100P7M1} & 3.2  & 1   & 10 &100 & 10 &  7.2 &  1 & B & \ref{sec:pres}, \ref{sec:discuss}\\
       \run{R1D100P18M1} & 5.5 & 1   & 3.3 &100 & 30 &  18.4&  1 & B & \ref{sec:pres}, \ref{sec:discuss} \\
       \run{R1D100P30M1} & 7.1 & 1   & 2   &100 & 50 &  29.5&  1 & B & \ref{sec:pres}, \ref{sec:discuss} \\
      \run{R1D300P1M1}  & 1  & 1   & 300  & 300 & 1 &  1   &  1 & D & \ref{sec:pres} \\
       \run{R1D300P7M1} & 3.2  & 1   & 30 &300 & 10 &  7.2 &  1 & D & \ref{sec:pres} \\
       \run{R1D300P18M1} & 5.5 & 1   & 10 &300 & 30 &  18.4&  1 & D & \ref{sec:pres} \\
      \run{R10D300P100M1} & 143.1 & 10   & 1.5 &300 & 205 &  100&  1 & D & \ref{sec:pres} \\
       \multicolumn{10}{c}{\textbf{Mach Numbers}} \\
       \run{R1D50P7M05} & 3.2 & 1 & 5  &50  & 10 & 7.2 &  0.5 & C  & \ref{sec:shear}, \ref{sec:discuss}\\
       \run{R1D50P7M1} & 3.2  & 1 & 5  &50  & 10 &  7.2 &  1 & C & \ref{sec:shear}, \ref{sec:discuss}\\
       \run{R1D50P7M2} & 3.2 & 1 & 5  &50  & 10 & 7.2 &  2   & B & \ref{sec:shear}, \ref{sec:discuss}\\
       \run{R1D100P7M01} & 3.2 & 1   & 10 &100 & 10 & 7.2 &  0.1 & F & \ref{sec:shear}, \ref{sec:discuss} \\
       \run{R1D100P7M05} & 3.2 & 1   & 10 &100 & 10 & 7.2 &  0.5 & F & \ref{sec:shear}, \ref{sec:discuss} \\
       \run{R1D100P7M1} & 3.2  & 1   & 10 &100 & 10 &  7.2 &  1 & B & \ref{sec:shear}, \ref{sec:discuss}\\
       \run{R1D100P7M2} & 3.2 & 1   & 10 &100 & 10 & 7.2 &  2   & D  & \ref{sec:shear}\\
       \run{R1D300P7M05} & 3.2 & 1   & 30 &300 & 10 & 7.2 &  0.5 & B  & \ref{sec:shear}, \ref{sec:discuss}\\
       \run{R1D300P7M1} & 3.2  & 1   & 30 &300 & 10 &  7.2 &  1 & D & \ref{sec:shear}\\
       \run{R1D300P7M2} & 3.2 & 1   & 30 &300 & 10 & 7.2 &  2   & D  & \ref{sec:shear}\\
       \hline
   \end{tabular*}
   \label{tab:sim_table}
\end{table*}


\section{Numerical methods}
\label{sec:methods}
In this section, we summarize the numerical simulations employed in this work. Most aspects of the setup, including radiative cooling, the clump-finding algorithm, and the imposed initial shape perturbations, closely follow \citet{Yao.etal.25}. We therefore provide only a concise overview here and refer the reader to \citet{Yao.etal.25} for a comprehensive description. The principal differences concern the boundary and initial conditions, which are described in detail in \cref{subsec:ICBC}.

\subsection{Simulation setup}
We perform three-dimensional idealized numerical simulations using the Eulerian adaptive mesh refinement (AMR) code \software{RAMSES}\footnote{Git commit hash: ebcb676} \citep{Teyssier.02}. The simulation domain is a cubic box of size \ifm{L_\mathrm{box} \approx 200\,\rx{s,f}}. The initial cold stream, extending along the \ifm{z}-direction within the box, has an initial radius \ifm{\rx{s,i}} and a temperature set by thermal equilibrium with the UVB. It is underpressurized relative to the surrounding hot CGM.

\smallskip
We employ a statically refined mesh with progressively increasing resolution toward a set of central cylindrical regions enclosing the initial cold stream. The maximum resolution is \ifm{L_\mathrm{box}/2048\simeq0.1\,\rx{s,f}} within \ifm{\rx{cyl}=9.5\,\rx{s,f}}. The cell size increases by a factor of
2 at \ifm{\rx{cyl}=[9.5, 19, 38, 57, 76]\,\rx{s,f}}, reaching a maximal value of \ifm{\sim3\,\rx{s,f}} at \ifm{\rx{cyl}>76\,\rx{s,f}}.

\smallskip
Radiative cooling and heating are modeled using the standard \software{RAMSES} cooling module, which includes atomic and fine-structure cooling, as well as photoheating and photoionization from a UVB \citep{Haardt.Madau.96} fixed at $z=2$ in all idealized simulations. Cooling is disabled for gas with temperatures above $0.8\,\Tx{h}$. This approximation is appropriate when the cooling time of virialized gas exceeds the Hubble time, or when radiative losses are offset by additional heating, either from feedback \citep[e.g.][]{Sharma.etal.12,Voit.etal.24a,Voit.etal.24b} or from compressive heating in a slowly inflowing atmosphere \citep{Stern.etal.19,Stern.etal.24}. We assume metallicities of $\Zx{h}=0.1\,Z_{\sun}$ for the background CGM and $\Zx{s}=0.03\,Z_{\sun}$ for the cold stream. Metallicity is passively advected, with no explicit metal-diffusion term, and the cooling rate is evaluated using the local, mixed metallicity. Cold clump properties are quantified using the built-in clump finder module \software{PHEW} \citep{Bleuler.etal.15}. The clump density threshold is set to the injection density of the stream, while the saddle density threshold is defined as the geometric mean of the injection density and the final stream density. To quantify the cold gas area, we interpolate the gas temperature onto a uniform grid at the highest refinement level using \software{yt} \citep{Turk.etal.11}, and then extract a two-dimensional surface mesh from the three-dimensional volume with the \software{Python} package \software{scikit-image} \citep{vanderWalt.etal.14}. We measure the area of the temperature isosurface corresponding to \ifm{T=\SI{5e4}{K}}.

\smallskip
Initial density perturbations are imposed within the cold stream following a Gaussian distribution with mean of unity and standard deviation 0.01, truncated at $3\sigma$. We use the same random seed across runs, giving the same density-perturbation realization for the same geometry and grid. In addition, we apply shape perturbations at the interface between the cold stream and the hot CGM. Such interface perturbations have
been shown to suppress the carbuncle instability \citep{Bustard.Gronke.22}, a numerical instability affecting strong shock fronts on the grid scale in multidimensional simulations. They do so by misaligning the interface with the grid and by generating additional vorticity and turbulence when the shock overtakes the interface \citep{Quirk.94}. The stream radius varies as
\begin{equation}
    \rx{s}=\rx{s,i}+\sqrt{\frac{2}{N_\mathrm{pert}}}\,\delta \rx{s}\sum_{j=1}^{N_\mathrm{pert}}\cos (k_j z + m_j \varphi + \phi_j)
\end{equation}
The perturbations have an rms amplitude of \ifm{\delta \rx{s} = 0.1\,\rx{s,i}}. The longitudinal wavenumbers are given by \ifm{k_j = 2\pi n_j/L_\mathrm{box}}, where \ifm{n_j} is an integer corresponding to a wavelength \ifm{\lambda_j = L_\mathrm{box}/n_j}. We include all modes in the range \ifm{n_j = 16\text{--}64}, corresponding to wavelengths between \ifm{\rx{s,i}} and \ifm{4\,\rx{s,i}}. The azimuthal mode number \ifm{m_j} characterizes the perturbation geometry, with \ifm{m=0} denoting axisymmetric modes, \ifm{m=1} helical modes, and \ifm{m \ge 2} higher-order fluting modes \citep{Mandelker.etal.16,Mandelker.etal.19}. For each longitudinal wavenumber \ifm{n_j}, we include two azimuthal modes, \ifm{m_j=0} and \ifm{m_j=1}, resulting in a total of \ifm{N_\mathrm{pert}=2\times49=98} modes. Each mode is assigned a random phase \ifm{\phi_j \in [0,2\pi)}. In Appendix~\ref{sec:perturb}, we show that the cold-gas line density, mass flux, and interfacial area of the representative coagulation run \run{R1D50P7M1} are insensitive to the perturbation amplitude, the random phases, and the range of azimuthal modes, so the results presented below do not depend on the specific perturbation realization.

\subsection{Boundary and initial conditions}
\label{subsec:ICBC}

To capture the effects of the shock-heated CGM and stream penetration, we impose a two-zone injection boundary condition at the upstream \ifm{z}-boundary. The interface is located at the cylindrical radius \ifm{\rx{inj}}, measured from the stream axis. Consistent with the initial shape perturbations, this radius depends on azimuth and time:
\begin{equation}
\label{eq:rinj}
    \rx{inj}(\varphi,t)=\rx{s,i}+\sqrt{\frac{2}{N_\mathrm{pert}}}\,\delta \rx{s}\sum_{j=1}^{N_\mathrm{pert}}\cos (k_j \vx{s,i}t + m_j \varphi + \phi_j),
\end{equation}
where the perturbation parameters are identical to those used in the initial conditions. Inside this perturbed aperture, \ifm{\rx{cyl}<\rx{inj}(\varphi,t)}, we continuously inject cold gas with properties \ifm{(\rhox{s,i},\Tx{s},\vx{s,i})}. Outside it, \ifm{\rx{cyl}>\rx{inj}(\varphi,t)}, we impose hot background gas with properties \ifm{(\rhox{h},\Tx{h},0)}. Thus, the two zones are adjacent regions of the same upstream boundary, separated by the perturbed stream surface. The injected stream is underpressurized by a factor \ifm{\mathcal{P}=\eta\mux{s}/\mux{h}}, where \ifm{\eta\equiv\rhox{s,f}/\rhox{s,i}} and $\mu_{\rm s}$ and $\mu_{\rm h}$ are the mean molecular weights in the cold stream and hot CGM, respectively, which can differ by up to a factor of $\sim 2$ due to the different ionization states and metallicities. The stream enters with velocity \ifm{\vx{s,i}}, corresponding to a Mach number relative to the background hot gas of \ifm{\Max{s}\equiv\vx{s,i}/\cx{h}}. This configuration is designed to mimic the halo virial shock in the shock's rest frame. The pre-existing cold stream and the surrounding hot gas within the simulation domain are initialized with the same properties as the injected gas.

\smallskip
At the downstream \ifm{z}-boundary, where the stream exits the domain, and at all four lateral boundaries, we apply inflow/outflow boundary conditions with zero gradients in all fluid properties. The system is evolved for two box-crossing times, \ifm{2\,\tx{box}=2\,L_\mathrm{box}/\vx{s,i}}, and we construct spatial profiles in the selected analysis regions (defined below) by averaging the final ten snapshots over \ifm{1.8\text{--}2\,\tx{box}}. In terms of the sound-crossing time of the cold stream, we have
\begin{equation}
\label{eq:tcross}
    \tx{box}=\frac{L_\mathrm{box}}{\vx{s,i}}\sim\frac{200\,\rx{s,f}}{\Max{s}\chix{f}^{1/2}\cx{s}}\sim\frac{200}{\Max{s}\chix{f}^{1/2}}\tx{sc}.
\end{equation}
Note that in runs with efficient mixing and entrainment, the cold-stream velocity declines rapidly, such that portions of the pre-existing stream remain inside the box even after \ifm{2\,\tx{box}}. These residual components are excluded from the analysis. We refer to the remaining portion of the domain, which excludes the residual pre-existing stream material at large \ifm{z}, as the selected analysis regions. Appendix~\ref{sec:steadystate} shows that most runs approach a statistically steady state within the selected analysis regions.

\smallskip
The full set of simulation parameters is summarized in \cref{tab:sim_table}. In all runs, we fix the final hydrogen number density of the cold stream to \ifm{\nx{s,f}=0.01\,\mathrm{cm}^{-3}}. As a baseline, we define three fiducial runs with \ifm{\rx{s,f}=\SI{1}{kpc}}, \ifm{\mathcal{P}=7}, and \ifm{\Max{s}=1}, spanning three final density contrasts, \ifm{\chix{f}=50,100,300}. Around this fiducial set, we then perform controlled parameter variations by systematically changing the final stream radius \ifm{\rx{s,f}} (\cref{sec:size}), the pressure contrast \ifm{\mathcal{P}} (\cref{sec:pres}), and the stream Mach number \ifm{\Max{s}} (\cref{sec:shear}), while keeping all other parameters fixed.

\smallskip
Movies visualizing the numerical evolution of all simulations are available on our \href{https://zhiyuan-21.github.io/HomePage/stream_penetration.html}{companion webpage}.

\begin{figure*}
    \phantomsection
    \centering	
    \includegraphics[width=\textwidth]{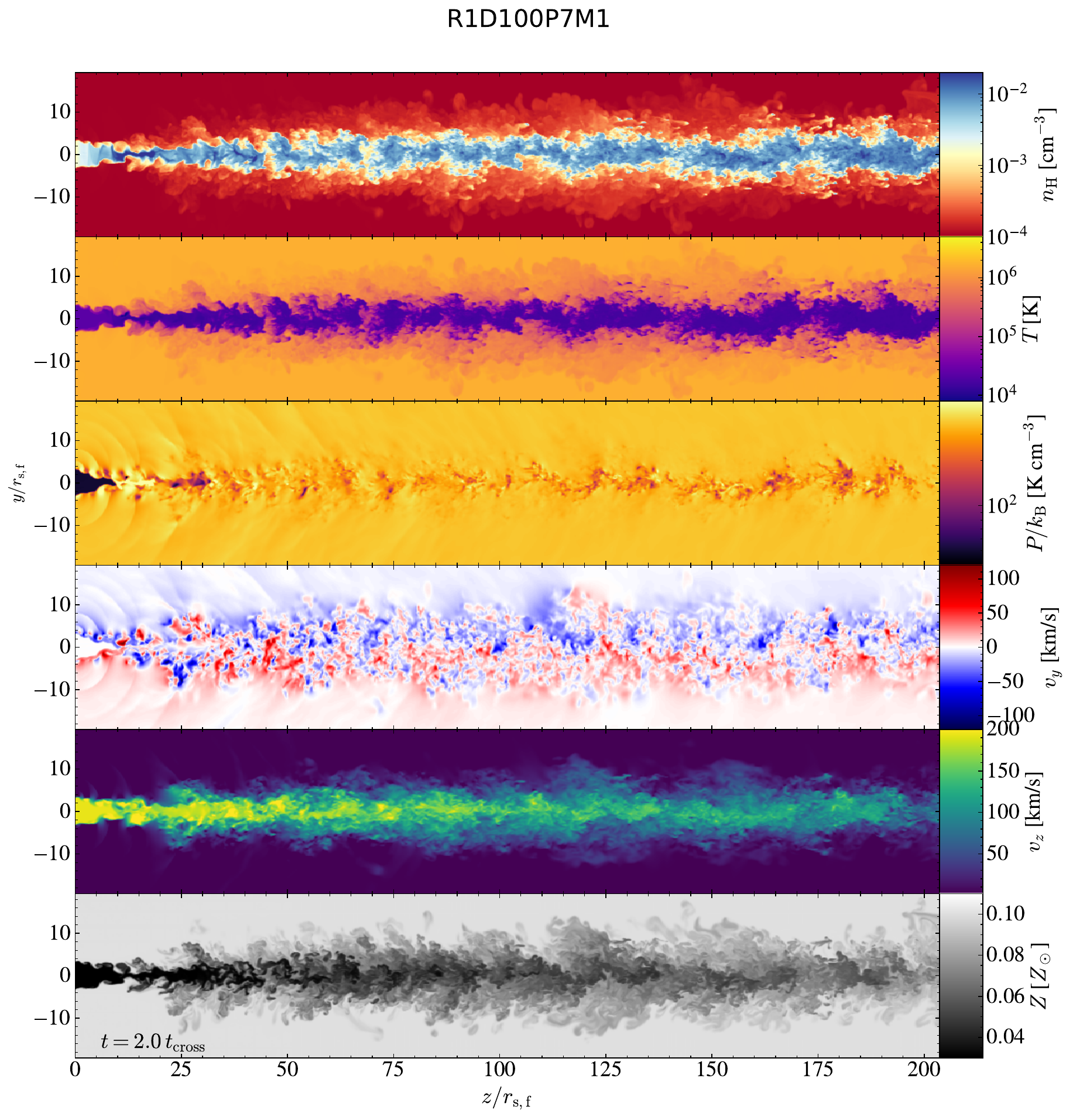}
    \caption{Projection and slice maps along the \ifm{x}-axis at \ifm{t = 2\,\tx{box}} for the fiducial run \run{R1D100P7M1}. From top to bottom, the panels show the hydrogen number density \ifm{\nx{H}}, temperature, pressure, velocity along the \ifm{y}-axis, velocity along the \ifm{z}-axis, and metallicity. We show the maximum and minimum number density and temperature along the whole box, respectively, to highlight the cold clumps. The remaining four panels are slices taken at \ifm{x = 0}. In each panel, the visualization domain spans the full box length in the stream direction (\ifm{z}-axis) and \ifm{0.2\,L_\mathrm{box}} in the perpendicular direction (\ifm{y}-axis).}
    \label{fig:proj_fid}
\end{figure*}

\section{Three regimes of the cold-stream mass and morphology}
\label{sec:size}

\subsection{The fiducial case}
To give a first impression of the simulation, we present projections and slices of several key quantities at the final snapshot, \ifm{t = 2\,\tx{box}}, for the fiducial run \run{R1D100P7M1} in \cref{fig:proj_fid}. The cold stream is continuously injected from the left boundary, with its injection radius perturbed in time, and exits through the right boundary. The injected stream is underpressurized relative to the hot background to mimic the virial shock at the left boundary, as seen in the pressure panel. The resulting inward-propagating cylindrical shock in the Lagrangian frame (or oblique shock in the lab frame), visible in the \ifm{\vx{y}} panel, compresses the stream radially until it reaches a radius \ifm{\rx{s,f}=\rx{s,i}\eta^{-1/2}}, corresponding to restored pressure equilibrium. The associated penetration distance to reach pressure equilibrium can be estimated as 
\begin{equation}
\label{eq:zeq}
    \zx{eq} \sim \frac{\vx{s,i}\rx{s,i}}{\vx{cyl}} \sim \frac{\Max{s}\cx{h}\eta^{1/2}\rx{s,f}}{\mathcal{P}^{1/2}\cx{s}} \sim \parfrac{\eta\chix{f}}{\mathcal{P}}^{1/2}\Max{s}\rx{s,f}, 
\end{equation}
where we assume the initial cylindrical shock speed is \ifm{\vx{cyl}\sim\mathcal{P}^{1/2}\cx{s}}. For the fiducial parameters \ifm{\eta = 10, \mathcal{P} = 7, \chix{f} = 100, \Max{s} = 1}, this yields \ifm{\zx{eq} \sim 12\,\rx{s,f}}, in good agreement with \cref{fig:proj_fid}.

\smallskip
After reaching pressure equilibrium, the stream continues to contract due to inertia until the cylindrical shock converges at the center, and the central stream pressure overshoots the background by a factor of several tens, in the range \ifm{z\simeq10\text{--}20\,\rx{s,f}}. This generates a reflected shock, which reverses the contraction and excites RMI upon impacting the stream surface. In the absence of shear, the stream would therefore be expected to expand and fragment until its cylindrical radius reaches \ifm{\rx{cyl} \sim 6\,\rx{s,f}} so that the number of clumps saturates, as found in the static-stream analysis of \citet{Yao.etal.25}. In contrast, we do not observe significant fragmentation in \cref{fig:proj_fid}, owing to the presence of velocity shear. As shown in \cref{fig:diag_size_chi}, this fiducial case lies in the borderline regime: the cold stream can marginally survive, whereas the fragments cannot. Consequently, the cold stream remains confined to a cylinder of radius \ifm{\rx{cyl} \sim 6\,\rx{s,f}} over a substantial distance, until it decelerates sufficiently so that the survival radius also decreases. We will quantify this behavior in more detail in the following sections.

\subsection{Coagulation, fragmentation, and disruption regimes in cold-stream penetration}

\begin{figure*}
    \centering	
    \includegraphics[width=0.88\textwidth]{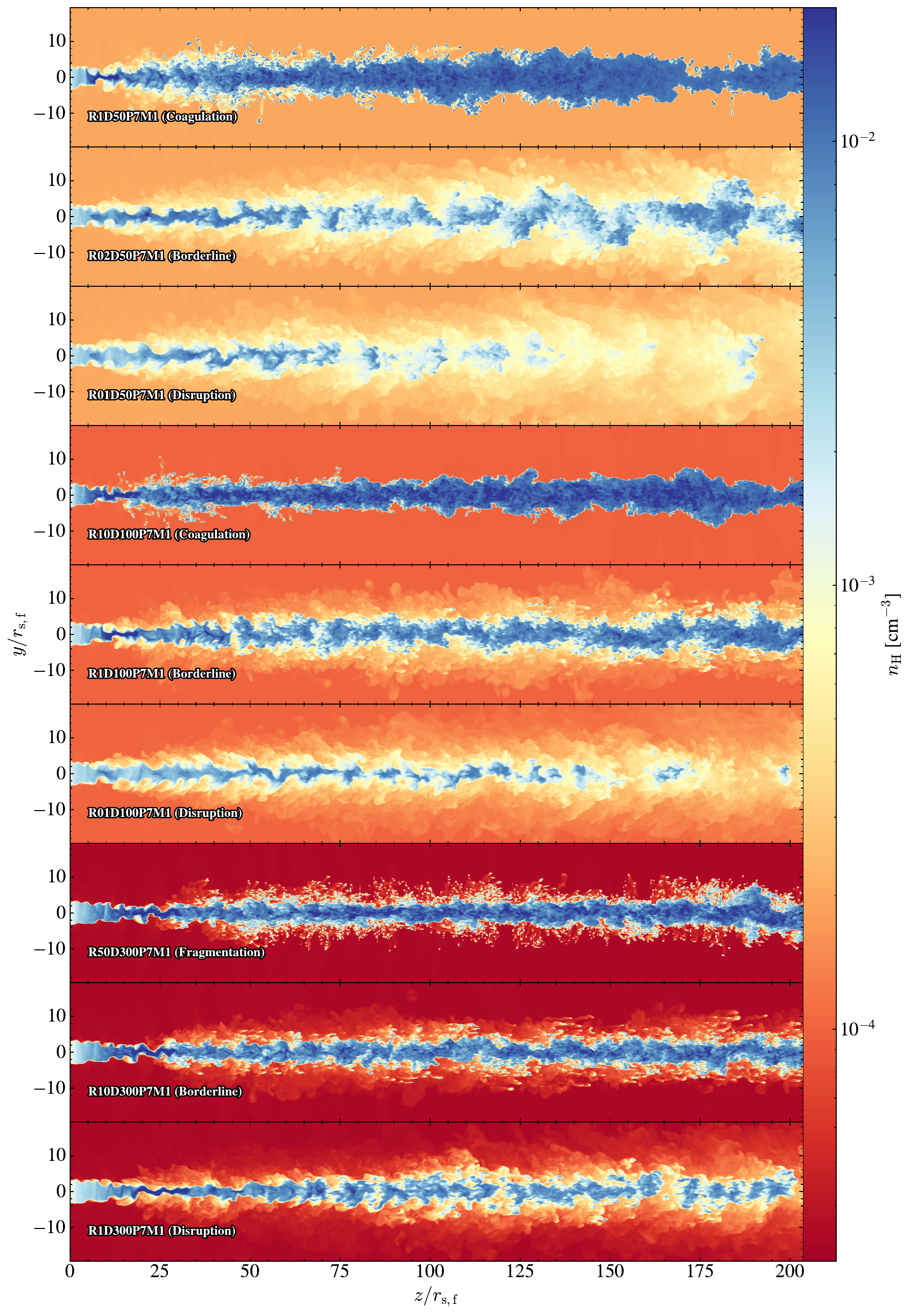}
    \caption{Similar to the number-density projection in \cref{fig:proj_fid}, but shown here for the nine simulations highlighted in \cref{fig:diag_size_chi}, spanning a range of \ifm{\chix{f}} and \ifm{\rx{s,f}}. All simulations are performed at a fixed pressure contrast of \ifm{\mathcal{P}=7} and a stream Mach number of \ifm{\Max{s}=1}.}
    \label{fig:proj_size}
\end{figure*}


\begin{figure*}
    \centering	
    \includegraphics[width=\textwidth]{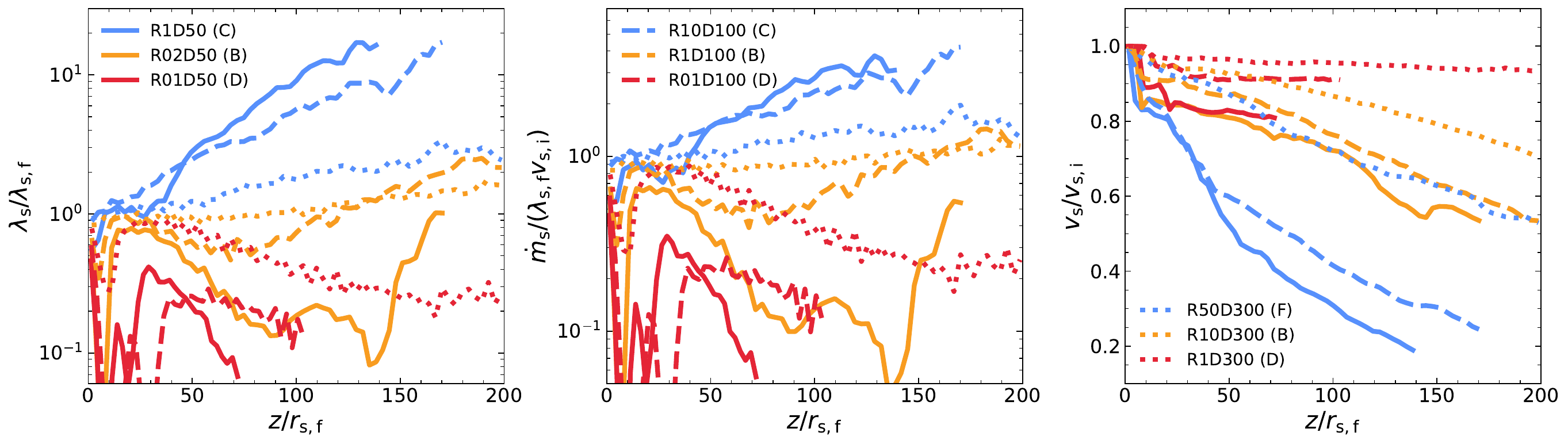}
    \caption{Dynamical properties of cold streams as a function of penetration depth in the hot background for simulations with varying final stream radii \ifm{\rx{s,f}} and density contrasts \ifm{\chix{f}} at \ifm{\mathcal{P}=7} and \ifm{\Max{s}=1}. From left to right, the panels show the cold-stream line density \ifm{\lambdax{s}}, the streamwise cold-gas mass flux \ifm{\dotm{s}\equiv\lambdax{s} \vx{s}}, and the mass-weighted velocity \ifm{\vx{s}} along the stream direction. The line density is normalized by \ifm{\lambdax{s,f}\equiv\rhox{s,f}\pi\rx{s,f}^2}, the mass flux by \ifm{\lambdax{s,f} \vx{s,i}}, and the velocity by \ifm{\vx{s,i}}. The profiles are obtained by averaging over the final 10 snapshots in the interval \ifm{1.8\text{--}2\,\tx{box}}, where each snapshot is binned into 64 uniform bins along the \ifm{z}-axis. Blue, orange, and red denote the largest, intermediate, and smallest \ifm{\rx{s,f}} at each \ifm{\chix{f}}, while solid, dashed, and dotted line styles correspond to density contrasts \ifm{\chix{f}=50,100,300}, respectively. The legend labels ``C'', ``F'', ``B'', and ``D'' denote coagulation, fragmentation, borderline, and disruption, respectively. Cold-stream gas is selected using a temperature threshold of \ifm{T<\SI{5e4}{K}} (see Appendix~\ref{sec:phasediag}). At large \ifm{z}, regions with very low velocities are excluded, as the stream propagation there is slow and the profiles are dominated by initial conditions rather than the evolved flow.}
    \label{fig:profile_mass_size}
\end{figure*}


\begin{figure*}
    \centering	
    \includegraphics[width=\textwidth]{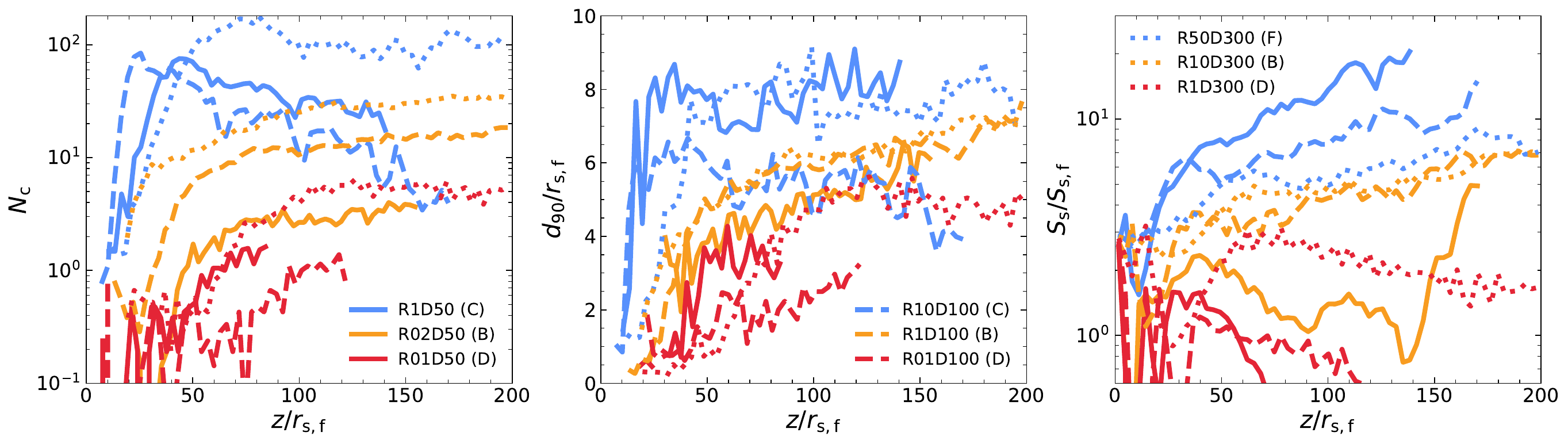}
    \caption{Morphological properties of cold streams as a function of penetration depth in the hot background for simulations spanning a range of final stream radii \ifm{\rx{s,f}} and density contrasts \ifm{\chix{f}} at \ifm{\mathcal{P}=7} and \ifm{\Max{s}=1}. From left to right, the panels display the number of clumps \ifm{\Nx{c}}, the 90th-percentile cylindrical distance of clumps from the stream's central axis (\ifm{x=y=0}) \ifm{d_{90}}, and the cold-stream surface area per unit length \ifm{\Sx{s}\equiv\Delta\Ax{s}/\Delta z}. Clumps are identified using the \software{RAMSES} clump finder, while the surface area is measured from the iso-temperature contour at \ifm{T=\SI{5e4}{K}}. The area per unit length is normalized by the analytical cylindrical circumference expected after the stream reaches pressure equilibrium, \ifm{\Sx{s,f} \equiv 2\pi \rx{s,f}}. As in \cref{fig:profile_mass_size}, the profiles are constructed by averaging over the final 10 snapshots in the interval \ifm{1.8\text{--}2\,\tx{box}}, with each snapshot binned into 64 uniform bins along the \ifm{z}-axis. Therefore the number of clumps can be lower than unity. Colors encode different \ifm{\rx{s,f}} as in \cref{fig:profile_mass_size}, while line styles distinguish different density contrasts \ifm{\chix{f}}. The legend labels ``C'', ``F'', ``B'', and ``D'' denote coagulation, fragmentation, borderline, and disruption, respectively.
    }
    \label{fig:profile_morph_size}
\end{figure*}

\cref{fig:diag_size_chi} presents nine simulations that sample a range of stream sizes and density contrasts, all at a fixed pressure contrast \ifm{\mathcal{P}=7} and stream Mach number \ifm{\Max{s}=1}. These runs correspond to the stream-size subset listed in \cref{tab:sim_table}. The corresponding hydrogen number density projections at \ifm{t=2\,\tx{box}} for all nine simulations are shown in \cref{fig:proj_size}. As shown in \cref{eq:zeq}, the penetration distance required to reach pressure equilibrium scales as \ifm{\zx{eq}\propto \chix{f}^{1/2}}, implying that the oblique shock angle decreases with increasing \ifm{\chix{f}}. 

\smallskip
In the two coagulation cases, \run{R1D50P7M1} and \run{R10D100P7M1}, the cold stream follows a similar morphological evolution along the stream direction. After reaching its equilibrium radius and briefly expanding during the rebound phase, the stream undergoes vigorous fragmentation. However, the resulting clumps do not remain isolated. Instead, they subsequently coagulate back onto the main body of the stream as it propagates downstream. The characteristic stream width increases with \ifm{z}, indicating efficient mass growth in these two cases. 

\smallskip
The fragmentation case \run{R50D300P7M1} exhibits sustained, pervasive fragmentation, with numerous clumps distributed around the stream along its entire length. However, the maximum clump distance is artificially limited to \(\ifm{\sim10\,\rx{s,f}}\) where the resolution drops. Clumps reaching this region are therefore often numerically disrupted, as discussed below.

\smallskip
In contrast, the three borderline cases, \run{R02D50P7M1}, \run{R1D100P7M1}, and \run{R10D300P7M1}, produce substantially fewer clumps despite the survival of the stream core. In these runs, strong velocity shear efficiently disrupts overdense fragments, consistent with their parameters satisfying \ifm{\rx{s,f}\lesssim \rx{crit,surv,sph}}. In both the fragmentation and borderline regimes, clumps preferentially survive at larger \ifm{z}, where stream deceleration weakens the velocity shear, allowing a greater fraction of fragments to persist.

\smallskip
For the three disrupted cases \run{R01D50P7M1}, \run{R01D100P7M1}, and \run{R1D300P7M1}, the cold stream initially fragments into elongated segments and subsequently disperses into a morphology resembling that of discrete spherical clouds. As expected, a lower density contrast \ifm{\chix{f}} results in a shorter disruption timescale, consistent with the classical cloud-crushing scaling \ifm{\tx{cc}\propto\chix{f}^{1/2}}. In the highest-\ifm{\chix{f}} run, \run{R1D300P7M1}, the stream is not completely disrupted before exiting the computational domain. We note that the mixing-layer region is considerably more extended in the borderline and disruption regimes than it is in the coagulation and fragmentation regimes. This trend is consistent with radiative mixing-layer theory, in which more efficient cooling reduces the thickness of the mixing layer \citep[e.g.][]{Ji.etal.19,Tan.etal.21}.

\smallskip
More quantitatively, we compare the cold-stream line density, streamwise mass flux, and velocity profiles of the nine simulations in \cref{fig:profile_mass_size}. Blue, orange, and red curves denote decreasing stream sizes at each \ifm{\chix{f}}, while solid, dashed, and dotted lines correspond to density contrasts of \ifm{\chix{f}=50,100,300}, respectively. We define the line density as \ifm{\lambdax{s}=\Delta \mx{s}/\Delta z}, where \ifm{\Delta \mx{s}} is the cold-gas mass in a fixed spatial bin, and the streamwise mass flux as \ifm{\dotm{s}\equiv\lambdax{s} \vx{s}}. Thus, the middle panel measures the transport of cold gas along the stream, rather than the transverse mass entrainment rate across the cold--hot interface. Owing to the imposed pressure contrast, both the cold-gas line density and the inflow velocity exhibit a sharp drop immediately downstream of the injection boundary. This feature is caused by the oblique shock, which converts part of the inflow kinetic energy into thermal energy of the cold stream. As we demonstrate in \cref{sec:pres}, increasing the pressure contrast \ifm{\mathcal{P}} causes the oblique shock to transition into a bow shock, leading to enhanced shock heating and a stronger reduction of the inflow velocity, comparable to that experienced by the background CGM. Nevertheless, for most parameters studied, the majority of the heated gas subsequently cools efficiently, and shock heating does not affect whether the stream survives or disrupts. The exception is the early bow-shock regime at sufficiently large \ifm{\mathcal{P}}, where the post-shock cooling time can exceed the halo-crossing time and shock heating itself becomes the decisive factor for survival (\cref{sec:pres}).

\smallskip
After the cold stream reaches pressure equilibrium with the ambient medium, the subsequent evolution of its line density and velocity closely resembles that in simulations without a pressure contrast \citep{Mandelker.etal.20a}, as will be shown in \cref{sec:pres,sec:discuss}. In the survival regime, the mass-flux change rate is comparable to the transverse mass entrainment rate, which can be expressed as \ifm{\Delta \dotm{s}\sim\dotm{ent}=\rhox{h}\vx{mix}\Ax{s}}, where \ifm{\vx{mix}\propto \cx{s}} and \ifm{\Ax{s}} is the surface area of the cold gas. Therefore, streams with larger \ifm{\rx{s,f}} and lower \ifm{\chix{f}} experience stronger entrainment and grow more rapidly because of their larger surface area \ifm{\Ax{s}} and higher background density \ifm{\rhox{h}}, respectively. This enhanced entrainment increases the cold-gas line density and modifies the streamwise mass flux, while also promoting momentum exchange with the background and reducing the inflow velocity as shown in the right panel. For the disrupted streams with \ifm{\rx{s,f}=\SI{0.1}{kpc}}, a lower density contrast \ifm{\chix{f}} shortens the Kelvin--Helmholtz timescale, leading to greater losses of cold gas, momentum, and streamwise velocity.

\smallskip
We compare the morphology of the cold streams in these runs in \cref{fig:profile_morph_size}. The left and middle panels show, respectively, the number of clumps and the 90th-percentile cylindrical distance of clumps identified by the \software{RAMSES} clump finder. These distinguish between the coagulation and fragmentation regimes at our fiducial resolution. The two coagulation cases, \run{R10D100P7M1} and \run{R1D50P7M1}, exhibit pronounced peaks in the number of clumps at \ifm{\sim25} and \ifm{\sim50\,\rx{s,f}}, respectively, followed by a decline with increasing \ifm{z} due to efficient coagulation. The clump distance \ifm{d_{90}} in \run{R10D100P7M1} shows a similar behavior, first reaching a maximum of \ifm{\sim6\,\rx{s,f}} and then decreasing, in agreement with the coagulation picture. The other coagulation case, \run{R1D50P7M1}, which also shows the strongest cold-mass growth, displays more complex behavior. In this run, the stream size increases significantly along the stream direction, causing the clump distances to remain large even at high \ifm{z}. 

\begin{figure*}
    \centering	
    \includegraphics[width=\textwidth]{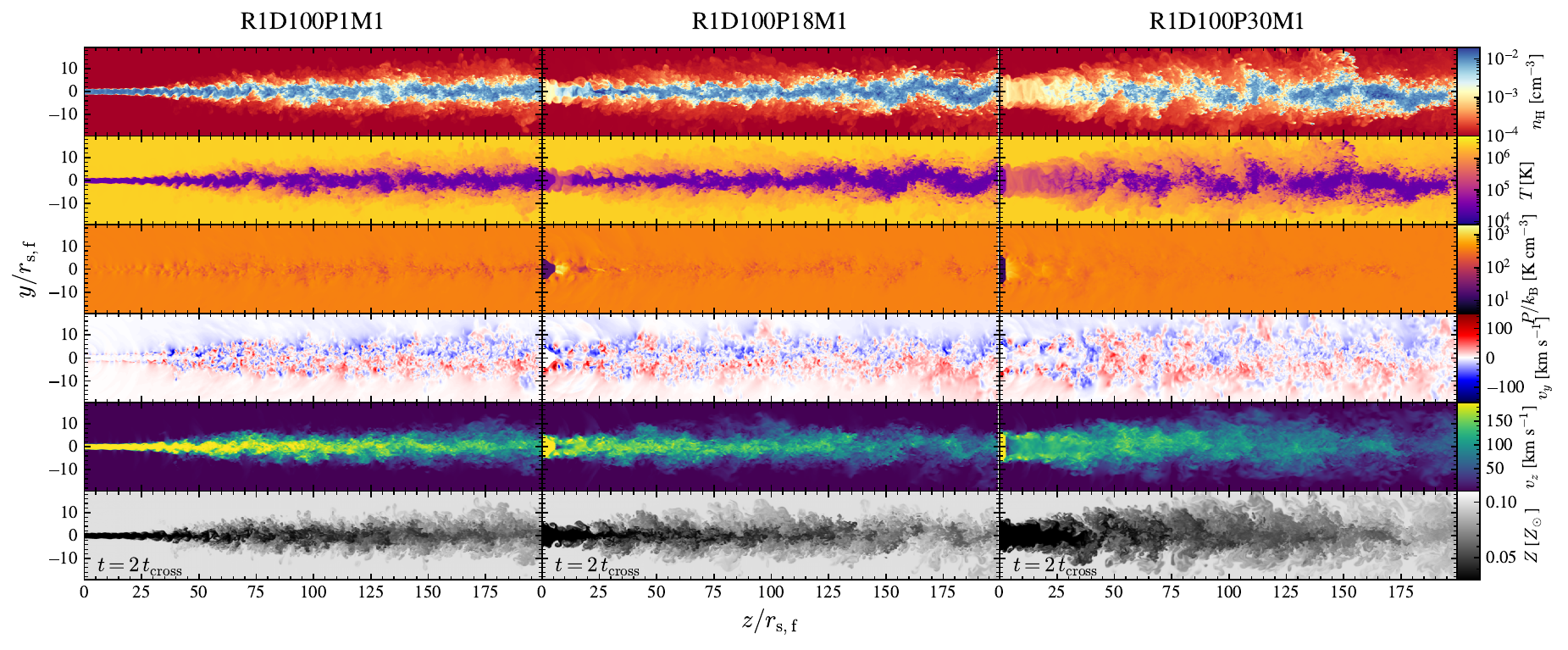}
    \caption{Projection and slice maps along the \ifm{x}-axis at \ifm{t = 2\,\tx{box}} for runs with pressure contrasts \ifm{\mathcal{P}=1} (\textit{left}), \ifm{\mathcal{P}=18} (\textit{middle}), and \ifm{\mathcal{P}=30} (\textit{right}). From top to bottom, the panels are the same as in \cref{fig:proj_fid}. As the pressure contrast increases, the oblique shock near the injection boundary becomes progressively steeper, eventually losing stability and transitioning into a bow shock in the rightmost panel.
    }
    \label{fig:proj_pres}
\end{figure*}


\smallskip
By contrast, in the fragmentation and borderline cases, both the number of clumps and their characteristic distances either continue to increase or remain approximately constant, rather than exhibiting a clear decline with increasing \ifm{z}. We caution that the measured clump distances should be interpreted as lower limits. Once fragments reach the lower-resolution regions of the computational domain, small clumps are more easily disrupted, which can suppress the measured growth of \ifm{d_{90}}. Because the stream decelerates and the shear weakens downstream, the number of clumps in the borderline cases keeps increasing with \ifm{z}, eventually overtaking the coagulation cases once their clumps have been depleted by coagulation. At small \ifm{z}, however, it remains well below the coagulation peaks. This reflects the stronger impact of velocity shear in such regimes, where streams typically have smaller sizes at fixed density contrast and are therefore more easily disrupted. Once the stream becomes sufficiently large, as in \run{R50D300P7M1}, many more clumps can survive at early times. However, fragmentation does not necessarily imply that each clump evolves as an isolated cloud. If the fragment stopping length, \ifm{\ell_\mathrm{stop}\sim\chix{f} r_\mathrm{clump}}, exceeds the typical separation between neighboring fragments, the fragments remain dynamically coupled through overlapping wakes and mixing layers. In this regime, the fragmented stream behaves as a collective multiphase structure, and its survival is controlled more by the scale of the parent stream than by the sizes of individual clumps. 

\smallskip
We stress that the absolute clump statistics shown here should not be over-interpreted. Both the clump counts and the maximum clump distances are affected by numerical resolution as shown in \cref{fig:profile_morph_res}, and the counts represent the number of clumps per \ifm{z}-axis bin rather than the total number of fragments in the stream. When integrated over the full penetration length, the total number of clumps can be orders of magnitude larger. Moreover, we find a shift from coagulation toward sustained fragmentation in the clump-number profile of \run{R1D50P7M1} at higher resolution in \cref{fig:profile_morph_res}, which may reflect an additional shear-driven fragmentation mechanism. This mechanism is distinct from the pressure-driven fragmentation associated with the initial rebound and can continue to produce small interface fragments downstream, potentially offsetting their removal through coagulation. Higher resolution may be needed to capture this continuing production of fragments. Strong resolution requirements for fragmentation have also been reported in radiative cloud--wind simulations \citep{Sparre.etal.19}. The coexistence of these mechanisms complicates the competition between fragmentation and coagulation, and we leave a quantitative separation of their contributions to future work. Appendix~\ref{sec:convergence} presents a more detailed discussion.

\smallskip
The right panel of \cref{fig:profile_morph_size} shows the cold--hot interfacial area per unit stream length, \ifm{\Sx{s}\equiv\Delta\Ax{s}/\Delta z}, where \ifm{\Ax{s}} is estimated from the temperature isosurface at \ifm{T=\SI{5e4}{K}}. Unlike the clump statistics, this measure does not depend on the clump-finding algorithm. The evolution of \ifm{\Sx{s}} broadly follows that of the cold-gas streamwise mass flux \ifm{\dotm{s}}, indicating that the growth or decline of cold gas mass flux is closely linked to the expansion or contraction of the cold--hot interface. As discussed in \cref{sec:discuss}, this connection can be understood from the approximate momentum conservation of the cold gas. In this limit, \ifm{\rx{s}\vx{s}} remains roughly constant, implying that, for a nearly cylindrical cold stream, \ifm{\dotm{s}/\Sx{s}\propto \rx{s}\vx{s}} should also vary only weakly along the stream.

\smallskip
In summary, we identify three distinct regimes governing the mass and morphological evolution of a cold stream penetrating a halo with an initial pressure contrast. In the \emph{coagulation regime}, the cold mass grows most efficiently. The stream undergoes vigorous early fragmentation, but the resulting clumps quickly merge, reassembling into a single, coherent cold stream that extends along the flow direction. In the \emph{fragmentation regime}, the cold mass growth is modest, accompanied by an increasing number of clumps that are more widely dispersed about the stream axis. In the \emph{disruption regime}, the cold mass decreases: the stream first breaks into elongated cold segments and is then gradually disrupted, ultimately resembling the morphology and evolution of a disrupted spherical cloud.

\section{Effects of pressure contrasts on cold-stream penetration}
\label{sec:pres}

\begin{figure*}
    \centering	
    \includegraphics[width=\textwidth]{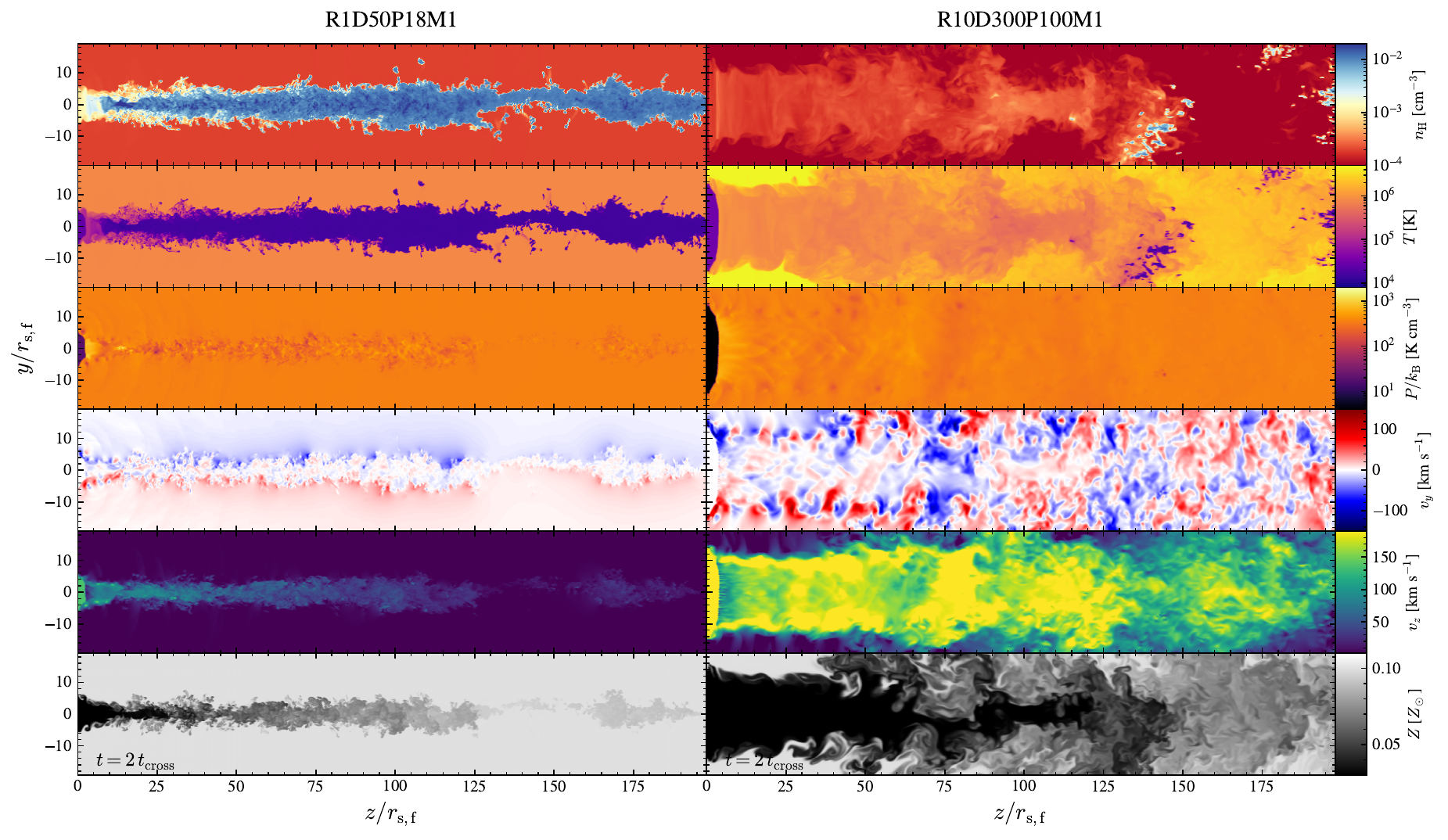}
    \caption{Projection and slice maps along the \ifm{x}-axis at \ifm{t = 2\,\tx{box}} for runs that undergo early bow-shock transitions: \run{R1D50P18M1} (\textit{left}) and \run{R10D300P100M1} (\textit{right}). From top to bottom, the panels are the same as in \cref{fig:proj_fid}. As the density contrast increases, the pressure contrast required for an early bow-shock transition scales approximately as \ifm{\chix{f}^{1/2}} (\cref{eq:bowshocktransition}). In this regime, cold-stream survival is governed by whether the post-shock gas can cool within the crossing time, rather than by the Kelvin--Helmholtz timescale entering the shear criterion in \cref{eq:Rshear}.}
    \label{fig:proj_pres_bowshock}
\end{figure*}


The parameter survey in \cref{sec:size} identified three regimes of cold-stream evolution at a fixed pressure contrast of \ifm{\mathcal{P}=7}. However, the pressure contrast introduces physics absent from the classical stream-stability problem, most notably the initial contraction of the stream and shock heating at the injection boundary. It therefore remains to be tested whether this regime classification is robust when \ifm{\mathcal{P}} varies, and how the pressure contrast modulates the quantitative evolution of the cold-stream mass. In this section, we address these questions by varying \ifm{\mathcal{P}} at a fixed final stream size of \ifm{\rx{s,f}=\SI{1}{kpc}}. As we show below, the regime classification is robust at moderate pressure contrasts, but a qualitatively new behavior emerges once \ifm{\mathcal{P}} exceeds a threshold that scales with the density contrast: the oblique shock at the injection boundary transitions into a detached bow shock, and stream survival becomes governed by post-shock cooling rather than by the shear criterion of \cref{eq:Rshear}. \cref{fig:proj_pres} compares projection and slice maps for three runs with different pressure contrasts at \ifm{\chix{f}=100}, ranging from \ifm{\mathcal{P}=1} on the left to \ifm{\mathcal{P}=30} on the right. For context, the cosmological model in \cref{eq:pratio} gives a mean-pressure contrast of \ifm{\mathcal{P}\sim35} for \ifm{\Max{s}=1} at \ifm{\Mx{v}=\SI{1e12}{\Msun}} and \ifm{z=2}. All panels are shown at the final snapshot of \ifm{t=2\,\tx{box}}. 

\smallskip
The leftmost panel corresponds to the case without a pressure contrast, for which \ifm{\rx{s,i}=\rx{s,f}}, and closely resembles the setup studied by \citet{Mandelker.etal.20a}. However, while \citet{Mandelker.etal.20a} adopted periodic boundary conditions and focused on the time evolution of cold streams, here we impose injection and outflow boundary conditions and examine the spatial evolution along the stream direction at a fixed time. As the cold stream is injected from the left boundary, it begins to wobble vertically at \ifm{z\sim10\text{--}20\,\rx{s,f}}, characteristic of the sinuous (S-mode, \ifm{m=1}) body mode of the Kelvin--Helmholtz instability. For \ifm{\Max{s}=1}, the flow lies near the threshold for exciting unstable body modes. The linear growth timescale of three-dimensional body modes is estimated as \ifm{\tx{body}\sim\tx{sc}} from Table~1 of \citet{Mandelker.etal.19}, implying an onset distance \ifm{z_{\mathrm{body}}\sim \tx{body}\vx{s,i}\sim\chix{f}^{1/2}\Max{s}\rx{s,f}\sim10\,\rx{s,f}}. Owing to the combined effects of relatively large surface shape perturbations, \ifm{\delta r=0.1\,\rx{s,i}=0.1\,\rx{s,f}}, and a comparable numerical resolution, \ifm{\Delta\sim0.1\,\rx{s,f}}, high-$m$ surface modes are suppressed, allowing the cold stream to remain intact over a long distance until the low-$m$ surface modes, which are still linearly unstable, develop the turbulent mixing layer. 

\smallskip
For \ifm{\mathcal{P}=7} (\cref{fig:proj_fid}) and \ifm{\mathcal{P}=18} (the middle column of \cref{fig:proj_pres}), the axisymmetric contraction due to the pressure contrast initially enhances the pinch (P-mode, \ifm{m=0}) body mode of the KHI. However, as pressure equilibrium is re-established, surface modes likely begin to dominate, producing a prominent turbulent mixing layer. 

\smallskip
A notable feature of the high-pressure-contrast runs is the transition from an oblique shock to a detached bow shock. In the \ifm{\mathcal{P}=18} case, this transition occurs at \ifm{z\sim7\,\rx{s,f}}, while in the \ifm{\mathcal{P}=30} case the bow shock forms shortly after injection. Empirically, our simulations therefore suggest that the transition occurs once the pressure contrast reaches \ifm{\mathcal{P}\sim18\text{--}30} for the fiducial parameters explored here.

\smallskip
The transition from an attached oblique shock to a detached bow shock has been studied extensively in compressible-fluid dynamics. The Rankine--Hugoniot jump conditions for an oblique shock give the standard \ifm{\theta\text{--}\beta\text{--}\mathcal{M}} relation \citep{NACA1135},
\begin{equation}
\label{eq:deflectionangle}
\tan\theta(\beta) = 2\cot\beta\,\frac{\Max{1}^2\sin^2\beta-1}{\Max{1}^2(\gamma+\cos2\beta)+2},
\end{equation}
where \ifm{\theta} is the flow deflection angle, \ifm{\beta} is the oblique shock angle, \ifm{\gamma} is the adiabatic index, and \ifm{\Max{1}=\Max{s}\chix{f}^{1/2}} is the Mach number of the cold stream relative to its own sound speed. An attached oblique shock exists only up to a maximum deflection angle, beyond which the shock detaches and becomes a bow shock. For \ifm{\gamma=5/3} and \ifm{\Max{1}\gtrsim5}, this maximum is \ifm{\thetax{max}\simeq35\degr}, corresponding to a critical shock angle of \ifm{\beta_\mathrm{c}\simeq63\degr}.

\smallskip
A naive estimate of the oblique-shock angle in the stream-penetration scenario is
\begin{equation}
\label{eq:shockangle}
\tan\beta \sim \frac{\vx{cyl}}{\vx{s,i}}
\sim \frac{\mathcal{P}^{1/2}}{\chix{f}^{1/2}\Max{s}} .
\end{equation}
However, if we require an early bow-shock transition through \ifm{\tan\beta>\tan\beta_\mathrm{c}\simeq2}, this estimate implies \ifm{\mathcal{P}\gtrsim380} for \ifm{\chix{f}=100} and \ifm{\Max{s}=1}, much larger than the pressure contrast in the early bow-shock run \run{R1D100P30M1} (\cref{fig:proj_pres}, right column). The acceleration of the converging cylindrical shock as it propagates toward the stream axis, discussed in \citet{Yao.etal.25}, can in principle increase the shock angle. In practice, however, the self-similar steepening is weak: for \ifm{\gamma=5/3}, \ifm{\vx{cyl}\propto \rx{cyl}^{-0.19}} \citep{Modelevsky.Sari.21}. It therefore becomes important only very close to the axis, after the shock has already swept through most of the stream. This is the opposite of what is needed for a prompt transition near the virial-shock radius.

\smallskip
Inspecting the time-dependent visualizations\footnote{Available on our \href{https://zhiyuan-21.github.io/HomePage/stream_penetration.html}{companion webpage}.} of the early bow-shock development, we find that the ``early'' bow shock does not form directly at the injection boundary. Instead, it first appears downstream, near the point where the cylindrical shock converges onto the stream axis, and subsequently propagates upstream until it settles close to the injection boundary. The pressure maps in \cref{fig:proj_pres} show the central pressure enhancement. Related pressure focusing during cylindrical implosion is illustrated by the radial profiles in \citet[][Fig.~2]{Yao.etal.25}, although that study did not include the axial shear flow considered here. This behavior suggests the following physical picture. When the cylindrical shock converges on the axis, the resulting stagnated, overpressured core acts as an effective obstacle to the incoming stream. A bow shock can then propagate upstream ahead of this obstacle provided that the core pressure exceeds the incoming stream ram pressure, \ifm{\rhox{s,i}\vx{s,i}^2}. To identify the physics setting this core pressure, we have rerun several bow-shock cases with radiative cooling disabled and find essentially the same transitions at similar positions. A cooling-independent estimate of the core pressure follows from the work performed by the ambient pressure during the cylindrical collapse. Per unit length along the stream, the lateral work is
\begin{equation}
    \frac{W_{\perp}}{L}\sim \Px{h}\,\pi\left(\rx{s,i}^{2}-\rx{core}^{2}\right)\sim \Px{h}\,\pi\rx{s,i}^{2},
\end{equation}
where \ifm{\rx{core}\ll\rx{s,i}} is the radius of the stagnated core. If a fraction \ifm{\fx{focus}} of this work is retained and focused into the core, whose cross-section is comparable to the final equilibrium cross-section of the stream, \ifm{\rx{core}\sim\rx{s,f}}, the resulting core pressure is
\begin{equation}
    \Px{core}\sim \fx{focus}\,(\gamma-1)\,\Px{h}\parfrac{\rx{s,i}}{\rx{s,f}}^{2} = \fx{focus}\,(\gamma-1)\,\eta\,\Px{h},
\end{equation}
where \ifm{\eta\equiv\rhox{s,f}/\rhox{s,i}=(\rx{s,i}/\rx{s,f})^{2}} is the stream compression factor. In our setup, \ifm{\eta} is proportional to \ifm{\mathcal{P}} up to mean-molecular-weight factors, \ifm{\mathcal{P}=\eta\,\mux{s}/\mux{h}} (\cref{subsec:ICBC}), so with \ifm{\Px{h}=\mathcal{P}\Px{s,i}} the core pressure scales as \ifm{\Px{core}/\Px{s,i}\sim\eta\,\mathcal{P}\sim\mathcal{P}^{2}}. The bow shock should then appear once this focused core pressure becomes comparable to the ram pressure required to stall the incoming stream, \ifm{\Px{core}\sim\rhox{s,i}\vx{s,i}^{2}}, which yields
\begin{equation}
\label{eq:bowshocktransition}
    \mathcal P\gtrsim C_{\rm bow}\,\frac{\vx{s,i}}{\cx{s}} = C_{\rm bow}\,\chix{f}^{1/2}\Max{s},
\end{equation}
where \(C_{\rm bow}\) is an order-unity coefficient that absorbs the focusing efficiency \ifm{\fx{focus}}, shock-jump coefficients, and mean-molecular-weight factors. Guided by our simulations, we adopt \(C_{\rm bow}\simeq2\). For \ifm{\chix{f}=100} and \ifm{\Max{s}=1}, this gives \ifm{\mathcal{P}\gtrsim20}. This estimate is consistent with the delayed transition in the \ifm{\mathcal{P}=18} run (middle column of \cref{fig:proj_pres}) and the prompt transition in the \ifm{\mathcal{P}=30} run (right column of \cref{fig:proj_pres}). The two additional simulations that exhibit prompt bow-shock transitions in \cref{fig:proj_pres_bowshock}, \run{R1D50P18M1} and \run{R10D300P100M1}, also satisfy \cref{eq:bowshocktransition}. By contrast, \run{R1D100P7M05} (central panel of \cref{fig:proj_mach}) does not satisfy this criterion and does not undergo a prompt transition, despite having an initial shock angle comparable to that of \run{R1D100P30M1} (\cref{eq:shockangle}). We emphasize that \cref{eq:bowshocktransition} should be interpreted as an empirically motivated guide, calibrated on our limited set of simulations, rather than as a strict analytic threshold. A systematic parameter study would be required to establish a robust quantitative transition criterion, which we leave to future work.

\begin{figure*}
    \centering	
    \includegraphics[width=\textwidth]{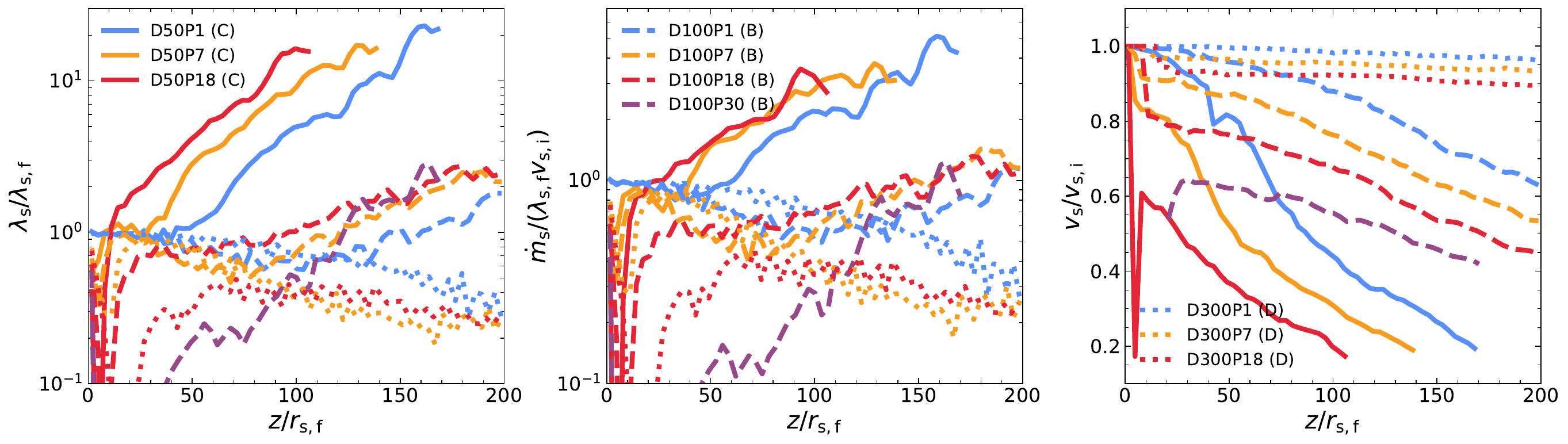}
    \caption{Similar to \cref{fig:profile_mass_size}, but for runs with different pressure contrasts \ifm{\mathcal{P}} and density contrasts \ifm{\chix{f}} at \ifm{\rx{s,f}=\SI{1}{kpc}} and \ifm{\Max{s}=1}. Blue, orange, red, and purple curves correspond to \ifm{\mathcal{P}=1,7,18,30}, respectively, while solid, dashed, and dotted line styles denote \ifm{\chix{f}=50,100,300}. The legend labels ``C'', ``B'', and ``D'' denote coagulation, borderline, and disruption, respectively. The pressure contrast acts mainly as an amplifier, boosting the amplitude of the cold-gas line-density and mass flux evolution by a factor of a few without altering the underlying survival or disruption outcome, provided the post-shock gas cools well within the crossing time, as is the case for all runs shown here (see text).
    }
    \label{fig:profile_mass_pres}
\end{figure*}


\begin{figure*}
    \centering	
    \includegraphics[width=\textwidth]{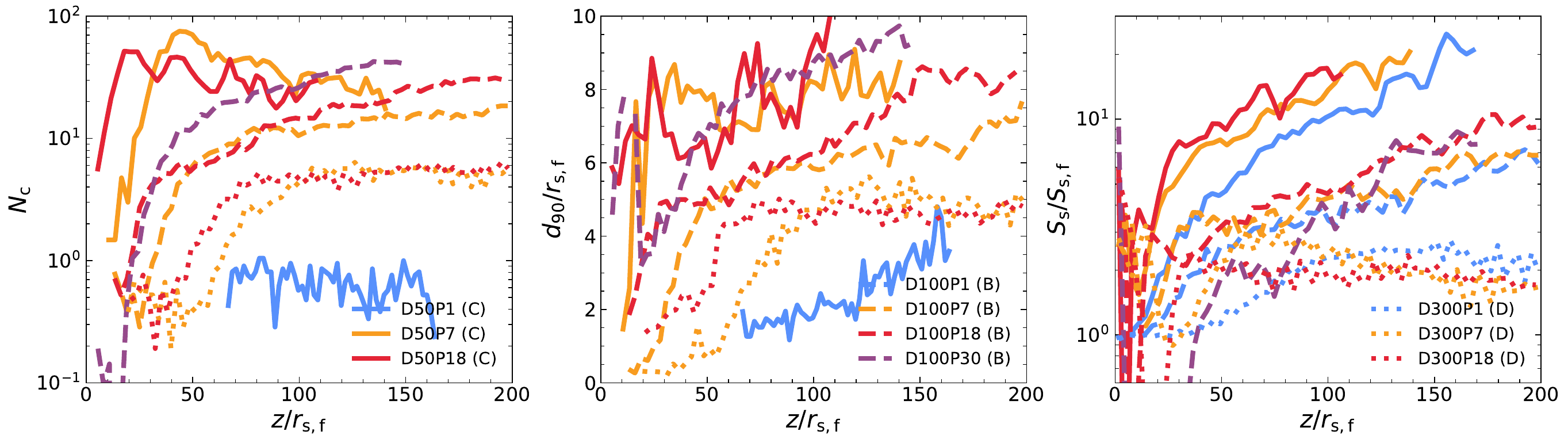}
    \caption{Similar to \cref{fig:profile_morph_size}, but for runs with varying pressure contrasts \ifm{\mathcal{P}} and density contrasts \ifm{\chix{f}} at \ifm{\rx{s,f}=\SI{1}{kpc}} and \ifm{\Max{s}=1}. Blue, orange, red, and purple curves correspond to \ifm{\mathcal{P}=1,7,18,30}, respectively, while solid, dashed, and dotted line styles denote \ifm{\chix{f}=50,100,300}. The legend labels ``C'', ``B'', and ``D'' denote coagulation, borderline, and disruption, respectively.
    }
    \label{fig:profile_morph_pres}
\end{figure*}


\smallskip
In the run \run{R1D100P30M1} (\cref{fig:proj_pres}, right column), due to the formation of a bow shock prior to a significant increase in the stream density during contraction, radiative cooling is initially inefficient, and the stream is heated nearly globally. At the same time, the inflow velocity is substantially reduced, which weakens the Kelvin--Helmholtz instability and allows the post-shock warm gas to cool before significant mixing. In the strong-shock limit, the post-shock density is
\begin{equation}
\label{eq:rho_post}
    \rhox{\rm post}\sim\frac{\gamma+1}{\gamma-1}\rhox{s,i},
\end{equation}
and the post-shock temperature is given by
\begin{equation}
\label{eq:temp_post}
    \Tx{\rm post}\sim \Tx{s}\,\frac{2\gamma(\gamma-1)}{(\gamma+1)^2}\,\Max{s}^2\,\chix{f}.
\end{equation}
For \ifm{\gamma=5/3}, adopting a hydrogen number density of \ifm{\nx{H,s,i}\sim0.01/\eta\,\mathrm{cm}^{-3}\sim2\times10^{-4}\,\mathrm{cm}^{-3}}, where $\eta=\rho_{\rm s,f}/\rho_{\rm s,i}=50$ is the value used in the \run{R1D100P30M1} simulation, and \ifm{\Tx{s}\sim\SI{1.5e4}{K}}, we obtain a post-shock hydrogen number density of \ifm{\nx{H,post}\sim\SI{8e-4}{cm}^{-3}} and a temperature of \ifm{\Tx{\rm post}\sim\SI{4.6e5}{K}}. The corresponding cooling timescale can be estimated as
\begin{equation}
\label{eq:tcool_post}
\tx{cool,post}=\frac{\kb\Tx{post}}{(\gamma-1)\mu X\nx{H,post}\Lambda(\Tx{post})}
\sim0.7\text{--}1.4\,\SI{}{Gyr},
\end{equation}
where $X=0.76$ is the hydrogen mass fraction and the quoted range corresponds to metallicities in the range \ifm{Z=0.03}--\ifm{0.1\,Z_{\sun}}. Given the post-shock velocity
\begin{equation}
    \vx{post}\sim\frac{\gamma-1}{\gamma+1}\vx{s,i},
\end{equation}
the associated penetration distance before significant cooling occurs is then
\begin{equation}
\zx{cool}\sim \vx{post}\tx{cool,post}\sim \frac{\gamma-1}{\gamma+1}\,\Max{s}\,\cx{h}\,\tx{cool,post}\sim32\text{--}70\,\si{\kpc}=32\text{--}70\,\rx{s,f},
\end{equation}
which is in broad agreement with the simulation results. Similarly, we estimate \ifm{\zx{cool}=6\text{--}10\,\rx{s,f}} for \run{R1D50P18M1} and \ifm{\zx{cool}=194\text{--}248\,\rx{s,f}} for \run{R10D300P100M1}. These estimates are consistent with the penetration distances shown in \cref{fig:proj_pres_bowshock}. Note that the parameters \ifm{\rx{s,f}=\SI{10}{kpc}}, \ifm{\chix{f}=300}, and \ifm{\Max{s}=1} would place the stream in the survival regime under the shear-dominated criterion shown in \cref{fig:diag_size_chi}. Therefore, in the early bow-shock regime, cold-gas survival is not controlled primarily by the Kelvin--Helmholtz timescale entering the shear criterion in \cref{eq:Rshear}. Instead, it is governed by whether the shocked gas can cool within the crossing time, i.e. whether \ifm{\tx{cool,post}<\tx{v}}. Another important point is that disruption of the cold stream does not immediately erase the overdense structure. As shown in the right panel of \cref{fig:proj_pres_bowshock}, the overdense stream persists and penetrates to distances of \ifm{\sim150\,\rx{s,f}} before becoming fully disrupted.

\smallskip
We compare the cold-stream line density, streamwise mass flux, and velocity profiles for different pressure contrasts \ifm{\mathcal{P}} in \cref{fig:profile_mass_pres}, considering three density contrasts \ifm{\chix{f}=50,100,300} at a fixed final stream size \ifm{\rx{s,f}=1\,\mathrm{kpc}}. As shown in \cref{sec:size}, the fiducial runs for these three density contrasts with $\mathcal{P}=7$ correspond to the coagulation, fragmentation/borderline, and disruption regimes, respectively, and this remains true for all pressure contrasts explored at this stream size, \ifm{\mathcal{P}\le30}. We find that the qualitative behavior of \ifm{\lambdax{s}}, \ifm{\dotm{s}}, and \ifm{\vx{s}} is primarily set by the regime, while variations in the pressure contrast at the injection boundary introduce only a secondary modulation. In the coagulation and fragmentation cases, both of which lie within the survival regime, increasing the pressure contrast enhances the growth of cold mass and momentum, whereas in the disruption case it leads to a further reduction. 
The pressure contrast thus acts mainly as an amplifier, boosting the amplitude of the cold-mass evolution before pressure equilibrium is reached by a factor of a few without altering the underlying survival or disruption outcome. We stress that this insensitivity to \ifm{\mathcal{P}} relies on rapid post-shock cooling. As demonstrated by \run{R10D300P100M1} above, once an early bow shock forms with \ifm{\tx{cool,post}>\tx{v}}, the pressure contrast itself changes the survival outcome, disrupting a stream that would otherwise satisfy the shear criterion.

\smallskip
More specifically, increasing the pressure contrast initially suppresses both the cold-gas line density and the streamwise mass flux immediately downstream of the injection point, with larger contrasts producing a deeper and more extended suppression. At sufficiently large distances, however, the cold-gas line density recovers to comparable levels in all runs. In the survival cases, this recovered line density exceeds that in the no-contrast run by a factor of \ifm{2\text{--}3}, whereas in the disruption case it remains lower by a factor of \ifm{\sim2}, indicating that pressure perturbations amplify the competition between mixing and radiative cooling. The recovery distance increases with both density contrast and pressure contrast due to longer cooling timescales. 

\smallskip
The inflow velocity exhibits systematic deceleration in all cases as a result of mixing with the initially static hot background gas. Stronger deceleration occurs at higher pressure contrasts. Larger contrasts more abruptly reduce the cold-stream velocity immediately after penetration by increasing the shock angle (see \cref{eq:shockangle}) and by inducing an earlier transition to bow shocks, which more efficiently remove the velocity component aligned with the stream direction.

\smallskip
We compare the morphological properties of cold streams for this same set of simulations
in \cref{fig:profile_morph_pres}. The left and middle panels show, respectively, the number of clumps and their 90th-percentile distances from the stream center, as identified by the \software{RAMSES} clump finder. In the coagulation regime (\ifm{\chix{f}=50}) with \ifm{\mathcal{P}>1}, the clump number exhibits an initial peak caused by pressure-induced fragmentation close to the injection boundary, followed by a systematic decline downstream as clumps merge efficiently through coagulation. This coagulation process is less apparent in the clump-distance statistic, because the stream radii increase rapidly with \ifm{z} in these efficient mass growth cases; as a result, clumps located near the stream surface can still reside at large distances from the center. In the borderline regime (\ifm{\chix{f}=100}), the number of clumps increases monotonically with both \ifm{z} and \ifm{\mathcal{P}}, eventually surpassing that in the coagulation cases. The clump distances display a similar trend, indicating a more violent fragmentation process and a broader spatial distribution of clumps at higher pressure contrasts. 

\smallskip
The right panel of \cref{fig:profile_morph_pres} shows the cold-stream surface area per unit length, normalized to its analytical value, measured along the stream direction using the temperature isosurface \ifm{T=\SI{5e4}{K}}. In simulations with a pressure contrast at injection, the normalized areas initially exceed unity but rapidly decline as the streams are compressed by the ambient medium. At larger distances, the areas recover and rise above unity again for \ifm{z>50\,\rx{s,f}}. Consistent with the behavior of the cold-gas line density and mass-flux profiles, the areas increase systematically with decreasing density contrasts. Pressure contrasts primarily act to enhance the subsequent growth or decrease of the area, amplifying the overall evolution rather than changing its qualitative trend.

\smallskip
In summary, a finite pressure contrast at injection drives an oblique shock at small \ifm{\mathcal{P}}, which steepens into a bow shock once \ifm{\mathcal{P}} becomes sufficiently large. This transition increases the conversion efficiency of inflowing kinetic energy into thermal energy and changes the relevant stream-survival condition: instead of being set only by the Kelvin--Helmholtz evolution, survival in the early bow-shock regime requires the post-shock gas to cool faster than the halo-crossing timescale, i.e. \ifm{t_{\rm cool,post}<t_{\rm v}}. In the oblique-shock regime, the pressure contrast mainly acts as an \emph{amplifier} of the mass evolution, increasing the amplitude of the cold-mass variation by a factor of a few without altering the final outcome of survival or disruption. In the coagulation regime, pressure-induced fragmentation produces an early peak in the clump number, followed by a downstream decline through coagulation, whereas in the borderline regime both the number of clumps and their characteristic distances grow monotonically with \ifm{\mathcal{P}}.

\begin{figure*}
    \centering	
    \includegraphics[width=\textwidth]{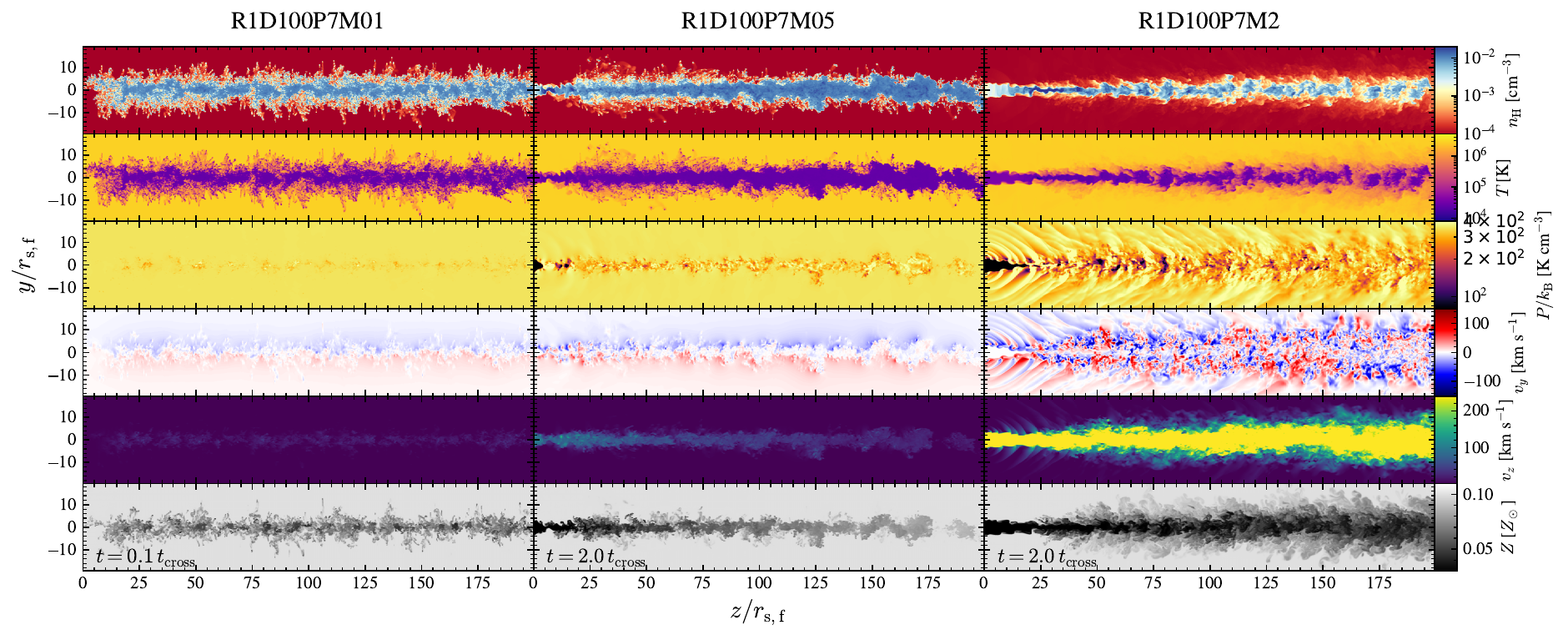}
    \caption{Similar to \cref{fig:proj_pres}, but for runs with different stream Mach numbers: \ifm{\Max{s}=0.1} (\textit{left}), \ifm{\Max{s}=0.5} (\textit{middle}), and \ifm{\Max{s}=2} (\textit{right}), all with \ifm{\chix{f}=100}, $r_{\rm s,f}=1\,{\rm kpc}$, and $\mathcal{P}=7$. The \ifm{\Max{s}=0.1} case (left) fails to penetrate the hot CGM; we therefore show this run at \ifm{t=0.1\,\tx{box}}.}
    \label{fig:proj_mach}
\end{figure*}


\section{Effects of velocity shear on cold-stream fragmentation}
\label{sec:shear}

\begin{figure*}
    \centering	
    \includegraphics[width=\textwidth]{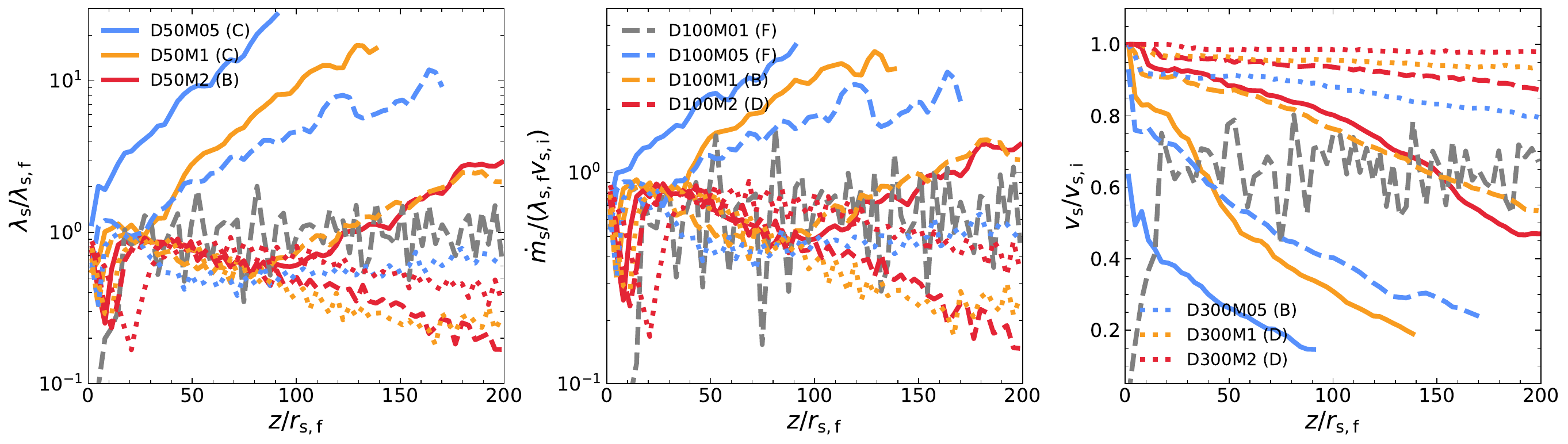}
    \caption{Similar to \cref{fig:profile_mass_size}, but for runs with varying stream Mach numbers \ifm{\Max{s}} and density contrasts \ifm{\chix{f}} at \ifm{\rx{s,f}=\SI{1}{kpc}} and \ifm{\mathcal{P}=7}. Blue, orange, and red curves correspond to \ifm{\Max{s}=0.5, 1, 2}, respectively, while solid, dashed, and dotted line styles denote \ifm{\chix{f}=50, 100, 300}. The legend labels ``C'', ``F'', ``B'', and ``D'' denote coagulation, fragmentation, borderline, and disruption, respectively. The gray dashed curve shows the \ifm{\Max{s}=0.1} case at \ifm{t=0.1\,\tx{box}}, since the stream fails to penetrate the halo.}
    \label{fig:profile_mass_Mach}
\end{figure*}


Previous sections have highlighted the disruption of cold streams by strong velocity shear, particularly during the early stages of fragmentation-dominated cases. In all simulations discussed above, we fixed the injection velocity to \ifm{\Max{s}=1}. Here, we relax this assumption and explore the impact of varying the Mach number at three density contrasts, \ifm{\chix{f}=50,100,300}, focusing on how velocity shear influences the evolution of the cold stream and the formation of fragmented clumps.

\smallskip
\cref{fig:proj_mach} presents projection and slice maps for three runs with different stream Mach numbers \ifm{\Max{s}=0.1,0.5,2} at \ifm{\chix{f}=100}, $r_{\rm s,f}=1\,{\rm kpc}$, and $\mathcal{P}=7$. Penetration of the virial shock requires the ram pressure of the stream to be larger than the ambient thermal pressure in the halo, \ifm{\Px{ram,s}\gtrsim \Px{h}}, which implies
\begin{equation}
\Max{s}\gtrsim \chix{i}^{-1/2}=(\eta/\chix{f})^{1/2}.
\end{equation}
Accordingly, the cold stream cannot penetrate the virial shock at the left boundary in run \run{R1D100P7M01}, where \ifm{\chix{i}=10} but \ifm{\Max{s}=0.1}. Consistent with this expectation, the leftmost panel shows no penetration and closely resembles the static, underpressurized stream studied in \citet{Yao.etal.25}. For this case, we show the map at \ifm{t=0.1\,\tx{box}\sim20\,\tx{sc,i}/(\Max{s}\chix{f}^{1/2}\eta^{1/2})\sim6.4\,\tx{sc,i}} using \cref{eq:tcross}, where \ifm{\tx{sc,i}=\rx{s,i}/\cx{s}=\eta^{1/2}\tx{sc}}. This corresponds to the time shortly after the cold-stream mass reaches its transient peak following the pressure-driven implosion.

\begin{figure*}
    \centering	
    \includegraphics[width=\textwidth]{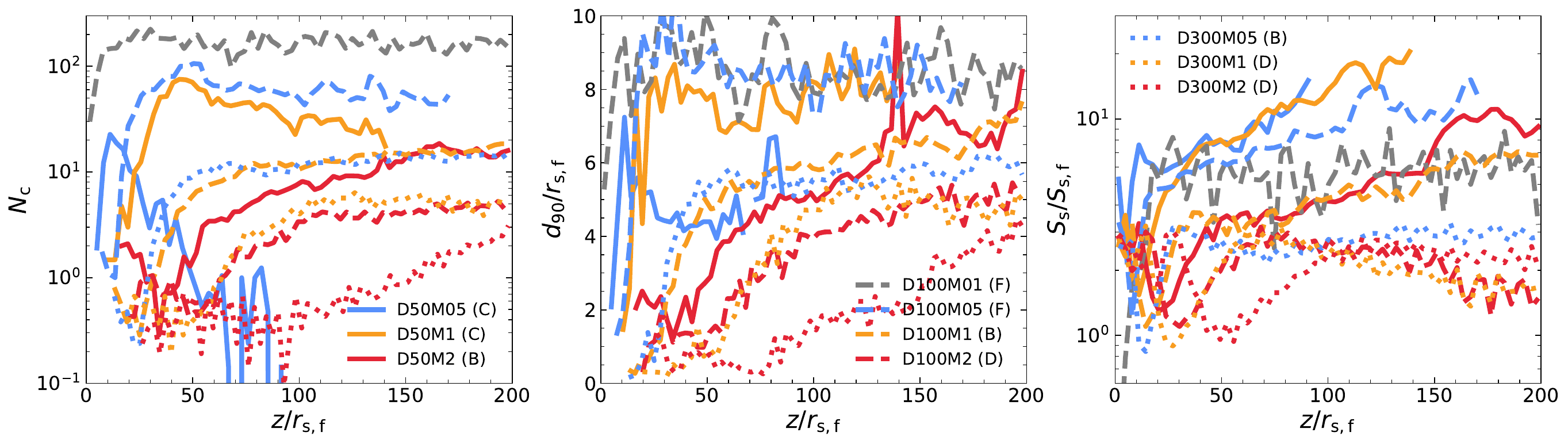}
    \caption{Similar to \cref{fig:profile_morph_size}, but varying the Mach number of the cold stream \ifm{\Max{s}} and density contrasts \ifm{\chix{f}} at \ifm{\rx{s,f}=\SI{1}{kpc}} and \ifm{\mathcal{P}=7}.
    }
    \label{fig:profile_morph_Mach}
\end{figure*}


\begin{figure*}
    \centering	
    \includegraphics[width=\textwidth]{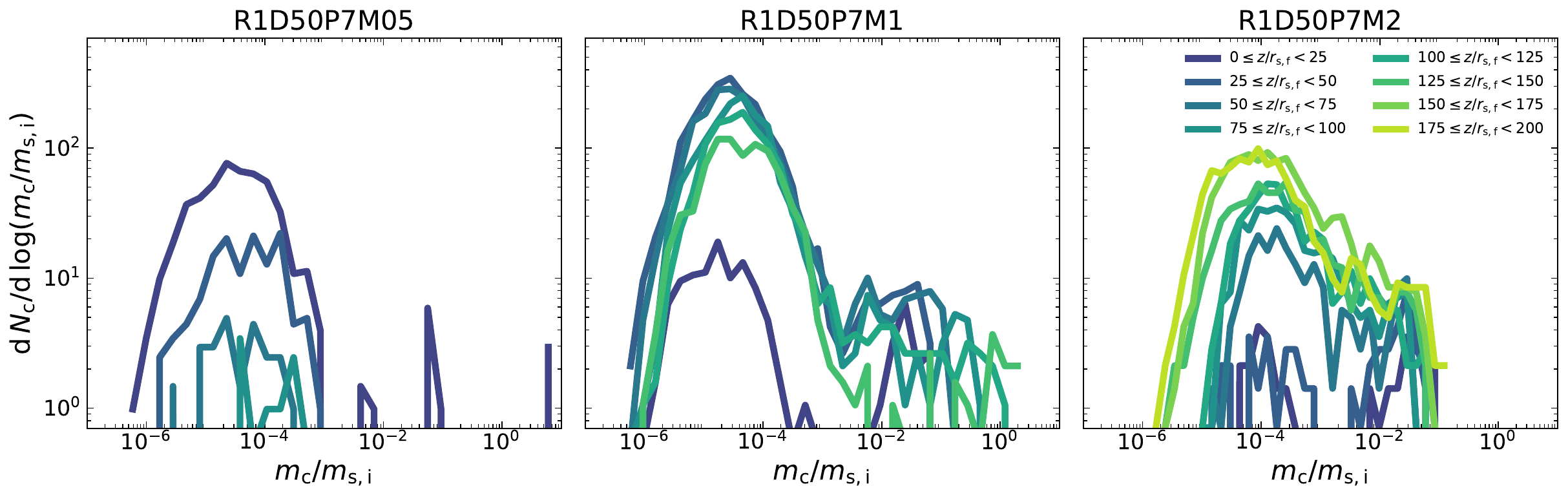}
    \caption{Clump mass distributions along the stream direction for three simulations with different Mach numbers at fixed \ifm{\chix{f}=50}, averaging over the final 10 snapshots in the interval \ifm{1.8\text{--}2\,\tx{box}}. Clumps are identified using the \software{RAMSES} clump finder, and their masses are normalized by \ifm{\mx{s,i}}, the total initial cold-stream mass within the simulation box. The vertical axis \ifm{\mathrm{d}\,N_\mathrm{c}/\mathrm{d}\log (\mx{c}/\mx{s,i})} is the number of clumps per logarithmic mass bin. Dark curves correspond to distributions measured at small \ifm{z}, while bright curves correspond to large \ifm{z}. In the left and central panels, distributions at large \ifm{z} are omitted to avoid contamination from initial-condition effects.
    }
    \label{fig:clump_mass_dist}
\end{figure*}


\smallskip
In contrast, the simulations shown in the central and right columns satisfy \ifm{\Max{s}>\chix{i}^{-1/2}} and therefore exhibit successful penetration; these maps are shown at the final snapshot, \ifm{t=2\,\tx{box}}. Despite the steep shock angle in \run{R1D100P7M05}, no transition from an oblique shock to a bow shock is observed. This is consistent with its pressure contrast lying below the focused-core pressure criterion in \cref{eq:bowshocktransition}. In the supersonic run \run{R1D100P7M2}, a clear Mach wave is visible in the pressure panel with an opening angle given by \ifm{\thetax{M}=\arcsin{\Max{s}^{-1}}=30\degr}, trailing the cold stream as a consequence of compressive motion. The vertical velocity component \ifm{\vx{y}} is also significantly enhanced compared to subsonic cases, indicating stronger body-mode KHI. Nevertheless, only limited fragmentation is observed around the stream, reflecting the more efficient shear-driven disruption and the shorter \ifm{\tx{box}} available for fragmentation to develop.

\smallskip
\cref{fig:profile_mass_Mach} presents the cold-stream line density, streamwise mass flux, and velocity profiles for varying Mach numbers and density contrasts, at fixed stream size \ifm{\rx{s,f}=1\,\mathrm{kpc}} and pressure contrast \ifm{\mathcal{P}=7}. The gray curves correspond to the run \run{R1D100P7M01} at \ifm{t=0.1\,\tx{box}}. As this case shows no stream penetration, the spatial profiles are not physically meaningful, and we therefore focus only on the mean values averaged along \ifm{z}. Varying the Mach number leaves the fragmentation criterion unchanged but significantly affects the survival criterion and the corresponding borderline region. In line with the survival criterion derived for cases without a pressure contrast, increasing the Mach number generally suppresses the growth of cold-gas line density. Both \run{R1D50P7M2} and \run{R1D100P7M2} exhibit initial stream disruption; however, in \run{R1D50P7M2} the velocity declines rapidly, allowing the cold-gas line density to recover and grow at \ifm{z>100\,\rx{s,f}}, similarly to \run{R1D100P7M1}. Notably, for disruption cases with \ifm{\chix{f}=300}, a supersonic inflow with \ifm{\Max{s}=2} leads to a smaller reduction in cold-gas line density than the \ifm{\Max{s}=1} case, indicating a less efficient disruption mechanism in the supersonic Kelvin--Helmholtz regime. The streamwise mass flux broadly follows the line-density evolution, while multiplication by the velocity reduces the apparent differences among simulations. In the right panel, higher Mach numbers result in weaker velocity decay, reflecting the shorter timescale available for momentum exchange between the hot and cold phases.

\smallskip
\cref{fig:profile_morph_Mach} presents the cold-stream morphological profiles for these same simulations.
The largest number of fragmented clumps occurs in \run{R1D100P7M01}, where the velocity shear is minimal. In this case, the clump distances approach the numerical limit discussed above, so we do not over-interpret their absolute values. For cases with stream penetration, increasing velocity shear drives a clear transition in behavior. In the coagulation regime, represented by \run{R1D50P7M05} and \run{R1D50P7M1}, the number of clumps first rises to a maximum and then declines. This gives way to the borderline case \run{R1D50P7M2}, in which most early-time fragments are destroyed by velocity shear, but the clump number subsequently increases as the stream decelerates, closely resembling the behavior of \run{R1D100P7M1}. The run \run{R1D100P7M05} exhibits clear fragmentation, with both the number and the characteristic distance of clumps increasing rapidly after penetration before reaching an apparent saturation around \ifm{z=50\,\rx{s,f}}. The run \run{R1D300P7M05} lies at the threshold of survival, whereas in \run{R1D100P7M2}, \run{R1D300P7M1}, and \run{R1D300P7M2} all three morphological measures remain small due to the progressive loss of cold mass along the stream.

\smallskip
\cref{fig:clump_mass_dist} extends the analysis by examining the clump mass distributions along the stream direction in three simulations with increasing Mach number, all with \ifm{\chix{f}=50}, $r_{\rm s,f}=1\,{\rm kpc}$, and $\mathcal{P}=7$. In the lowest Mach-number case, \run{R1D50P7M05}, the number of clumps decreases substantially with increasing \ifm{z}, indicating efficient coagulation along the stream. The apparent absence of very massive clumps at large \ifm{z} arises because most of the cold gas remains connected as a coherent stream. As a result, the clump finder identifies it as a single extended structure and does not count it repeatedly once it has already been detected at smaller \ifm{z}.

\smallskip
In the intermediate Mach-number case, \run{R1D50P7M1}, the clump mass distribution first broadens from \ifm{0\leq z/\rx{s,f}<25} to \ifm{25\leq z/\rx{s,f}<50}, and then becomes slightly narrower at larger distances. This non-monotonic behavior, in contrast to the low-Mach case, reflects the faster motion of the cold stream and the shorter box-crossing time. Fragmentation is therefore not fully developed in the first \ifm{z} bin, becomes prominent in the second bin, and is subsequently overtaken by coagulation. In both cases, the low-mass end of the distribution is limited by numerical resolution. 

\smallskip
In the highest Mach-number case, \run{R1D50P7M2}, the distribution is initially biased toward relatively massive clumps and shifts gradually toward lower masses with increasing \ifm{z}. Here, the low-mass end lies above the resolution limit, suggesting that strong velocity shear efficiently destroys small fragments at early times. As the cold stream decelerates, smaller clumps can survive for longer, producing the observed downstream extension of the distribution toward lower clump masses.

\smallskip
In summary, velocity shear plays a central role in shaping both the mass budget and morphology of cold streams. Shear directly influences the survival of the stream and its fragments, and this survivability evolves with penetration distance as the inflow velocity of the stream changes.

\section{Discussion}
\label{sec:discuss}
\subsection{Mass loading and momentum fluxes in all survival cases}

In \cref{fig:profile_mass_theory}, we examine the streamwise transport of cold gas in all survival cases. The left panel shows the normalized streamwise mass flux, \ifm{\dotm{s}=\lambdax{s} \vx{s}}, where \ifm{\lambdax{s}} is the cold-gas line density and \ifm{\vx{s}} is the mass-weighted streamwise velocity. By definition, all survival cases retain a growing or approximately sustained cold component along the stream, although temporary decreases in \ifm{\dotm{s}} can occur in borderline cases. The coagulation cases generally show the strongest increase in \ifm{\dotm{s}}, followed by the fragmentation and borderline cases. This trend reflects the decreasing efficiency of radiative cooling and boundary entrainment from coagulation to borderline regimes.

\begin{figure*}
    \centering	
    \includegraphics[width=\textwidth]{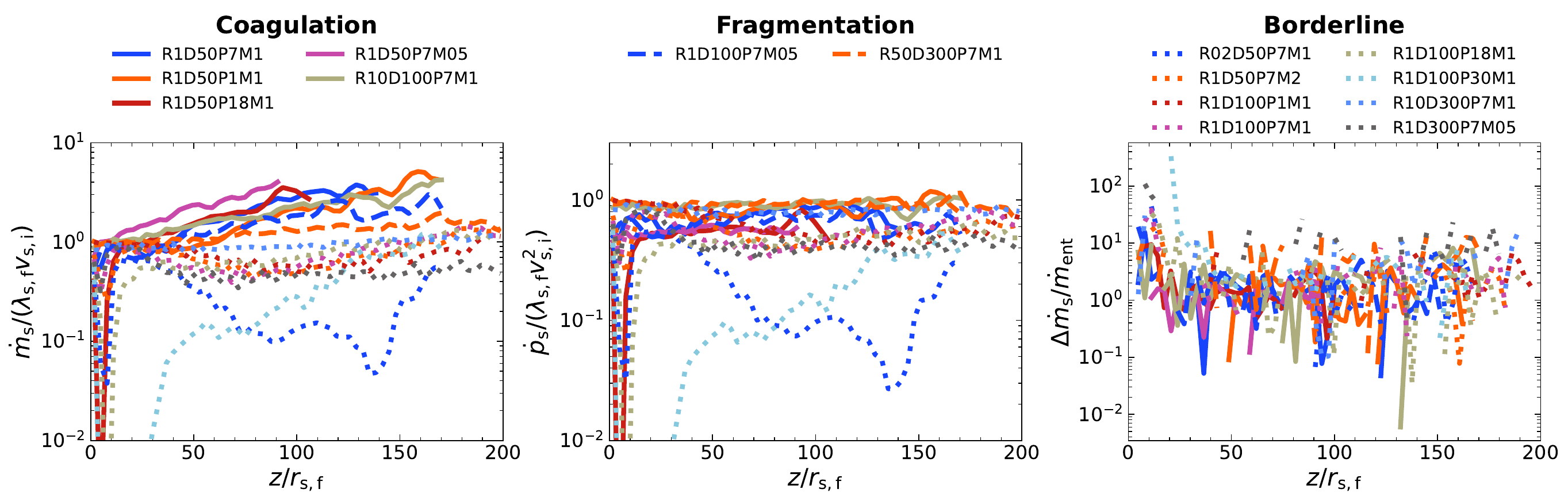}
    \caption{Profiles of cold-gas transport quantities along the stream direction for all survival cases. The left and middle panels show the streamwise mass flux, \ifm{\dotm{s}=\lambdax{s} \vx{s}}, and the mean streamwise momentum flux, \ifm{\dot{p}_\mathrm{s}=\lambdax{s} \vx{s}^2}, each normalized by \ifm{\lambdax{s,f}\vx{s,i}} and \ifm{\lambdax{s,f}\vx{s,i}^2}, respectively. The mean streamwise momentum flux retains only the mean-flow (ram) term of the full momentum flux, which we have verified dominates over the velocity-dispersion (Reynolds-stress) and thermal-pressure contributions. The right panel shows the local mass-flux change rate, normalized by the entrainment-rate prediction. Solid, dashed, and dotted lines denote coagulation, fragmentation, and borderline regimes, respectively. Bins with negative mass changes are excluded in the right panel to allow logarithmic scaling. In all survival cases, \ifm{\dotm{s}} generally increases downstream due to boundary entrainment, while \ifm{\dot{p}_\mathrm{s}} remains approximately conserved after pressure equilibrium is restored. The mass-flux change rates initially exceed the entrainment prediction immediately after injection because of pressure-driven transients, but subsequently relax toward the theoretical expectation once the stream re-establishes pressure equilibrium.}
    \label{fig:profile_mass_theory}
\end{figure*}

\smallskip
To quantify this interpretation, the right panel compares the measured mass-flux change rate across bins, \ifm{\Delta \dotm{s}}, with the entrainment rate expected from radiative mixing-layer theory \citep{Ji.etal.19,Gronke.Oh.20a,Fielding.etal.20,Mandelker.etal.20a,Tan.etal.21}. In a steady flow with no explicit time dependence of the density, the continuity equation tells us that the divergence of density times velocity equals the local rate of mass addition or removal by sources or sinks. In this case, this implies that $\partial \dotm{s}/\partial z = \dotm{ent}/\Delta z$, where 
\begin{equation}
\dotm{ent}=\rhox{h}\vx{mix}\Ax{s} 
\end{equation}
is the mass entrainment rate through the cold-gas surface\footnote{More accurately, through the turbulent mixing layer at the interface of cold and hot gas.} in the bin of width $\Delta z$. Here, \ifm{\Ax{s}} is the cold--hot interfacial area within each bin, and
\ifm{\vx{mix}\sim0.4\cx{s}(\rx{s,f}/\lsh)^{1/4}} \citep{Gronke.Oh.20a,Yao.etal.25}. The measured mass-flux growth initially exceeds the radiative mixing-layer prediction in most survival cases. This excess reflects the transient restoring phase driven by the imposed pressure contrast: compression and shock heating temporarily modify both the cold-gas line density and the streamwise velocity before the stream re-establishes pressure equilibrium with the ambient medium. At larger distances, the measured growth rates relax toward the mixing-layer expectation, supporting the interpretation that pressure contrast mainly amplifies the early transient response, while the subsequent long-term mass loading is governed by boundary entrainment.

\smallskip
The middle panel shows the normalized \emph{mean} streamwise momentum flux, \ifm{\dot{p}_\mathrm{s}=\lambdax{s} \vx{s}^2}. Solid, dashed, and dotted curves denote the coagulation, fragmentation, and borderline regimes, respectively. The full streamwise momentum flux transported across a stream cross-section is \ifm{\Pi=\int\left(\rho\,\vx{z}^2+P\right)\mathrm{d}A}, which we decompose as
\begin{equation}
\Pi\simeq\lambdax{s}\vx{s}^2+\int\rho\left(\vx{z}-\vx{s}\right)^2\mathrm{d}A+\int P\,\mathrm{d}A,
\end{equation}
the sum of a mean-flow (ram) term, a velocity-dispersion (Reynolds-stress) term, and a thermal-pressure term, where \ifm{\vx{z}} is the local streamwise velocity and \ifm{\vx{s}} its mass-weighted mean. We have measured all three terms directly and find that the mean-flow term \ifm{\lambdax{s}\vx{s}^2} dominates over both the velocity-dispersion and thermal-pressure contributions throughout the survival cases. We therefore retain only this term and refer to \ifm{\dot{p}_\mathrm{s}=\lambdax{s}\vx{s}^2} as the mean streamwise momentum flux. 

\smallskip
As shown in the middle panel, \ifm{\dot{p}_\mathrm{s}} is approximately conserved in the survival regime, especially in the coagulation and fragmentation cases. This does not imply that the streamwise mass flux \ifm{\dotm{s}} is conserved. Instead, boundary entrainment increases \ifm{\dotm{s}} by adding cold gas to the stream, while momentum exchange with the initially static background reduces \ifm{\vx{s}}. If the entrained gas carries little initial streamwise momentum and pressure-gradient forces are subdominant after equilibration, the increase in \ifm{\lambdax{s} \vx{s}} can be offset by the decrease in \ifm{\vx{s}}, leaving \ifm{\lambdax{s} \vx{s}^2} approximately constant. The larger deviations in the borderline cases indicate stronger cold-gas loss and less complete recovery to a coherent, momentum-conserving stream.

\smallskip
The close correlation between the cold--hot interfacial area per unit stream length and the streamwise mass flux mentioned at the end of \cref{sec:size} can be understood from the approximate conservation of the mean streamwise momentum flux. For a cylindrical stream, \ifm{\Sx{s}\simeq2\pi\rx{s}} and 
\begin{equation} 
\frac{\dotm{s}}{\Sx{s}} \simeq \frac{\rhox{s}\pi\rx{s}^2\vx{s}}{2\pi\rx{s}}=
\frac{\rhox{s}\rx{s}\vx{s}}{2}.
\end{equation}
Using the approximate momentum-flux conservation,
\begin{equation}
\dot{p}_\mathrm{s}\equiv\rhox{s}\pi\rx{s}^2\vx{s}^2\simeq\dot{p}_0,
\end{equation}
this becomes
\begin{equation}
\frac{\dotm{s}}{\Sx{s}}
\simeq
\left(\frac{\rhox{s}\dot{p}_0}{4\pi}\right)^{1/2}.
\end{equation}
After the stream reaches pressure equilibrium, \ifm{\rhox{s}\simeq\rhox{s,f}} varies only weakly among the survival cases. Therefore, \ifm{\dotm{s}/\Sx{s}} is expected to remain approximately constant, explaining why the evolution of \ifm{\Sx{s}} closely follows that of the streamwise mass flux.

\subsection{Applications to galaxy evolution}
\label{sec:galaxy}

We now connect our idealized stream-penetration experiments to the cosmological evolution of cold streams in massive halos. The central question is whether a cold stream penetrating a virial shock around a dark matter halo remains coherent, fragments into long-lived clumps, or is disrupted before reaching the central galaxy. This can be assessed by comparing the characteristic stream radius in the CGM, \(r_\mathrm{s,f}\), with two critical scales: the shear-survival radius, \(r_\mathrm{crit,surv}\), and the fragmentation radius, \(r_\mathrm{crit,frag}\). The first determines whether the stream survives against shear-driven disruption, while the second determines whether fragmented cold structures coagulate back into a coherent stream or remain as separated clumps.

\smallskip
As shown in Eq.~(42) of \citet{Yao.etal.25}, the cold stream radius in the CGM can be written as
\begin{equation}
\label{eq:Rs_halo}
r_\mathrm{s,f}\simeq \SI{7}{kpc}(1+z)_3^{-1}\left(\frac{\fx{acc,0.3}\fx{cold,0.2}\Thetax{s}}{\fx{h,0.3}\Max{s}\Thetax{h}^{3/2}}\right)^{1/2},
\end{equation}
where \ifm{(1+z)_3\equiv(1+z)/3}, \ifm{\fx{acc,0.3}} is the fraction of total accretion onto the halo carried by a single stream normalized to a fiducial value of 0.3, \ifm{\fx{cold,0.2}} is the cold gas fraction in the cosmic web filament normalized to 0.2,
\ifm{\fx{h,0.3}} is the hot gas fraction in the halo normalized to 0.3, \ifm{\Thetax{s}\equiv \Tx{s}/10^4\,\mathrm{K}} is the normalized cold-stream temperature, \ifm{\Thetax{h}\equiv \Tx{h}/\Tx{v}} is the hot-CGM temperature normalized by the virial temperature, and $\mathcal{M}_{\rm s}\simeq1.34\Max{v}\Thetax{h}^{-1/2}$ is the stream velocity normalized by the sound speed of the hot CGM \citep{Mandelker.etal.16}. Here, \ifm{\Max{v}} denotes the stream velocity normalized by the virial velocity, as defined in \citet{Yao.etal.25}. This expression gives the post-equilibration stream radius after the stream has adjusted to pressure balance with the ambient hot CGM, and therefore specifies \(r_\mathrm{s,f}\) as a function of halo mass and redshift. Note that the fiducial stream size is roughly a factor of two smaller than that in \citet{Yao.etal.25}, since we adopt a fiducial cold gas fraction of 0.2, estimated in \citet{Lu.etal.24} by averaging over ten filaments at \ifm{\Mx{v}\simeq\SI{1e12}{\Msun}} and \ifm{z\simeq4}. This parameter is highly uncertain and may depend on halo mass and redshift \citep{Aung.etal.24}, but properly characterizing it requires more detailed work on cold streams within intergalactic filaments, which is beyond the scope of the current study. We leave this to future work.

\smallskip
For the survival criterion in this framework, \citet{Mandelker.etal.20b} expressed the critical stream radius as
\begin{equation}
\label{eq:Rshear_halo}
r_\mathrm{crit,surv}\simeq0.9\,\mathrm{kpc}\,
M_{12}^{1/3}(1+z)_3^{-5/2}
\frac{\alpha_{0.1}\Max{s}\Theta_\mathrm{s}^{1/2}\Theta_\mathrm{h}}
{\Delta_{1/6}f_\mathrm{h,0.3}\Lambda_\mathrm{mix,-22.5}},
\end{equation}
where \ifm{M_{12}\equiv \Mx{v}/\SI{1e12}{\Msun}}, and \ifm{\Delta_{1/6}\equiv \Delta(\Rx{v})/(1/6)} is the ratio of the halo density at  $R_{\rm v}$ to the mean density in the halo normalized by \ifm{1/6}. Here \ifm{\alpha_{0.1}=\alpha/0.1} normalizes the dimensionless shear-layer growth coefficient defined below \cref{eq:Rshear}. Cold streams are expected to survive shear disruption when
\begin{align}
\label{eq:Rs_over_Rshear_halo}
\frac{r_\mathrm{s,f}}{r_\mathrm{crit,surv}}&\simeq8M_{12}^{-1/3}(1+z)_3^{3/2}\frac{\Delta_{1/6}\Lambdax{mix,-22.5}(\fx{acc,0.3}\fx{cold,0.2}\fx{h,0.3})^{1/2}}{\alpha_{0.1}\Max{s}^{3/2}\Thetax{h}^{7/4}} \notag\\
&>1.
\end{align}
This condition is therefore harder to satisfy in massive halos and toward lower redshift, where the hot CGM is more extended, hotter, and less efficient at radiative cooling.

\begin{figure*}
    \centering
    \includegraphics[width=\textwidth]{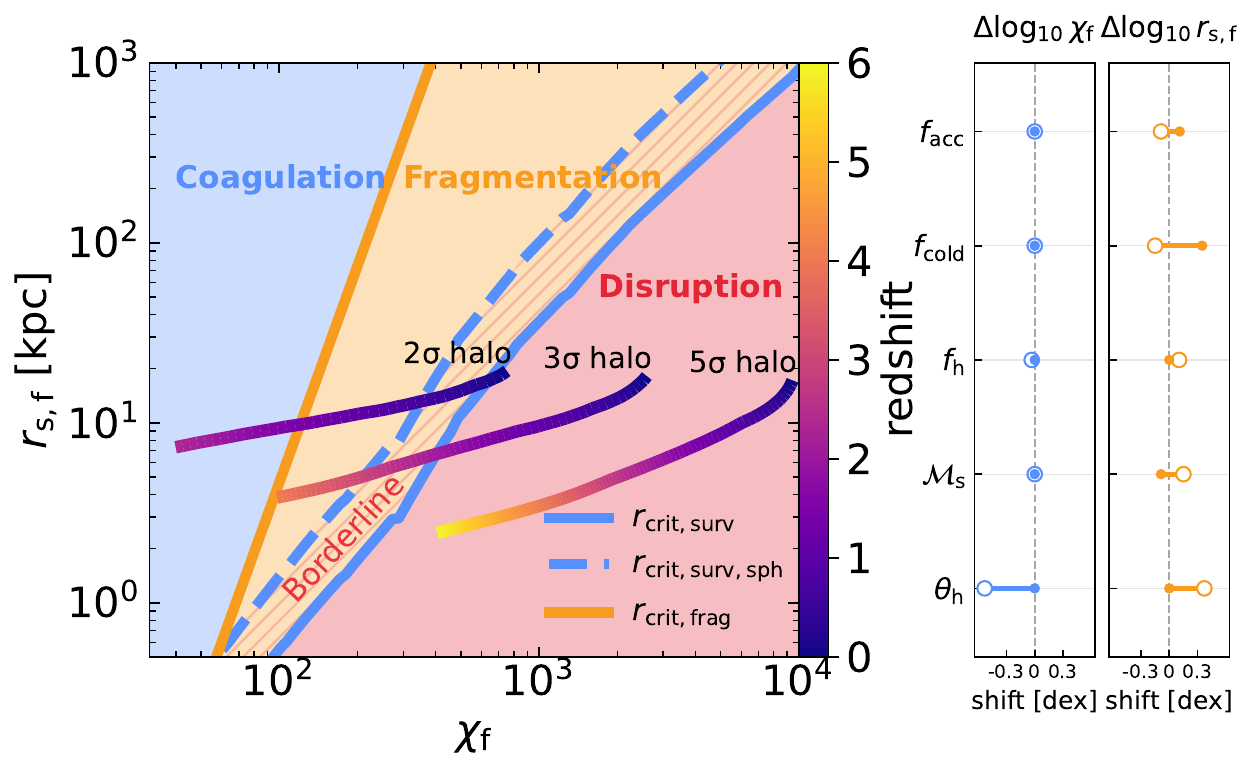}
    \caption{\textit{Left:} Similar to \cref{fig:diag_size_chi}, but including evolutionary tracks of halos with different peak heights, color-coded by redshift, for the fiducial model parameters shown in the text. We require halos to remain above \ifm{\Mx{shock}\sim\SI{6e11}{\Msun}} in order to sustain a hot CGM environment \citep{Dekel.Birnboim.06}. Along these evolutionary tracks, halos progressively transition from the coagulation regime to fragmentation, and eventually to disruption toward lower redshift. \textit{Right:} Changes in \ifm{\log_{10}\chix{f}} (blue) and \ifm{\log_{10}\rx{s,f}} (orange) relative to the fiducial model when varying one input parameter at a time over its adopted range, with the remaining inputs held fixed and the equilibrium temperature determined self-consistently. Open and filled circles denote the lower and upper parameter values, respectively; connecting segments indicate the corresponding span of shifts. Vertical dashed lines mark zero change.}
    \label{fig:diag_size_chi_halo}
\end{figure*}

\smallskip
A second criterion becomes relevant once the cold stream survives shear-driven disruption: whether fragmented cold structures coagulate back into a coherent stream or remain as separated clumps. This depends on the shattering length scale in the streams, which varies with the stream density, equilibrium temperature, and cooling function. The stream density as a function of halo mass and redshift was evaluated by \citet{Mandelker.etal.20b} to be  
\begin{equation}
\label{eq:ns_halo}
\nx{s,f} = 5.1\times 10^{-3}\,{\rm cm^{-3}}\,M_{12}^{2/3}(1+z)_3^4\frac{\Theta_{\rm h}\Delta_{1/6}f_{\rm h,0.3}}{\Theta_{\rm s}}.
\end{equation}
Unlike the fixed final density in the idealized simulation suite, this density varies with halo mass and redshift. The cosmological application therefore uses the density and temperature dependence of the physical criteria, rather than assigning \ifm{\nx{s,f}=\SI{0.01}{cm^{-3}}} to every halo. Combining this with the definition of the shattering length scale shown below \cref{eq:Rcoag} yields
\begin{equation}
\lsh\simeq\SI{93}{pc}M_{12}^{-2/3}(1+z)_3^{-4}
\frac{\Thetax{s}^{5/2}}
{\Delta_{1/6}\fx{h,0.3}\mux{s,0.6}^{3/2}\Thetax{h}\Lambda_\mathrm{1.5T_\mathrm{s},-22.8}},
\end{equation}
where $\Lambda_\mathrm{1.5T_\mathrm{s},-22.8}$ is the cooling function evaluated at a temperature of $T=1.5T_{\rm s}$, where the cooling time tends to be at its minimum \citep{Mandelker.etal.20a}, normalized to a value of $10^{-22.8}\,{\rm erg\,s^{-1}\,cm^3}$. 
We combine this with the expression for the density contrast \citep{Mandelker.etal.20b} 
\begin{equation}
\label{eq:chif}
    \chi_{\rm f} = 100\,M_{12}^{2/3}\,(1+z)_3\,\frac{\Theta_{\rm h}}{\Theta_{\rm s}},
\end{equation}
and substitute them into \cref{eq:Rcoag} to obtain 
\begin{equation}
\label{eq:Rcoag_halo}
r_\mathrm{crit,frag}\simeq2.3\,\mathrm{kpc}\,
M_{12}^2
\frac{\Theta_\mathrm{h}^3}
{\Theta_\mathrm{s}^{3/2}\mu_\mathrm{s,0.6}^{3/2}\Delta_{1/6}
f_\mathrm{h,0.3}\Lambda_\mathrm{1.5T_\mathrm{s},-22.8}} ,
\end{equation}
which has no explicit redshift factor and scales as \ifm{\rx{crit,frag}\propto M_\mathrm{v}^2} when the cooling and dimensionless factors are held fixed. The coagulation regime corresponds to
\begin{align}
\label{eq:Rs_over_Rcoag_halo}
\frac{\rx{s,f}}{\rx{crit,frag}}&\simeq3M_{12}^{-2}(1+z)_3^{-1} \notag\\
&\times\frac{\Delta_{1/6}\mux{s,0.6}^{3/2}\Thetax{s}^2\Lambdax{1.5\Tx{s},-22.8}(\fx{acc,0.3}\fx{cold,0.2}\fx{h,0.3})^{1/2}}{\Max{s}^{1/2}\Thetax{h}^{15/4}}>1,
\end{align}
where fragmented cold structures can efficiently merge back into a coherent stream. This condition is therefore most difficult to satisfy in massive halos, where \(\ifm{r_\mathrm{crit,frag}}\) increases steeply with halo mass. In this regime, cold gas is more likely to remain fragmented, making stream fragmentation especially important for rare, massive halos.

\smallskip
These criteria are summarized in \cref{fig:diag_size_chi_halo}, where the critical radii are compared with halo evolutionary tracks of different peak heights. The figure assumes a fixed, mild pressure contrast at penetration and does not include the bow-shock transition, thereby isolating the effects of stream size and density contrast. Halos associated with \(1\sigma\) peaks mostly remain in the cold-accretion regime and are typically embedded within filaments. They are therefore not expected to undergo the stream-penetration scenario considered here and are not shown. In contrast, higher-\(\sigma\) halos pass through the hot-CGM regime, in which cold streams penetrate the virial shock. Along these evolutionary tracks, the expected sequence is coagulation, fragmentation, and eventual disruption at low redshift. Thus, \cref{fig:diag_size_chi_halo} suggests that stream fragmentation is most relevant for rare, massive halos at \ifm{z\sim2\text{--}4}, where a hot halo has already formed but shear has not yet fully disrupted the penetrating stream.

\smallskip
The right-hand panels of \cref{fig:diag_size_chi_halo} illustrate the sensitivity of this cosmological mapping to the assumed model parameters. Motivated by the parameter ranges discussed in \citet{Mandelker.etal.20b}, we explore \ifm{\fx{acc}\in[0.2,0.5]}, \ifm{\fx{cold}\in[0.1,1]}, \ifm{\fx{h}\in[0.2,0.3]}, \ifm{\Max{s}\in[0.75,2.25]}, and \ifm{\Thetax{h}\in[3/8,1]}. The fiducial values are chosen to be \ifm{\fx{acc}=0.3}, \ifm{\fx{cold}=0.2}, \ifm{\fx{h}=0.3}, \ifm{\Max{s}=1.5}, and \ifm{\Thetax{h}=1}. Each input parameter is varied individually while the other inputs remain at their fiducial values. For each choice, \ifm{\Thetax{s}} is determined from thermal equilibrium with the UVB at the corresponding redshift and a metallicity of \ifm{0.03\,Z_{\sun}}, giving \ifm{\Thetax{s}\simeq0.7\text{--}5} across the plotted range. Varying metallicity could introduce additional scatter, but we do not consider it here given the many parameters already involved. The resulting shifts, \ifm{\Delta\log_{10}Q\equiv\log_{10}(Q/Q_\mathrm{fid})} for \ifm{Q=\chix{f}} or \ifm{\rx{s,f}}, quantify the horizontal and vertical displacements of the halo tracks and hence their sensitivity to the adopted parameter choices. We therefore treat \cref{fig:diag_size_chi_halo} as illustrative of the qualitative trend rather than a quantitatively accurate prediction.

\smallskip
A rough estimate of the cold-gas covering fraction can be obtained by assuming that three cold streams feed the central galaxy and that each stream has a cone-like geometry, narrowing toward the galaxy center. In the absence of fragmentation, this gives
\begin{equation}
\label{eq:fc}
\fx{c}\sim\frac{3\rx{s,f}\Rx{v}}{\pi \Rx{v}^2}\sim0.07M_{12}^{-1/3},
\end{equation}
where we have adopted the virial-radius scaling from \citet{Dekel.etal.13},
\begin{equation}
\label{eq:rv}
\Rx{v}\simeq\SI{100}{kpc}M_{12}^{1/3}(1+z)_3^{-1}.
\end{equation}
At fixed dimensionless parameters, the covering fraction in \cref{eq:fc} has no explicit redshift dependence and decreases weakly with halo mass. In the fragmentation regime, however, cold clumps can spread to distances of at least \(\ifm{10\,\rx{s,f}}\) from the stream axis\footnote{In our simulations, this value is limited by the transverse extent of the highest-resolution region.}, potentially increasing the covering fraction by an order of magnitude and approaching \(\ifm{\fx{c}\sim0.7}\) at \ifm{\Mx{v}=\SI{1e12}{\Msun}}. We defer a more quantitative assessment of this covering-fraction enhancement to future simulations with more realistic setups.

\smallskip
We caution that the direct combination of these two criteria is valid only within the oblique-shock penetration scenario. It also implicitly assumes that fragmentation, coagulation, and disruption occur after the stream has reached pressure equilibrium over a distance much shorter than the halo radius. Applying the same galaxy-evolution framework to \cref{eq:zeq,eq:rv}, we obtain
\begin{align}
\label{eq:zeq_rv}
\frac{\zx{eq}}{\Rx{v}}&\simeq\frac{\eta^{1/2}\chix{f}^{1/2}\Max{s}\rx{s,f}}{\SI{100}{kpc}\mathcal{P}^{1/2}M_{12}^{1/3}(1+z)_3^{-1}}\notag\\
&\simeq(1+z)_3^{1/2}\left(\frac{\fx{acc,0.3}\fx{cold,0.2}\Max{s}}{\fx{h,0.3}\Thetax{h}^{1/2}}\right)^{1/2}.
\end{align}
This ratio may be reduced if the virial-shocked region is larger, if the cylindrical shock propagates faster in a stratified CGM, or if angular momentum delays the infall of the cold stream toward the galaxy. Nevertheless, the back-of-the-envelope estimate \(\zx{eq}\sim\Rx{v}\) indicates that directly combining the two criteria is not always straightforward. In particular, there may be cases in which the cold stream remains underpressurized throughout most of its journey through the halo.

\begin{figure*}
    \centering	
    \includegraphics[width=\textwidth]{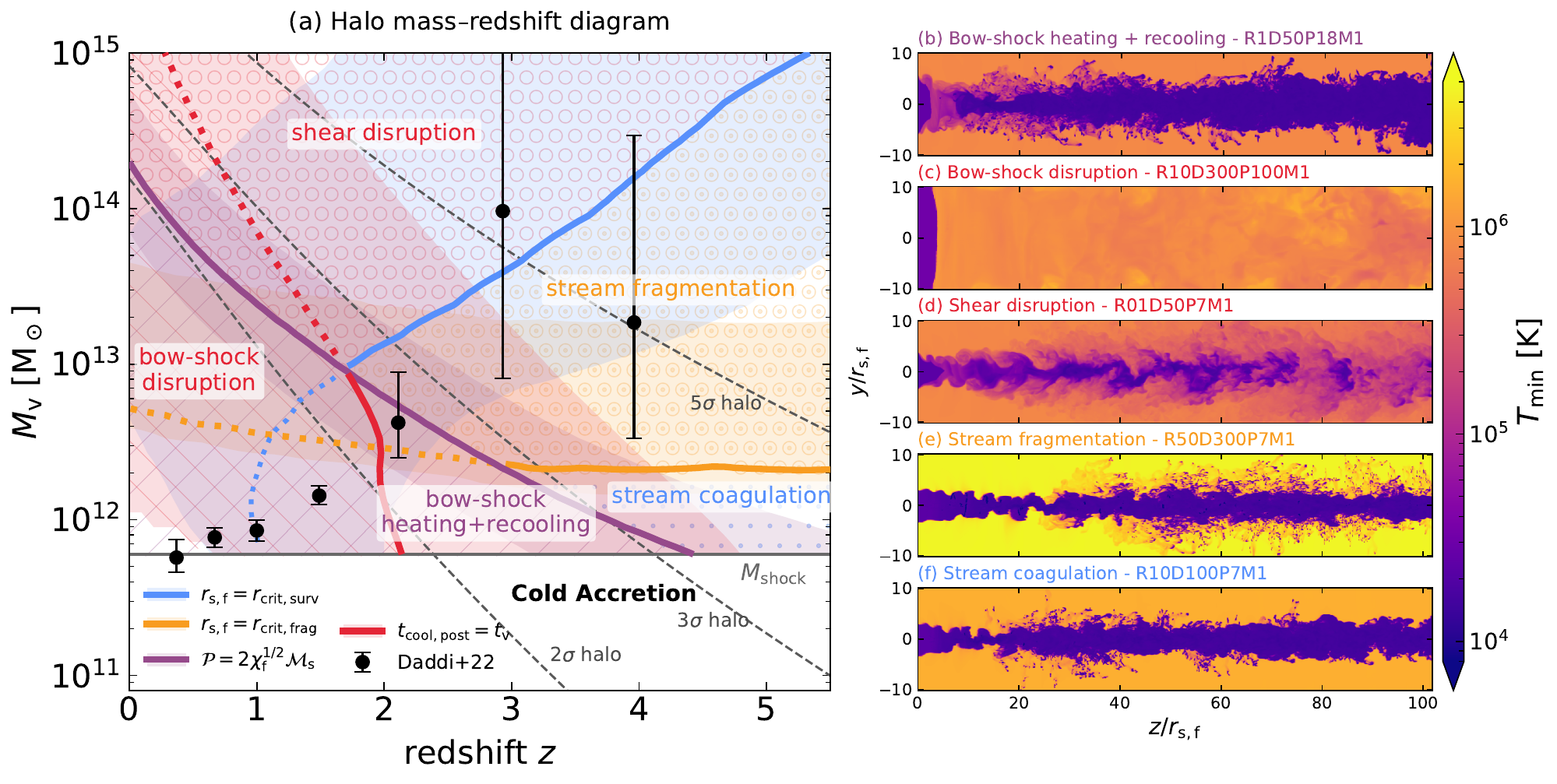}
    \caption{Cold-stream penetration regimes in the halo mass--redshift plane (panel a) and representative simulations of each regime (panels b--f). \textit{Panel (a)}: curves show thresholds for the fiducial parameters, and shaded bands show the sampled parameter sensitivity by varying the parameters in the adopted range. The horizontal gray line marks the appearance of a stable virial shock, \ifm{\Mx{shock}\simeq\SI{6e11}{\Msun}} \citep{Dekel.Birnboim.06}. The blue curve marks the shear-survival threshold, \ifm{\rx{s,f}=\rx{crit,surv}} (\cref{eq:Rs_over_Rshear_halo}), and the orange curve marks the fragmentation threshold, \ifm{\rx{s,f}=\rx{crit,frag}} (\cref{eq:Rs_over_Rcoag_halo}). The purple curve marks the bow-shock heating criterion, \ifm{\mathcal{P}=2\chix{f}^{1/2}\Max{s}} (\cref{eq:bowshock_halo}). The red curve marks \ifm{\tx{cool,post}=\tx{v}} (\cref{eq:tcoolpost2tv}), separating shock-heated streams that can cool within a virial crossing time from those that remain hot. Hatching marks the fiducial regimes: blue dots for stream coagulation, orange circles with dots for stream fragmentation, red circles for shear disruption, purple slashes for bow-shock heating, and red backslashes for bow-shock disruption (\ifm{\tx{cool,post}>\tx{v}}). The annotated ``bow-shock heating+recooling'' region, where purple slashes appear without red backslashes (\ifm{\tx{cool,post}<\tx{v}}), marks streams that are shock heated but recool within a virial crossing time. Dotted curve segments indicate where the corresponding criterion is outside its adopted domain of applicability (see text for details). Fragmentation and coagulation are shown only for streams that survive shear disruption in a hot halo and lie outside the bow-shock heating regime. Black points with error bars show halo masses inferred from the bending of the SF main-sequence by \citet{Daddi.etal.22b}, providing observational context. Black dashed lines show halo evolutionary tracks for \ifm{2\sigma}, \ifm{3\sigma}, and \ifm{5\sigma} peaks. \textit{Panels (b)--(f)}: minimum-temperature projections along the \ifm{x}-axis at \ifm{t=2\,\tx{box}} for one simulation from each regime: bow-shock heating+recooling (\run{R1D50P18M1}), bow-shock disruption (\run{R10D300P100M1}), shear disruption (\run{R01D50P7M1}), stream fragmentation (\run{R50D300P7M1}), and stream coagulation (\run{R10D100P7M1}). Each panel spans \ifm{0\le z\le100\,\rx{s,f}} along the stream and \ifm{|y|\le10\,\rx{s,f}} across it, and all panels share the color scale shown on the right. The panel titles are colored to match the regime labels in panel (a). The bow-shock heated stream in panel (b) recools within the domain, whereas in panel (c) it remains hot. The stream in panel (d) is mixed away by shear, while the streams in panels (e) and (f) survive, either as a persistently fragmented stream or after coagulating into a coherent stream.}
    \label{fig:diag_Mv_z}
\end{figure*}

\smallskip
In the above discussion and in \cref{fig:diag_size_chi_halo}, we neglect the effect of a varying pressure contrast. An estimate based on the mean pressures of cosmological halos and the cosmic web filaments that feed them gives the halo-mass and redshift dependence \citep[][Eq.~40]{Yao.etal.25} 
\begin{equation}
\label{eq:pratio}
    \mathcal{P}\simeq35M_{12}^{-0.11}(1+z)_3^{-1}\frac{\Delta_\mathrm{v,h}}{10\Delta_\mathrm{v,f}}\fx{acc,0.3}^{-1}\Max{s}\Thetax{h}^{1/2},
\end{equation}
where \ifm{\Delta_\mathrm{v,h}\sim200} is the halo virial overdensity in the spherical collapse model, and \ifm{\Delta_\mathrm{v,f}\sim20} is the virial overdensity in cosmic web filaments assuming cylindrical collapse. As emphasized by \citet{Yao.etal.25}, neither the CGM nor the cosmic-web filament is perfectly isobaric. The cold stream at the filament center is expected to have a pressure somewhat higher than the virial value, whereas the hot CGM near the halo outskirts should have a pressure somewhat lower than the virial value. Indeed, \citet{Lu.etal.24} find that the pressure increases by a factor of \ifm{\sim2} from the outer filament to the central cold stream. If the CGM exhibits a comparable pressure variation, then \ifm{\mathcal{P}} in \cref{eq:pratio} may be overestimated by a factor of \ifm{\gtrsim4}. A particularly interesting scenario occurs at high pressure contrast, as in \(\run{R1D100P30M1}\), where the stream undergoes an early bow-shock transition which directly heats it as it penetrates the virial shock. Although the cold stream eventually recovers in this specific simulation, this is not guaranteed across the full cosmological parameter space. A necessary condition for the recovery of cold accretion is that the post-shock cooling time be shorter than the virial crossing time, \(\ifm{\tx{cool,post}<\tx{v}}\) \citep{Dekel.Birnboim.06,Cornuault.etal.18}. This motivates \cref{fig:diag_Mv_z}, where we combine the shock-heating, survival, and fragmentation constraints in the more conventional halo mass--redshift diagram.

\smallskip
\cref{fig:diag_Mv_z} provides a global view of cold-stream penetration regimes in the halo mass--redshift plane, shown in panel (a), alongside minimum-temperature projections of one representative simulation from each regime in panels (b)--(f). We adopt the same fiducial parameters as in \cref{fig:diag_size_chi_halo}, showing parameter uncertainties as shaded regions. We fix \ifm{\fx{cold}=0.2} in this figure, since its large uncertainty would otherwise dominate over the others and make the plot less informative, and vary the remaining four parameters over the ranges given above. At halo masses below the classical shock-heating threshold, \(\ifm{M_\mathrm{shock}\simeq\SI{6e11}{\Msun}}\) \citep{Dekel.Birnboim.06}, accretion proceeds primarily through the cold mode, and the stream-penetration problem considered here is not applicable. We caution that this threshold, \(\ifm{M_\mathrm{shock}}\), should be regarded as an order-of-magnitude estimate: feedback can lower the effective transition mass to \(\ifm{\sim\SI{1e11}{\Msun}}\) \citep{Fielding.etal.17,Stern.etal.20,Stern.etal.21,Pandya.etal.23,Aung.etal.24}. Above \(\ifm{M_\mathrm{shock}}\), cold streams can reach the central galaxy if they survive shear-driven disruption and, when shock heated, cool within the halo-crossing time. The purple curve marks the condition for an early transition from an oblique shock to a detached bow shock during stream penetration, given by combining \cref{eq:bowshocktransition,eq:chif,eq:pratio}, 
\begin{equation}
\label{eq:bowshock_halo}
\frac{\mathcal{P}}{2\chix{f}^{1/2}\Max{s}}
\simeq
1.75M_{12}^{-0.44}(1+z)_3^{-3/2}
\frac{\Delta_\mathrm{v,h}\Thetax{s}^{1/2}}
{10\Delta_\mathrm{v,f}\fx{acc,0.3}}=1 .
\end{equation}
As emphasized in \cref{sec:pres}, this threshold should be regarded as an empirical guide from a sparse parameter survey, rather than as a densely calibrated criterion. The purple band in \cref{fig:diag_Mv_z} shows how varying \ifm{\fx{acc}} shifts this threshold when \ifm{\Thetax{s}} is determined from thermal equilibrium at each redshift. For a given parameter choice, the ratio exceeds unity below the corresponding purple boundary; together with the requirement \ifm{\Mx{v}>\Mx{shock}}, this defines the bow-shock heating regime, marked by purple slashes.

\smallskip
To estimate the cooling time of the post-shock gas, we need an estimate of the pre-shock density in the cold stream (outside the halo virial shock). This is given by the density in the cores of cosmic web filaments \citep{Lu.etal.24,Aung.etal.24,Yao.etal.25}. We assume that the cores are in pressure equilibrium with the shock-heated outer regions of filaments, which are heated to the filament virial temperature by a cylindrical accretion shock \citep{Lu.etal.24}. The stream density outside the halo virial shock thus becomes \(\ifm{\rhox{s,i}=\rhox{v,f}\Tx{v,f}/\Tx{s}}\), where $T_{\rm s}=\Theta_{\rm s}\times 10^4{\rm K}$
is the characteristic cold-stream temperature and \(\ifm{\rhox{v,f}}\) and \(\ifm{\Tx{v,f}}\) are the filament virial density and temperature (after the cylindrical accretion shock), given by Eqs.~(35) and (36) of \citet{Yao.etal.25} \citep[see also][]{Lu.etal.24}. This gives
\begin{equation}
\label{eq:rhosi}
\rhox{s,i}\simeq\SI{9.4e-27}{g\,cm^{-3}}M_{12}^{0.77}(1+z)_3^5 \, \frac{\Delta_\mathrm{v,f,20}\fx{b,0.17}\fx{acc,0.3}}{\Max{s}\Theta_{\rm s}\Theta_{\rm h}^{1/2}},
\end{equation}
where $\Delta_\mathrm{v,f,20}=\Delta_{\rm v,f}/20$ and \(\ifm{\fx{b,0.17}\equiv\fx{b}/0.17}\) is the baryon fraction normalized to 0.17. Combining this density estimate with \cref{eq:rho_post,eq:temp_post,eq:tcool_post}, we obtain the post-shock cooling time
\begin{equation}
    \tx{cool,post}\simeq\SI{25}{Myr}M_{12}^{-0.11}(1+z)_3^{-4}\frac{\Max{s}^3\Thetax{s}\Thetax{h}^{3/2}}{\Delta_\mathrm{v,f,20}\fx{b,0.17}\fx{acc,0.3}\Lambdax{post,-23}},
\end{equation}
where \ifm{\Lambdax{post,-23}\equiv\Lambdax{post}/\SI{1e-23}{erg\,s^{-1}\,cm^3}} assuming a metallicity of \ifm{0.03\,Z_{\sun}}. We compare this timescale with the virial crossing time from Eq.~(47) of \citet{Yao.etal.25}
\begin{equation}
\tx{v}\simeq\SI{500}{Myr}(1+z)_3^{-3/2}\Max{s}^{-1}\Thetax{h}^{-1/2},
\end{equation}
accounting for the Mach number of the cold stream. Their ratio is then
\begin{equation}
\label{eq:tcoolpost2tv}
\frac{\tx{cool,post}}{\tx{v}}
\simeq0.05M_{12}^{-0.11}(1+z)_3^{-5/2}
\frac{\Max{s}^4\Thetax{s}\Thetax{h}^2}
{\Delta_\mathrm{v,f,20}\fx{b,0.17}\fx{acc,0.3}\Lambdax{post,-23}},
\end{equation}
where the red line marks where this ratio equals unity. For streams undergoing bow-shock heating, the condition \(\ifm{\tx{cool,post}>\tx{v}}\) defines the region of direct bow-shock disruption, shown with red backslashes in \cref{fig:diag_Mv_z} and illustrated by \run{R10D300P100M1} in panel (c). In this regime, the shock-heated stream gas cannot cool efficiently before reaching the central galaxy, and thus remains hot over most of its penetration through the halo. For the fiducial parameters, the red boundary places this regime predominantly at \ifm{z\lesssim2}, with a weak halo-mass dependence in the transition. The surrounding band is mainly set by the Mach number of the stream, given \ifm{\tx{cool,post}/\tx{v}\propto\Max{s}^4} for \ifm{\Max{s}\in[0.75,2.25]}. We further note that the criterion is only valid where a bow-shock transition occurs; the red curve is therefore shown dotted above the purple curve, indicating the range where it is  invalid. Conversely, where \ifm{\tx{cool,post}<\tx{v}}, the purple slashes appear without red backslashes: streams there are shock heated at penetration but recool within the halo, as in \run{R1D100P30M1} and \run{R1D50P18M1} (panel b), and we annotate this region as ``bow-shock heating+recooling''.

\smallskip
This picture is consistent with the lower prevalence of cold streams in \software{TNG50} at \(\ifm{z<1}\) reported by \citet{Medlock.etal.26}, as well as with recent zoom-in cosmological simulations of low-redshift clusters, which find that filamentary inflows are shock heated as they approach \(\ifm{\Rx{v}}\) \citep{Kotecha.etal.22,Vurm.etal.23,Rost.etal.24,Paste.etal.26}. We emphasize that this criterion concerns the survival of the cold (\(\ifm{<\SI{5e4}{K}}\)) phase: the red region marks where shock-heated gas cannot re-cool before reaching the central galaxy, and should not be read as the destruction of the overdense filament itself. 
The derived equilibrium range \ifm{\Thetax{s}\simeq0.7\text{--}5} corresponds to stream temperatures \ifm{\Tx{s}\simeq7\times10^3\text{--}5\times10^4\,\mathrm{K}}; if streams reach massive low-redshift halos already warm (\(\ifm{T\sim\SI{1e6}{K}}\)), both the density contrast \(\ifm{\chix{f}}\) and the pressure contrast \(\ifm{\mathcal{P}}\) are substantially smaller, the converging shock is correspondingly weaker, and the survival of a distinct cold phase is no longer the relevant question.

\smallskip
Assuming that the cold stream is not bow-shock heated, the blue curve in \cref{fig:diag_Mv_z} shows the shear-survival threshold \ifm{\rx{s,f}=\rx{crit,surv}} from \cref{eq:Rs_over_Rshear_halo}. On the side where \ifm{\rx{s,f}<\rx{crit,surv}}, shown with red circles, streams are expected to be disrupted by KHI, as illustrated by \run{R01D50P7M1} in panel (d). The black points show the halo masses associated with the bend in the star-forming main sequence, as inferred by \citet{Daddi.etal.22b}. Their broad correspondence with the predicted transition provides observational context, while the spread in the blue band illustrates the sensitivity of this comparison to stream properties. Similar transitions have also been inferred from the bending of the star-formation-rate relation \citep{Popesso.etal.23}, the evolution of Ly\ifm{\alpha} emission \citep{Daddi.etal.22a}, and the evolution of the quiescent fraction of galaxies \citep{Behroozi.etal.19,Elias.etal.26}. Together, these results support a picture in which the suppression of efficient cold-stream penetration due to stream disruption is associated with the emergence of a stable hot CGM.

\smallskip
For streams that survive shear disruption and avoid bow-shock heating, the orange curve in \cref{fig:diag_Mv_z} shows the fragmentation threshold \ifm{\rx{s,f}=\rx{crit,frag}} from \cref{eq:Rs_over_Rcoag_halo}. Below this curve, where \ifm{\rx{s,f}>\rx{crit,frag}}, fragmented cold structures coagulate back into a coherent stream (blue dots); above it, the stream remains fragmented (orange circles with dots). Panels (e) and (f) show representative examples of these two outcomes, \run{R50D300P7M1} and \run{R10D100P7M1}, respectively. Since \ifm{\rx{s,f}/\rx{crit,frag}\propto M_{12}^{-2}(1+z)_3^{-1}} at fixed dimensionless parameters, the threshold halo mass scales only as \ifm{(1+z)^{-1/2}}, and for the fiducial parameters the orange curve lies at nearly constant \ifm{\Mx{v}\approx\SI{2e12}{\Msun}}. The curve is shown dotted where it falls within the bow-shock heating regime, since the fragmentation criterion assumes oblique-shock penetration. Consequently, stream fragmentation is expected in halos above \ifm{\sim\SI{2e12}{\Msun}} at \ifm{z\gtrsim2}, bounded by bow-shock heating toward lower redshift and by shear disruption toward higher halo mass, consistent with the picture drawn from \cref{fig:diag_size_chi_halo}.

\smallskip
Using \cref{eq:tcc2tv}, we estimate the ratio between the stream disruption timescale and the virial crossing time as
\begin{equation}
\frac{\tx{cc}}{\tx{v}}\simeq0.7(1+z)_3^{1/2}\left(\frac{\fx{acc,0.3}\fx{cold,0.2}}{\fx{h,0.3}\Max{s}\Thetax{h}^{1/2}}\right)^{1/2}.
\end{equation}
This ratio decreases toward lower redshift, implying that shear-driven disruption becomes increasingly efficient at \(\ifm{z\lesssim2}\), along with the increased efficiency of bow-shock heating and inefficient post-shock cooling which may also suppress cold-stream penetration. Similarly, using \cref{eq:tent2tv}, the entrainment timescale relative to the virial crossing time is
\begin{equation}
\frac{\tx{ent}}{\tx{v}}\simeq24M_{12}^{1/2}(1+z)_3^{3/4}\left(\frac{\fx{acc,0.3}^3\fx{cold,0.2}^3\Max{s}^5\Thetax{h}^{11/2}}{\Delta_{1/6}^2\mux{s,0.6}^3\fx{h,0.3}^5\Thetax{s}^4}\right)^{1/8}.
\end{equation}
This estimate suggests that entrainment remains inefficient in massive halos across the redshift range considered here. We caution, however, that this neglects the halo gravitational potential, which can increase the gas density as the stream falls inward and dramatically enhance entrainment \citep{Aung.etal.24}.

\smallskip
Taken together, \cref{fig:diag_size_chi_halo,fig:diag_Mv_z} suggest the following physical picture. Cold streams in ordinary \(1\sigma\) halos are mostly associated with the cold-accretion regime and do not require penetration through a hot halo. In rarer \(2\sigma\) and \(3\sigma\) halos, a hot CGM forms while dense cold streams can still enter the halo. At high redshift, these streams are expected to survive and may undergo fragmentation, producing multiphase cold structures embedded in the hot medium. Toward lower redshift and higher halo mass, the combined effects of stronger shocks, longer cooling times, and shear-driven disruption progressively suppress cold-stream penetration. This transition provides a natural framework for connecting the morphology of simulated cold streams to the observed decline of efficient star formation in massive halos.

\subsection{Caveats}
\subsubsection{Radial dependence of the pressure contrast}
\label{sec:caveat_pratio}

The pressure contrast in \cref{eq:pratio} was derived in \citet{Yao.etal.25} from the \textit{mean} pressures of the CGM and of cosmic-web filaments. An independent, local estimate at the virial radius follows from comparing the stream densities on the two sides of the virial shock: \cref{eq:ns_halo,eq:rhosi} both assume pressure equilibrium with the local confining medium, namely the hot CGM at \ifm{\Rx{v}} \citep{Mandelker.etal.20b} and the shocked filament sheath \citep{Lu.etal.24,Aung.etal.24}, respectively, at the common stream temperature \ifm{\Tx{s}}. Their ratio, after converting \cref{eq:rhosi} to a hydrogen number density, therefore measures the pressure contrast at the virial radius,
\begin{equation}
\label{eq:pratio_rv}
\mathcal{P}(\Rx{v})\simeq1.2\,M_{12}^{-0.11}(1+z)_3^{-1}\Max{s}\,
\frac{\Thetax{h}^{3/2}\Delta_{1/6}\fx{h,0.3}}{\Delta_\mathrm{v,f,20}\fx{b,0.17}\fx{acc,0.3}},
\end{equation}
which has the same halo-mass and redshift scalings as \cref{eq:pratio} but a normalization a factor of \ifm{\sim30} smaller. This factor comprises the ratio between the mean CGM pressure and the CGM pressure at \ifm{\Rx{v}} for the assumed profiles, and the ratio between the mean pressure in the virialized filament (the circumfilamentary medium, CFM) and the pressure in the filament core which is the cold stream. Taken at face value, the two estimates therefore imply that the pressure contrast is radius dependent, increasing from \cref{eq:pratio_rv} at the virial radius to \cref{eq:pratio} at the radii that dominate the mean halo pressure (\ifm{\sim0.2\text{--}0.3\,\Rx{v}}).

\smallskip
An order-unity contrast at \ifm{\Rx{v}} would not by itself conflict with the presence of a virial shock, which propagates into the diffuse intergalactic medium (IGM) rather than into the filament. At \ifm{M_{12}=1} and \ifm{z=2}, both the hot CGM at \ifm{\Rx{v}} and the hot CFM have thermal pressures of \ifm{P/\kb\simeq100\,\mathrm{K\,cm^{-3}}} (\cref{eq:ns_halo,eq:rhosi}), comparable to the post-shock pressure required to balance the ram pressure of the infalling IGM, \ifm{(3/4)\rhox{IGM}\vx{v}^2/\kb\simeq110\,\mathrm{K\,cm^{-3}}} for infall at the virial velocity \ifm{\vx{v}}, whereas the diffuse IGM (at an overdensity of \ifm{\sim4} and \ifm{T\sim10^4\,\mathrm{K}}) has \ifm{P/\kb\sim1\,\mathrm{K\,cm^{-3}}}. In this picture, the virial shock would be strong over most of its surface but weak or absent along the filament, as envisioned in the cold-stream scenario \citep{Dekel.Birnboim.06,Dekel.etal.09}. However, \cref{eq:pratio_rv} may equally well underestimate the contrast at \ifm{\Rx{v}}: the gas density profile at \ifm{\Rx{v}} is shallower than the dark-matter-like profile assumed in \ifm{\Delta_{1/6}}; non-thermal (turbulent and ram) pressure contributes to the confinement of the stream; the post-shock pressure scales with the density and infall velocity of the pre-shock gas, each uncertain at the factor-of-two level; the contrast scales as \ifm{\Max{s}(1+z)_3^{-1}}, so faster streams or lower redshifts than assumed here would raise it further; and the filament properties entering \cref{eq:rhosi} are calibrated far from the halo, where the halo potential and tidal compression may modify them. Compounded, these effects could plausibly raise \ifm{\mathcal{P}(\Rx{v})} to a factor of several or more. We therefore cannot exclude a significant shock acting on the cold stream near \ifm{\Rx{v}}, even if \cref{eq:pratio} overestimates the contrast there.

\smallskip
The effective contrast experienced by the stream further depends on how it responds during infall. \cref{eq:zeq_rv} suggests \ifm{\zx{eq}\sim\Rx{v}}, in which case the stream cannot re-establish pressure equilibrium during its descent and instead continuously reacts to the rising confining pressure. Imposing the cumulative contrast of \cref{eq:pratio} at the injection boundary would then remain a reasonable idealization, with global shock heating developing within \ifm{\sim\Rx{v}} of entry rather than promptly at the virial shock; this is qualitatively consistent with our simulations, in which the bow shock first appears downstream, near the convergence point, before propagating upstream (\cref{sec:pres}). On the other hand, \ifm{\zx{eq}} itself carries order-unity uncertainties (see the discussion following \cref{eq:zeq_rv}), and the converging shock propagates faster at higher pressure contrast, so the stream may respond more promptly than this estimate implies. We therefore regard both the normalization and the effective value of \ifm{\mathcal{P}} as uncertain at the factor-of-several level. Conditional on the stream undergoing bow-shock heating, the post-shock cooling criterion in \cref{eq:tcoolpost2tv} has no explicit dependence on \ifm{\mathcal{P}}. However, the pressure contrast directly affects whether that heating occurs, shifting the purple boundary in \cref{fig:diag_Mv_z} and hence the region where bow shocks can suppress cold accretion. A direct calibration of the radius-dependent pressure contrast \ifm{\mathcal{P}(r)} along filaments in cosmological simulations, together with a systematic parameter survey of the bow-shock transition, is left for future work \citep[see also][]{Lu.etal.24,Aung.etal.24}.

\subsubsection{Missing physics}
Despite incorporating velocity shear and self-consistently imposing the post-shock pressure contrast, making our setup more realistic than that of \citet{Yao.etal.25} for modeling stream penetration, most of the caveats discussed there remain applicable.

\begin{enumerate}[label=(\roman*),leftmargin=*]

\item \textit{Halo potential.} The halo potential can stratify the CGM as the stream falls inward. The resulting rise in ambient pressure may further enhance the pressure contrast, strengthening the cylindrical shock and potentially promoting fragmentation. Furthermore, our assumption that the stream regains pressure equilibrium after traveling a distance $z_{\rm eq}$ is only valid if \ifm{\zx{eq}<H_\mathrm{P}}, where \ifm{H_\mathrm{P}} is the pressure scale height of the stratified CGM; otherwise, our picture for subsequent stream evolution may need to be altered. In addition, the stream continues to accelerate down the potential well and this dominates over shear- and entrainment-induced deceleration \citep{Aung.etal.24}, so velocity shear can disrupt a substantial fraction of fragments even long after penetration. A higher inflow velocity also shortens the virial crossing time, leaving less time for the cold stream to develop a fragmented structure. As the stream descends, its radius decreases and its density increases. The denser ambient medium then enhances mass entrainment during the post-equilibrium phase \citep{Aung.etal.24}.

\item \textit{Turbulent CGM.} In our present setup, the background medium is static, and radiative cooling is artificially disabled in the hot phase. This idealization is not fully realistic, and turbulence may help the hot CGM offset radiative losses and maintain a quasi-steady state without suppressing cooling artificially. Turbulence may be driven by cosmological accretion \citep{Goldner.etal.25}, feedback \citep{Stern.etal.21,Pandya.etal.23}, or galaxy interactions \citep[see reviews by][]{Tumlinson.etal.17,Faucher-Giguere.Oh.23}. As shown by \citet{Gronke.etal.22}, cold gas is generally more difficult to preserve in a turbulent environment than in a coherent wind under identical control parameters. This implies a more stringent survival criterion for cold structures embedded in a turbulent CGM. This destructive effect of turbulence may become more pronounced at high pressure contrasts, where the cold streams are fully shock heated by the bow shock. In our simulations, the shock-heated stream subsequently cools without altering the survival outcome only when \ifm{\tx{cool,post}<\tx{v}}. However, in a turbulent CGM, the shock-heated stream may mix more efficiently with the ambient gas instead of remaining shielded by the leading cold material, thereby increasing the effective cooling time. Conversely, turbulence can enhance the fragmentation of the stream and its clumps, while the surviving fragments may grow more rapidly due to their increased surface area. In this respect, turbulence may act in a manner analogous to a moderate pressure contrast, amplifying the evolution of mass and surface area without altering the qualitative outcome of survival and disruption.

\item \textit{Magnetic fields.} Most astrophysical plasmas are magnetized. Although the outer CGM and the high-redshift cosmic web are not expected to be strongly magnetized, magnetic fields can be amplified during the cold stream penetration and fragmentation. As a result, small cloudlets produced by fragmentation may become magnetically dominated and fall out of thermal pressure equilibrium with their surroundings \citep{Nelson.etal.20}. This non-thermal pressure support can drive streaming motions along magnetic field lines and produce filamentary clumps, because the total pressure gradient becomes unbalanced along the field direction \citep{Wang.etal.25}. Magnetic draping can further modify the kinematics and morphology of clumps \citep{Hidalgo-Pineda.etal.24,Ramesh.etal.24}. Magnetic fields may also suppress mixing in turbulent mixing layers, thereby reducing the entrainment rate \citep{Ji.etal.19,Sparre.etal.20,Gronke.Oh.20a,Gronnow.etal.22,Ledos.etal.23,Kaul.etal.25}. However, when turbulence is externally driven, hydrodynamic and magnetohydrodynamic mixing rates can become comparable at fixed rms turbulent velocity \citep{Das.Gronke.24}.

\end{enumerate}

\section{Summary}
\label{sec:summary}

We investigated the evolution of cold streams after penetrating the virial shock, focusing on how pressure contrast and velocity shear jointly regulate stream morphology (coherent versus clumpy/fragmented) and cold-gas mass evolution and survival. Using idealized three-dimensional simulations of cold streams flowing into and traversing an overpressurized hot medium extending over 200 times the post-equilibrium stream radius, we systematically varied the stream size, initial pressure contrast, and Mach number across three density contrasts while controlling the post-equilibration stream properties. We evolved the system for two box-crossing times and then analyzed time-averaged properties along the stream direction within the selected analysis regions. Our main results are as follows.

\begin{enumerate}[label=(\roman*),leftmargin=*]
\item The pressure contrast at penetration drives an oblique shock within the cold stream. At sufficiently large contrasts \ifm{\mathcal{P}\gtrsim2\chix{f}^{1/2}\Max{s}} or equivalently \ifm{\mathcal{P}\gtrsim4\chix{i}\Max{s}^2}, this shock steepens into a bow shock, enhancing the conversion of inflowing kinetic energy into thermal energy and temporarily heating the cold stream. In this regime, cold-stream survival is no longer controlled primarily by the shear criterion. Instead, it requires the post-shock gas to cool within the virial crossing time, i.e. \ifm{\tx{cool,post}<\tx{v}} (\cref{fig:proj_pres,fig:proj_pres_bowshock}).

\item In the oblique-shock regime, the evolution of cold streams falls into three regimes: coagulation, fragmentation, and disruption, with a borderline case in which the stream marginally survives while the clumps do not (\cref{fig:diag_size_chi,fig:proj_size,fig:profile_mass_size,fig:profile_morph_size}).

\item Below the bow-shock transition, \ifm{\mathcal{P}\lesssim2\chix{f}^{1/2}\Max{s}}, the pressure contrast does not determine the survival regime. Instead, it amplifies the transient mass evolution during the restoring phase prior to pressure equilibrium. Higher pressure contrasts strengthen the shock and enhance the temporary growth or loss of cold mass and surface area, after which the evolution converges toward the shear-dominated state (\cref{fig:profile_mass_pres,fig:profile_morph_pres}).

\item Velocity shear is the primary regulator of clump survival. Strong shear efficiently disrupts small fragments and shortens the survival distance of cold gas, whereas weaker shear permits the persistence of long-lived clumps and produces a broader clump size spectrum (\cref{fig:proj_mach,fig:profile_mass_Mach,fig:profile_morph_Mach,fig:clump_mass_dist}).

\item For the survival cases, the downstream change in mass flux is well described by the entrainment model once the stream has reached pressure equilibrium. In the same regime, the cold-gas mean streamwise momentum flux remains approximately conserved (\cref{fig:profile_mass_theory}).

\item Applying our idealized stream-penetration experiments to the galaxy-evolution framework, we find that cold streams in \ifm{2\sigma} or higher halos at \ifm{z>2} are expected to survive penetration and may undergo fragmentation at \ifm{\Mx{v}\gtrsim\SI{2e12}{\Msun}}, producing multiphase cold structures embedded in the hot CGM. Toward lower redshift, bow-shock heating and inefficient post-shock cooling can suppress cold-stream penetration.
\end{enumerate}

These results clarify the respective roles of pressure imbalance and shear in cold-stream penetration. Pressure contrast governs the transient response at penetration, while velocity shear and cooling determine long-term survival. Streams that penetrate a hotter CGM with a larger radius and experience a stronger pressure contrast at the virial shock tend to develop a more fragmented sheath of clumps surrounding the cold core, with the most distant clumps reaching distances up to ten times the stream radius; this maximum measured distance is limited by the radial extent of the highest-resolution region. The combined effect naturally produces a multiphase CGM in massive halos at high redshift and provides a physical framework for interpreting the large covering fractions of cool gas inferred from observations. We caution that the galaxy-evolution mapping above assumes the stream reaches pressure equilibrium with the hot CGM over a distance much smaller than the virial radius. Equation~(\ref{eq:zeq_rv}) gives \ifm{\zx{eq}\sim\Rx{v}}, however, so this assumption is only marginally satisfied, and the diagnostics in \cref{fig:diag_size_chi_halo} apply only in the limit of efficient equilibration. Verifying whether streams equilibrate before reaching the central galaxy will require global cosmological simulations.

\section*{Acknowledgements}
We thank the anonymous referee for helpful comments that have improved the quality of this manuscript. We thank Andrei Antipov, Yuval Birnboim, Frank van den Bosch, Sharon Lapiner, Zhaozhou Li, Elisha Modelevsky, Daisuke Nagai, Yotam Ohad, Nadav Shoval, Jonathan Stern, Romain Teyssier, and Bocheng Zhu for helpful comments and interesting discussions. ZY and NM acknowledge support from BSF grant 2022736 (from the NSF-BSF program) and from BSF grant 2022281. SPO acknowledges support from NSF AST240752. Artificial intelligence tools were used in a limited capacity during the preparation of this manuscript to assist with code development, language editing, and refinement of schematic figures. These tools were used solely for technical and editorial support and did not contribute to the scientific analysis, physical interpretation, or conclusions of this work, which remain entirely the responsibility of the authors.

\textit{Software:} matplotlib \citep{Hunter.2007}, numpy \citep{Harris.2020}, scipy \citep{Scipy.20}, astropy \citep{astropy.22}, yt \citep{Turk.etal.11}, scikit-image \citep{vanderWalt.etal.14}, colossus \citep{Diemer.18}, grackle \citep{Smith.etal.17}

\section*{Data Availability}

The simulation data underlying this article will be shared on reasonable request to the corresponding author. Simulation movies for all runs are publicly available at \href{https://zhiyuan-21.github.io/HomePage/stream_penetration.html}{this page}.



\bibliographystyle{mnras}
\bibliography{StreamPenetration} 

@techreport{NACA1135,
  author      = {{Ames Research Staff}},
  title       = {Equations, Tables, and Charts for Compressible Flow},
  institution = {National Advisory Committee for Aeronautics},
  year        = {1953},
  number      = {Report 1135},
  url         = {https://www.grc.nasa.gov/WWW/K-12/airplane/Images/naca1135.pdf}
}

@article{Richtmyer.60,
  title={Taylor instability in shock acceleration of compressible fluids},
  author={Robert D. Richtmyer},
  journal={Communications on Pure and Applied Mathematics},
  year={1960},
  volume={13},
  pages={297-319},
  doi={10.1002/cpa.3160130207}
}

@article{Meshkov.69,
  title={Instability of the interface of two gases accelerated by a shock wave},
  author={Evgeny E. Meshkov},
  journal={Fluid Dynamics},
  year={1969},
  volume={4},
  pages={101-104},
  url={https://api.semanticscholar.org/CorpusID:123494913}
}

@incollection{Dimotakis.91,
  title={Turbulent free shear layer mixing and combustion},
  author={Dimotakis, Paul E.},
  booktitle={High-Speed Flight Propulsion Systems},
  series={Progress in Astronautics and Aeronautics},
  volume={137},
  editor={Murthy, S. N. B. and Curran, E. T.},
  pages={265--340},
  year={1991},
  publisher={AIAA},
  address={Washington, DC},
  doi={10.2514/5.9781600866104.0265.0340}
}

@ARTICLE{Quirk.94,
       author = {{Quirk}, James J.},
        title = "{A contribution to the great Riemann solver debate}",
      journal = {International Journal for Numerical Methods in Fluids},
         year = 1994,
        month = mar,
       volume = {18},
       number = {6},
        pages = {555-574},
          doi = {10.1002/fld.1650180603},
       adsurl = {https://ui.adsabs.harvard.edu/abs/1994IJNMF..18..555Q}
}

@ARTICLE{Haardt.Madau.96,
       author = {{Haardt}, Francesco and {Madau}, Piero},
        title = "{Radiative Transfer in a Clumpy Universe. II. The Ultraviolet Extragalactic Background}",
      journal = {\apj},
         year = 1996,
        month = apr,
       volume = {461},
        pages = {20},
          doi = {10.1086/177035},
archivePrefix = {arXiv},
       eprint = {astro-ph/9509093},
 primaryClass = {astro-ph},
       adsurl = {https://ui.adsabs.harvard.edu/abs/1996ApJ...461...20H}
}

@ARTICLE{Teyssier.02,
       author = {{Teyssier}, R.},
        title = "{Cosmological hydrodynamics with adaptive mesh refinement. A new high resolution code called RAMSES}",
      journal = {\aap},
         year = 2002,
        month = apr,
       volume = {385},
        pages = {337-364},
          doi = {10.1051/0004-6361:20011817},
archivePrefix = {arXiv},
       eprint = {astro-ph/0111367},
 primaryClass = {astro-ph},
       adsurl = {https://ui.adsabs.harvard.edu/abs/2002A&A...385..337T}
}

@ARTICLE{Birnboim.Dekel.03,
       author = {{Birnboim}, Yuval and {Dekel}, Avishai},
        title = "{Virial shocks in galactic haloes?}",
      journal = {\mnras},
         year = 2003,
        month = oct,
       volume = {345},
       number = {1},
        pages = {349-364},
          doi = {10.1046/j.1365-8711.2003.06955.x},
archivePrefix = {arXiv},
       eprint = {astro-ph/0302161},
 primaryClass = {astro-ph},
       adsurl = {https://ui.adsabs.harvard.edu/abs/2003MNRAS.345..349B}
}

@ARTICLE{Keres.etal.05,
       author = {{Kere{\v{s}}}, Du{\v{s}}an and {Katz}, Neal and {Weinberg}, David H. and {Dav{\'e}}, Romeel},
        title = "{How do galaxies get their gas?}",
      journal = {\mnras},
         year = 2005,
        month = oct,
       volume = {363},
       number = {1},
        pages = {2-28},
          doi = {10.1111/j.1365-2966.2005.09451.x},
archivePrefix = {arXiv},
       eprint = {astro-ph/0407095},
 primaryClass = {astro-ph},
       adsurl = {https://ui.adsabs.harvard.edu/abs/2005MNRAS.363....2K}
}

@ARTICLE{Dekel.Birnboim.06,
       author = {{Dekel}, Avishai and {Birnboim}, Yuval},
        title = "{Galaxy bimodality due to cold flows and shock heating}",
      journal = {\mnras},
         year = 2006,
        month = may,
       volume = {368},
       number = {1},
        pages = {2-20},
          doi = {10.1111/j.1365-2966.2006.10145.x},
archivePrefix = {arXiv},
       eprint = {astro-ph/0412300},
 primaryClass = {astro-ph},
       adsurl = {https://ui.adsabs.harvard.edu/abs/2006MNRAS.368....2D}
}

@Article{Hunter.2007,
  Author    = {Hunter, J. D.},
  Title     = {Matplotlib: A 2D graphics environment},
  Journal   = {Computing in Science \& Engineering},
  Volume    = {9},
  Number    = {3},
  Pages     = {90--95},
  publisher = {IEEE COMPUTER SOC},
  doi       = {10.1109/MCSE.2007.55},
  year      = 2007
}

@ARTICLE{Turk.etal.11,
       author = {{Turk}, Matthew J. and {Smith}, Britton D. and {Oishi}, Jeffrey S. and {Skory}, Stephen and {Skillman}, Samuel W. and {Abel}, Tom and {Norman}, Michael L.},
        title = "{yt: A Multi-code Analysis Toolkit for Astrophysical Simulation Data}",
      journal = {\apjs},
         year = 2011,
        month = jan,
       volume = {192},
       number = {1},
          eid = {9},
        pages = {9},
          doi = {10.1088/0067-0049/192/1/9},
archivePrefix = {arXiv},
       eprint = {1011.3514},
 primaryClass = {astro-ph.IM},
       adsurl = {https://ui.adsabs.harvard.edu/abs/2011ApJS..192....9T}
}

@ARTICLE{vanderWalt.etal.14,
       author = {{van der Walt}, Stefan and {Sch{\"o}nberger}, Johannes L. and {Nunez-Iglesias}, Juan and {Boulogne}, Fran{\c{c}}ois and {Warner}, Joshua D. and {Yager}, Neil and {Gouillart}, Emmanuelle and {Yu}, Tony and {scikit-image Contributors}},
        title = "{scikit-image: Image processing in Python}",
      journal = {PeerJ},
         year = 2014,
        month = jan,
       volume = {2},
          eid = {e453},
        pages = {e453},
          doi = {10.7717/peerj.453},
archivePrefix = {arXiv},
       eprint = {1407.6245},
 primaryClass = {cs.MS},
       adsurl = {https://ui.adsabs.harvard.edu/abs/2014PeerJ...2..453V}
}

@ARTICLE{Zhou.17a,
       author = {{Zhou}, Ye},
        title = "{Rayleigh-Taylor and Richtmyer-Meshkov instability induced flow, turbulence, and mixing. I}",
      journal = {\physrep},
         year = 2017,
        month = dec,
       volume = {720},
        pages = {1-136},
          doi = {10.1016/j.physrep.2017.07.005},
       adsurl = {https://ui.adsabs.harvard.edu/abs/2017PhR...720....1Z}
}

@ARTICLE{Zhou.17b,
       author = {{Zhou}, Ye},
        title = "{Rayleigh-Taylor and Richtmyer-Meshkov instability induced flow, turbulence, and mixing. II}",
      journal = {\physrep},
         year = 2017,
        month = dec,
       volume = {723},
        pages = {1-160},
          doi = {10.1016/j.physrep.2017.07.008},
       adsurl = {https://ui.adsabs.harvard.edu/abs/2017PhR...723....1Z}
}

@ARTICLE{Dekel.etal.09,
       author = {{Dekel}, A. and {Birnboim}, Y. and {Engel}, G. and {Freundlich}, J. and {Goerdt}, T. and {Mumcuoglu}, M. and {Neistein}, E. and {Pichon}, C. and {Teyssier}, R. and {Zinger}, E.},
        title = "{Cold streams in early massive hot haloes as the main mode of galaxy formation}",
      journal = {\nat},
         year = 2009,
        month = jan,
       volume = {457},
       number = {7228},
        pages = {451-454},
          doi = {10.1038/nature07648},
archivePrefix = {arXiv},
       eprint = {0808.0553},
 primaryClass = {astro-ph},
       adsurl = {https://ui.adsabs.harvard.edu/abs/2009Natur.457..451D}
}

@ARTICLE{Steidel.etal.10,
       author = {{Steidel}, Charles C. and {Erb}, Dawn K. and {Shapley}, Alice E. and {Pettini}, Max and {Reddy}, Naveen and {Bogosavljevi{\'c}}, Milan and {Rudie}, Gwen C. and {Rakic}, Olivera},
        title = "{The Structure and Kinematics of the Circumgalactic Medium from Far-ultraviolet Spectra of z \raisebox{-0.5ex}\textasciitilde= 2-3 Galaxies}",
      journal = {\apj},
         year = 2010,
        month = jul,
       volume = {717},
       number = {1},
        pages = {289-322},
          doi = {10.1088/0004-637X/717/1/289},
archivePrefix = {arXiv},
       eprint = {1003.0679},
 primaryClass = {astro-ph.CO},
       adsurl = {https://ui.adsabs.harvard.edu/abs/2010ApJ...717..289S}
}

@ARTICLE{vandevoort.etal.11,
       author = {{van de Voort}, Freeke and {Schaye}, Joop and {Booth}, C.~M. and {Haas}, Marcel R. and {Dalla Vecchia}, Claudio},
        title = "{The rates and modes of gas accretion on to galaxies and their gaseous haloes}",
      journal = {\mnras},
         year = 2011,
        month = jul,
       volume = {414},
       number = {3},
        pages = {2458-2478},
          doi = {10.1111/j.1365-2966.2011.18565.x},
archivePrefix = {arXiv},
       eprint = {1011.2491},
 primaryClass = {astro-ph.CO},
       adsurl = {https://ui.adsabs.harvard.edu/abs/2011MNRAS.414.2458V}
}

@ARTICLE{Rudie.etal.12,
       author = {{Rudie}, Gwen C. and {Steidel}, Charles C. and {Trainor}, Ryan F. and {Rakic}, Olivera and {Bogosavljevi{\'c}}, Milan and {Pettini}, Max and {Reddy}, Naveen and {Shapley}, Alice E. and {Erb}, Dawn K. and {Law}, David R.},
        title = "{The Gaseous Environment of High-z Galaxies: Precision Measurements of Neutral Hydrogen in the Circumgalactic Medium of z \raisebox{-0.5ex}\textasciitilde 2-3 Galaxies in the Keck Baryonic Structure Survey}",
      journal = {\apj},
         year = 2012,
        month = may,
       volume = {750},
       number = {1},
          eid = {67},
        pages = {67},
          doi = {10.1088/0004-637X/750/1/67},
archivePrefix = {arXiv},
       eprint = {1202.6055},
 primaryClass = {astro-ph.CO},
       adsurl = {https://ui.adsabs.harvard.edu/abs/2012ApJ...750...67R}
}

@article{Sharma.etal.12,
  title = {Thermal {{Instability}} \& the {{Feedback Regulation}} of {{Hot Halos}} in {{Clusters}}, {{Groups}}, and {{Galaxies}}},
  author = {Sharma, Prateek and McCourt, Michael and Quataert, Eliot and Parrish, Ian J.},
  year = 2012,
  journal = {Monthly Notices of the Royal Astronomical Society},
  volume = {420},
  number = {4},
  eprint = {1106.4816},
  primaryclass = {astro-ph.CO},
  pages = {3174},
  issn = {00358711},
  doi = {10.1111/j.1365-2966.2011.20246.x},
  archiveprefix = {arXiv}
}

@article{Dekel.etal.13,
  title = {Toy Models for Galaxy Formation versus Simulations},
  author = {Dekel, A. and Zolotov, A. and Tweed, D. and Cacciato, M. and Ceverino, D. and Primack, J. R.},
  year = 2013,
  month = oct,
  journal = {Monthly Notices of the Royal Astronomical Society},
  volume = {435},
  number = {2},
  pages = {999--1019},
  issn = {1365-2966, 0035-8711},
  doi = {10.1093/mnras/stt1338}
}

@ARTICLE{Prochaska.etal.13,
       author = {{Prochaska}, J. Xavier and {Hennawi}, Joseph F. and {Lee}, Khee-Gan and {Cantalupo}, Sebastiano and {Bovy}, Jo and {Djorgovski}, S.~G. and {Ellison}, Sara L. and {Lau}, Marie Wingyee and {Martin}, Crystal L. and {Myers}, Adam and et al.},
        title = "{Quasars Probing Quasars. VI. Excess H I Absorption within One Proper Mpc of z \raisebox{-0.5ex}\textasciitilde 2 Quasars}",
      journal = {\apj},
         year = 2013,
        month = oct,
       volume = {776},
       number = {2},
          eid = {136},
        pages = {136},
          doi = {10.1088/0004-637X/776/2/136},
archivePrefix = {arXiv},
       eprint = {1308.6222},
 primaryClass = {astro-ph.CO},
       adsurl = {https://ui.adsabs.harvard.edu/abs/2013ApJ...776..136P}
}

@article{scikit-image,
 title = {scikit-image: image processing in {P}ython},
 author = {van der Walt, {S}t\'efan and {S}ch\"onberger, {J}ohannes {L}. and
           {Nunez-Iglesias}, {J}uan and {B}oulogne, {F}ran\c{c}ois and {W}arner,
           {J}oshua {D}. and {Y}ager, {N}eil and {G}ouillart, {E}mmanuelle and
           {Y}u, {T}ony and the scikit-image contributors},
 year = {2014},
 month = {6},
 volume = {2},
 pages = {e453},
 journal = {PeerJ},
 issn = {2167-8359},
 url = {https://doi.org/10.7717/peerj.453},
 doi = {10.7717/peerj.453}
}

@ARTICLE{Cantalupo.etal.14,
       author = {{Cantalupo}, Sebastiano and {Arrigoni-Battaia}, Fabrizio and {Prochaska}, J. Xavier and {Hennawi}, Joseph F. and {Madau}, Piero},
        title = "{A cosmic web filament revealed in Lyman-{\ensuremath{\alpha}} emission around a luminous high-redshift quasar}",
      journal = {\nat},
         year = 2014,
        month = feb,
       volume = {506},
       number = {7486},
        pages = {63-66},
          doi = {10.1038/nature12898},
archivePrefix = {arXiv},
       eprint = {1401.4469},
 primaryClass = {astro-ph.CO},
       adsurl = {https://ui.adsabs.harvard.edu/abs/2014Natur.506...63C}
}

@ARTICLE{Bleuler.etal.15,
       author = {{Bleuler}, Andreas and {Teyssier}, Romain and {Carassou}, S{\'e}bastien and {Martizzi}, Davide},
        title = "{PHEW: a parallel segmentation algorithm for three-dimensional AMR datasets. Application to structure detection in self-gravitating flows}",
      journal = {Computational Astrophysics and Cosmology},
         year = 2015,
        month = jun,
       volume = {2},
          eid = {5},
        pages = {5},
          doi = {10.1186/s40668-015-0009-7},
archivePrefix = {arXiv},
       eprint = {1412.0510},
 primaryClass = {astro-ph.IM},
       adsurl = {https://ui.adsabs.harvard.edu/abs/2015ComAC...2....5B}
}

@ARTICLE{Mandelker.etal.16,
       author = {{Mandelker}, Nir and {Padnos}, Dan and {Dekel}, Avishai and {Birnboim}, Yuval and {Burkert}, Andreas and {Krumholz}, Mark R. and {Steinberg}, Elad},
        title = "{Instability of supersonic cold streams feeding galaxies - I. Linear Kelvin-Helmholtz instability with body modes}",
      journal = {\mnras},
         year = 2016,
        month = dec,
       volume = {463},
       number = {4},
        pages = {3921-3947},
          doi = {10.1093/mnras/stw2267},
archivePrefix = {arXiv},
       eprint = {1606.06289},
 primaryClass = {astro-ph.GA},
       adsurl = {https://ui.adsabs.harvard.edu/abs/2016MNRAS.463.3921M}
}

@ARTICLE{Wisotzki.etal.16,
       author = {{Wisotzki}, L. and {Bacon}, R. and {Blaizot}, J. and {Brinchmann}, J. and {Herenz}, E.~C. and {Schaye}, J. and {Bouch{\'e}}, N. and {Cantalupo}, S. and {Contini}, T. and {Carollo}, C.~M. and et al.},
        title = "{Extended Lyman {\ensuremath{\alpha}} haloes around individual high-redshift galaxies revealed by MUSE}",
      journal = {\aap},
         year = 2016,
        month = mar,
       volume = {587},
          eid = {A98},
        pages = {A98},
          doi = {10.1051/0004-6361/201527384},
archivePrefix = {arXiv},
       eprint = {1509.05143},
 primaryClass = {astro-ph.GA},
       adsurl = {https://ui.adsabs.harvard.edu/abs/2016A&A...587A..98W}
}

@ARTICLE{Leclercq.etal.17,
       author = {{Leclercq}, Floriane and {Bacon}, Roland and {Wisotzki}, Lutz and {Mitchell}, Peter and {Garel}, Thibault and {Verhamme}, Anne and {Blaizot}, J{\'e}r{\'e}my and {Hashimoto}, Takuya and {Herenz}, Edmund Christian and {Conseil}, Simon and et al.},
        title = "{The MUSE Hubble Ultra Deep Field Survey. VIII. Extended Lyman-{\ensuremath{\alpha}} haloes around high-z star-forming galaxies}",
      journal = {\aap},
         year = 2017,
        month = dec,
       volume = {608},
          eid = {A8},
        pages = {A8},
          doi = {10.1051/0004-6361/201731480},
archivePrefix = {arXiv},
       eprint = {1710.10271},
 primaryClass = {astro-ph.GA},
       adsurl = {https://ui.adsabs.harvard.edu/abs/2017A&A...608A...8L}
}

@ARTICLE{Smith.etal.17,
   author = {{Smith}, B.~D. and {Bryan}, G.~L. and {Glover}, S.~C.~O. and
        {Goldbaum}, N.~J. and {Turk}, M.~J. and {Regan}, J. and {Wise}, J.~H. and
        {Schive}, H.-Y. and {Abel}, T. and {Emerick}, A. and {O'Shea}, B.~W. and
        {Anninos}, P. and {Hummels}, C.~B. and {Khochfar}, S.},
    title = "{GRACKLE: a chemistry and cooling library for astrophysics}",
  journal = {\mnras},
archivePrefix = "arXiv",
   eprint = {1610.09591},
     year = 2017,
    month = apr,
   volume = 466,
    pages = {2217-2234},
      doi = {10.1093/mnras/stw3291},
   adsurl = {http://adsabs.harvard.edu/abs/2017MNRAS.466.2217S}
}

@ARTICLE{Tumlinson.etal.17,
       author = {{Tumlinson}, Jason and {Peeples}, Molly S. and {Werk}, Jessica K.},
        title = "{The Circumgalactic Medium}",
      journal = {\araa},
         year = 2017,
        month = aug,
       volume = {55},
       number = {1},
        pages = {389-432},
          doi = {10.1146/annurev-astro-091916-055240},
archivePrefix = {arXiv},
       eprint = {1709.09180},
 primaryClass = {astro-ph.GA},
       adsurl = {https://ui.adsabs.harvard.edu/abs/2017ARA&A..55..389T}
}

@ARTICLE{Fielding.etal.17,
       author = {{Fielding}, Drummond and {Quataert}, Eliot and {McCourt}, Michael and {Thompson}, Todd A.},
        title = "{The impact of star formation feedback on the circumgalactic medium}",
      journal = {\mnras},
         year = 2017,
        month = apr,
       volume = {466},
       number = {4},
        pages = {3810-3826},
          doi = {10.1093/mnras/stw3326},
archivePrefix = {arXiv},
       eprint = {1606.06734},
 primaryClass = {astro-ph.GA},
       adsurl = {https://ui.adsabs.harvard.edu/abs/2017MNRAS.466.3810F}
}

@ARTICLE{Wisotzki.etal.18,
       author = {{Wisotzki}, L. and {Bacon}, R. and {Brinchmann}, J. and {Cantalupo}, S. and {Richter}, P. and {Schaye}, J. and {Schmidt}, K.~B. and {Urrutia}, T. and {Weilbacher}, P.~M. and {Akhlaghi}, M. and et al.},
        title = "{Nearly all the sky is covered by Lyman-{\ensuremath{\alpha}} emission around high-redshift galaxies}",
      journal = {\nat},
         year = 2018,
        month = oct,
       volume = {562},
       number = {7726},
        pages = {229-232},
          doi = {10.1038/s41586-018-0564-6},
archivePrefix = {arXiv},
       eprint = {1810.00843},
 primaryClass = {astro-ph.GA},
       adsurl = {https://ui.adsabs.harvard.edu/abs/2018Natur.562..229W}
}

@article{Gronke.Oh.18,
  title = {The Growth and Entrainment of Cold Gas in a Hot Wind},
  author = {Gronke, Max and Oh, S Peng},
  year = 2018,
  month = oct,
  journal = {Monthly Notices of the Royal Astronomical Society: Letters},
  volume = {480},
  number = {1},
  pages = {L111-L115},
  issn = {1745-3925, 1745-3933},
  doi = {10.1093/mnrasl/sly131}
}

@ARTICLE{Padnos.etal.18,
       author = {{Padnos}, Dan and {Mandelker}, Nir and {Birnboim}, Yuval and {Dekel}, Avishai and {Krumholz}, Mark R. and {Steinberg}, Elad},
        title = "{Instability of supersonic cold streams feeding galaxies-II. Non-linear evolution of surface and body modes of Kelvin-Helmholtz instability}",
      journal = {\mnras},
         year = 2018,
        month = jul,
       volume = {477},
       number = {3},
        pages = {3293-3328},
          doi = {10.1093/mnras/sty789},
archivePrefix = {arXiv},
       eprint = {1803.09105},
 primaryClass = {astro-ph.GA},
       adsurl = {https://ui.adsabs.harvard.edu/abs/2018MNRAS.477.3293P}
}

@ARTICLE{McCourt.etal.18,
       author = {{McCourt}, Michael and {Oh}, S. Peng and {O'Leary}, Ryan and {Madigan}, Ann-Marie},
        title = "{A characteristic scale for cold gas}",
      journal = {\mnras},
         year = 2018,
        month = feb,
       volume = {473},
       number = {4},
        pages = {5407-5431},
          doi = {10.1093/mnras/stx2687},
archivePrefix = {arXiv},
       eprint = {1610.01164},
 primaryClass = {astro-ph.GA},
       adsurl = {https://ui.adsabs.harvard.edu/abs/2018MNRAS.473.5407M}
}

@ARTICLE{Diemer.18,
       author = {{Diemer}, Benedikt},
        title = "{COLOSSUS: A Python Toolkit for Cosmology, Large-scale Structure, and Dark Matter Halos}",
      journal = {\apjs},
         year = 2018,
        month = dec,
       volume = {239},
       number = {2},
          eid = {35},
        pages = {35},
          doi = {10.3847/1538-4365/aaee8c},
archivePrefix = {arXiv},
       eprint = {1712.04512},
 primaryClass = {astro-ph.CO},
       adsurl = {https://ui.adsabs.harvard.edu/abs/2018ApJS..239...35D}
}

@article{Cornuault.etal.18,
  title = {Are Cosmological Gas Accretion Streams Multiphase and Turbulent?},
  author = {Cornuault, Nicolas and Lehnert, Matthew D. and Boulanger, Fran{\c c}ois and Guillard, Pierre},
  year = 2018,
  month = feb,
  journal = {Astronomy \& Astrophysics},
  volume = {610},
  pages = {A75},
  issn = {0004-6361, 1432-0746},
  doi = {10.1051/0004-6361/201629229}
}

@article{Stern.etal.19,
  title = {Cooling Flow Solutions for the Circumgalactic Medium},
  author = {Stern, Jonathan and Fielding, Drummond and {Faucher-Gigu{\`e}re}, Claude-Andr{\'e} and Quataert, Eliot},
  year = 2019,
  month = sep,
  journal = {Monthly Notices of the Royal Astronomical Society},
  volume = {488},
  number = {2},
  pages = {2549--2572},
  issn = {0035-8711, 1365-2966},
  doi = {10.1093/mnras/stz1859}
}

@ARTICLE{Mandelker.etal.19,
       author = {{Mandelker}, Nir and {Nagai}, Daisuke and {Aung}, Han and {Dekel}, Avishai and {Padnos}, Dan and {Birnboim}, Yuval},
        title = "{Instability of supersonic cold streams feeding Galaxies - III. Kelvin-Helmholtz instability in three dimensions}",
      journal = {\mnras},
         year = 2019,
        month = mar,
       volume = {484},
       number = {1},
        pages = {1100-1132},
          doi = {10.1093/mnras/stz012},
archivePrefix = {arXiv},
       eprint = {1806.05677},
 primaryClass = {astro-ph.GA},
       adsurl = {https://ui.adsabs.harvard.edu/abs/2019MNRAS.484.1100M}
}

@article{Ji.etal.19,
  title = {Simulations of Radiative Turbulent Mixing Layers},
  author = {Ji, Suoqing and Oh, S.Peng and Masterson, Phillip},
  year = 2019,
  month = jul,
  journal = {Monthly Notices of the Royal Astronomical Society},
  volume = {487},
  number = {1},
  eprint = {1809.09101},
  primaryclass = {astro-ph.GA},
  pages = {737--754},
  issn = {0035-8711, 1365-2966},
  doi = {10.1093/mnras/stz1248},
  archiveprefix = {arXiv}
}

@article{Behroozi.etal.19,
  title = {{{UniverseMachine}}: {{The}} Correlation between Galaxy Growth and Dark Matter Halo Assembly from z = 0-10},
  author = {Behroozi, Peter and Wechsler, Risa H. and Hearin, Andrew P. and Conroy, Charlie},
  year = 2019,
  month = sep,
  journal = {Monthly Notices of the Royal Astronomical Society},
  volume = {488},
  number = {3},
  pages = {3143--3194},
  issn = {0035-8711, 1365-2966},
  doi = {10.1093/mnras/stz1182}
}

@ARTICLE{Sparre.etal.19,
       author = {{Sparre}, Martin and {Pfrommer}, Christoph and {Vogelsberger}, Mark},
        title = "{The physics of multiphase gas flows: fragmentation of a radiatively cooling gas cloud in a hot wind}",
      journal = {\mnras},
         year = 2019,
        month = feb,
       volume = {482},
       number = {4},
        pages = {5401-5421},
          doi = {10.1093/mnras/sty3063},
archivePrefix = {arXiv},
       eprint = {1807.07971},
 primaryClass = {astro-ph.GA},
       adsurl = {https://ui.adsabs.harvard.edu/abs/2019MNRAS.482.5401S}
}

@ARTICLE{Stern.etal.20,
       author = {{Stern}, Jonathan and {Fielding}, Drummond and {Faucher-Gigu{\`e}re}, Claude-Andr{\'e} and {Quataert}, Eliot},
        title = "{The maximum accretion rate of hot gas in dark matter haloes}",
      journal = {\mnras},
         year = 2020,
        month = mar,
       volume = {492},
       number = {4},
        pages = {6042-6058},
          doi = {10.1093/mnras/staa198},
archivePrefix = {arXiv},
       eprint = {1909.07402},
 primaryClass = {astro-ph.GA},
       adsurl = {https://ui.adsabs.harvard.edu/abs/2020MNRAS.492.6042S}
}

@ARTICLE{Sparre.etal.20,
       author = {{Sparre}, Martin and {Pfrommer}, Christoph and {Ehlert}, Kristian},
        title = "{Interaction of a cold cloud with a hot wind: the regimes of cloud growth and destruction and the impact of magnetic fields}",
      journal = {\mnras},
         year = 2020,
        month = dec,
       volume = {499},
       number = {3},
        pages = {4261-4281},
          doi = {10.1093/mnras/staa3177},
archivePrefix = {arXiv},
       eprint = {2008.09118},
 primaryClass = {astro-ph.GA},
       adsurl = {https://ui.adsabs.harvard.edu/abs/2020MNRAS.499.4261S}
}

@ARTICLE{Gronke.Oh.20a,
       author = {{Gronke}, Max and {Oh}, S. Peng},
        title = "{How cold gas continuously entrains mass and momentum from a hot wind}",
      journal = {\mnras},
         year = 2020,
        month = feb,
       volume = {492},
       number = {2},
        pages = {1970-1990},
          doi = {10.1093/mnras/stz3332},
archivePrefix = {arXiv},
       eprint = {1907.04771},
 primaryClass = {astro-ph.GA},
       adsurl = {https://ui.adsabs.harvard.edu/abs/2020MNRAS.492.1970G}
}

@ARTICLE{Gronke.Oh.20b,
       author = {{Gronke}, Max and {Oh}, S. Peng},
        title = "{Is multiphase gas cloudy or misty?}",
      journal = {\mnras},
         year = 2020,
        month = may,
       volume = {494},
       number = {1},
        pages = {L27-L31},
          doi = {10.1093/mnrasl/slaa033},
archivePrefix = {arXiv},
       eprint = {1912.07808},
 primaryClass = {astro-ph.GA},
       adsurl = {https://ui.adsabs.harvard.edu/abs/2020MNRAS.494L..27G}
}

@ARTICLE{Fielding.etal.20,
       author = {{Fielding}, Drummond B. and {Ostriker}, Eve C. and {Bryan}, Greg L. and {Jermyn}, Adam S.},
        title = "{Multiphase Gas and the Fractal Nature of Radiative Turbulent Mixing Layers}",
      journal = {\apjl},
         year = 2020,
        month = may,
       volume = {894},
       number = {2},
          eid = {L24},
        pages = {L24},
          doi = {10.3847/2041-8213/ab8d2c},
archivePrefix = {arXiv},
       eprint = {2003.08390},
 primaryClass = {astro-ph.GA},
       adsurl = {https://ui.adsabs.harvard.edu/abs/2020ApJ...894L..24F}
}

@ARTICLE{Scipy.20,
  author  = {Virtanen, Pauli and Gommers, Ralf and Oliphant, Travis E. and
            Haberland, Matt and Reddy, Tyler and Cournapeau, David and
            Burovski, Evgeni and Peterson, Pearu and Weckesser, Warren and
            Bright, Jonathan and {van der Walt}, St{\'e}fan J. and
            Brett, Matthew and Wilson, Joshua and Millman, K. Jarrod and
            Mayorov, Nikolay and Nelson, Andrew R. J. and Jones, Eric and
            Kern, Robert and Larson, Eric and Carey, C J and
            Polat, {\.I}lhan and Feng, Yu and Moore, Eric W. and
            {VanderPlas}, Jake and Laxalde, Denis and Perktold, Josef and
            Cimrman, Robert and Henriksen, Ian and Quintero, E. A. and
            Harris, Charles R. and Archibald, Anne M. and
            Ribeiro, Ant{\^o}nio H. and Pedregosa, Fabian and
            {van Mulbregt}, Paul and {SciPy 1.0 Contributors}},
  title   = {{{SciPy} 1.0: Fundamental Algorithms for Scientific
            Computing in Python}},
  journal = {Nature Methods},
  year    = {2020},
  volume  = {17},
  pages   = {261--272},
  adsurl  = {https://rdcu.be/b08Wh},
  doi     = {10.1038/s41592-019-0686-2},
}

@ARTICLE{Nelson.etal.20,
       author = {{Nelson}, Dylan and {Sharma}, Prateek and {Pillepich}, Annalisa and {Springel}, Volker and {Pakmor}, R{\"u}diger and {Weinberger}, Rainer and {Vogelsberger}, Mark and {Marinacci}, Federico and {Hernquist}, Lars},
        title = "{Resolving small-scale cold circumgalactic gas in TNG50}",
      journal = {\mnras},
         year = 2020,
        month = oct,
       volume = {498},
       number = {2},
        pages = {2391-2414},
          doi = {10.1093/mnras/staa2419},
archivePrefix = {arXiv},
       eprint = {2005.09654},
 primaryClass = {astro-ph.GA},
       adsurl = {https://ui.adsabs.harvard.edu/abs/2020MNRAS.498.2391N}
}

@Article{Harris.2020,
 title         = {Array programming with {NumPy}},
 author        = {Charles R. Harris and K. Jarrod Millman and St{\'{e}}fan J.
                 van der Walt and Ralf Gommers and Pauli Virtanen and David
                 Cournapeau and Eric Wieser and Julian Taylor and Sebastian
                 Berg and Nathaniel J. Smith and Robert Kern and Matti Picus
                 and Stephan Hoyer and Marten H. van Kerkwijk and Matthew
                 Brett and Allan Haldane and Jaime Fern{\'{a}}ndez del
                 R{\'{i}}o and Mark Wiebe and Pearu Peterson and Pierre
                 G{\'{e}}rard-Marchant and Kevin Sheppard and Tyler Reddy and
                 Warren Weckesser and Hameer Abbasi and Christoph Gohlke and
                 Travis E. Oliphant},
 year          = {2020},
 month         = sep,
 journal       = {Nature},
 volume        = {585},
 number        = {7825},
 pages         = {357--362},
 doi           = {10.1038/s41586-020-2649-2},
 publisher     = {Springer Science and Business Media {LLC}},
 url           = {https://doi.org/10.1038/s41586-020-2649-2}
}

@ARTICLE{Mandelker.etal.20a,
       author = {{Mandelker}, Nir and {Nagai}, Daisuke and {Aung}, Han and {Dekel}, Avishai and {Birnboim}, Yuval and {van den Bosch}, Frank C.},
        title = "{Instability of supersonic cold streams feeding galaxies - IV. Survival of radiatively cooling streams}",
      journal = {\mnras},
         year = 2020,
        month = may,
       volume = {494},
       number = {2},
        pages = {2641-2663},
          doi = {10.1093/mnras/staa812},
archivePrefix = {arXiv},
       eprint = {1910.05344},
 primaryClass = {astro-ph.GA},
       adsurl = {https://ui.adsabs.harvard.edu/abs/2020MNRAS.494.2641M}
}

@ARTICLE{Mandelker.etal.20b,
       author = {{Mandelker}, Nir and {van den Bosch}, Frank C. and {Nagai}, Daisuke and {Dekel}, Avishai and {Birnboim}, Yuval and {Aung}, Han},
        title = "{Ly {\ensuremath{\alpha}} blobs from cold streams undergoing Kelvin-Helmholtz instabilities}",
      journal = {\mnras},
         year = 2020,
        month = oct,
       volume = {498},
       number = {2},
        pages = {2415-2427},
          doi = {10.1093/mnras/staa2421},
archivePrefix = {arXiv},
       eprint = {2003.01724},
 primaryClass = {astro-ph.CO},
       adsurl = {https://ui.adsabs.harvard.edu/abs/2020MNRAS.498.2415M}
}

@ARTICLE{Modelevsky.Sari.21,
       author = {{Modelevsky}, Elisha and {Sari}, Re'em},
        title = "{Revisiting the strong shock problem: Converging and diverging shocks in different geometries}",
      journal = {Physics of Fluids},
         year = 2021,
        month = may,
       volume = {33},
       number = {5},
          eid = {056105},
        pages = {056105},
          doi = {10.1063/5.0047518},
archivePrefix = {arXiv},
       eprint = {2102.07235},
 primaryClass = {astro-ph.HE},
       adsurl = {https://ui.adsabs.harvard.edu/abs/2021PhFl...33e6105M}
}

@article{Tan.etal.21,
  title = {Radiative Mixing Layers: Insights from Turbulent Combustion},
  author = {Tan, Brent and Oh, S Peng and Gronke, Max},
  year = 2021,
  month = feb,
  journal = {Monthly Notices of the Royal Astronomical Society},
  volume = {502},
  number = {3},
  pages = {3179--3199},
  issn = {0035-8711, 1365-2966},
  doi = {10.1093/mnras/stab053}
}

@ARTICLE{Ramsoy.etal.21,
       author = {{Rams{\o}y}, Marius and {Slyz}, Adrianne and {Devriendt}, Julien and {Laigle}, Clotilde and {Dubois}, Yohan},
        title = "{Rivers of gas - I. Unveiling the properties of high redshift filaments}",
      journal = {\mnras},
         year = 2021,
        month = mar,
       volume = {502},
       number = {1},
        pages = {351-368},
          doi = {10.1093/mnras/stab015},
archivePrefix = {arXiv},
       eprint = {2101.00844},
 primaryClass = {astro-ph.GA},
       adsurl = {https://ui.adsabs.harvard.edu/abs/2021MNRAS.502..351R}
}

@ARTICLE{Stern.etal.21,
       author = {{Stern}, Jonathan and {Faucher-Gigu{\`e}re}, Claude-Andr{\'e} and {Fielding}, Drummond and {Quataert}, Eliot and {Hafen}, Zachary and {Gurvich}, Alexander B. and {Ma}, Xiangcheng and {Byrne}, Lindsey and {El-Badry}, Kareem and {Angl{\'e}s-Alc{\'a}zar}, Daniel and et al.},
        title = "{Virialization of the Inner CGM in the FIRE Simulations and Implications for Galaxy Disks, Star Formation, and Feedback}",
      journal = {\apj},
         year = 2021,
        month = apr,
       volume = {911},
       number = {2},
          eid = {88},
        pages = {88},
          doi = {10.3847/1538-4357/abd776},
archivePrefix = {arXiv},
       eprint = {2006.13976},
 primaryClass = {astro-ph.GA},
       adsurl = {https://ui.adsabs.harvard.edu/abs/2021ApJ...911...88S}
}

@article{Banda-Barragan.etal.21,
  title = {Shock--Multicloud Interactions in Galactic Outflows -- {{II}}. {{Radiative}} Fractal Clouds and Cold Gas Thermodynamics},
  author = {{Banda-Barrag{\'a}n}, W E and Br{\"u}ggen, M and Heesen, V and Scannapieco, E and Cottle, J and Federrath, C and Wagner, A Y},
  year = 2021,
  month = aug,
  journal = {Monthly Notices of the Royal Astronomical Society},
  volume = {506},
  number = {4},
  pages = {5658--5680},
  issn = {0035-8711, 1365-2966},
  doi = {10.1093/mnras/stab1884}
}

@article{Daddi.etal.22a,
  title = {Evidence for {{Cold-stream}} to {{Hot-accretion Transition}} as {{Traced}} by {{Ly$\alpha$ Emission}} from {{Groups}} and {{Clusters}} at 2 {$<$} z {$<$} 3.3},
  author = {Daddi, E. and Rich, R.M. and Valentino, F. and others},
  year = 2022,
  month = feb,
  journal = {The Astrophysical Journal Letters},
  volume = {926},
  number = {2},
  pages = {L21},
  issn = {2041-8205, 2041-8213},
  doi = {10.3847/2041-8213/ac531f}
}

@ARTICLE{Daddi.etal.22b,
       author = {{Daddi}, E. and {Delvecchio}, I. and {Dimauro}, P. and {Magnelli}, B. and {Gomez-Guijarro}, C. and {Coogan}, R. and {Elbaz}, D. and {Kalita}, B.~S. and {Le Bail}, A. and {Rich}, R.~M. and {Tan}, Q.},
        title = "{The bending of the star-forming main sequence traces the cold- to hot-accretion transition mass over 0 < z < 4}",
      journal = {\aap},
         year = 2022,
        month = may,
       volume = {661},
          eid = {L7},
        pages = {L7},
          doi = {10.1051/0004-6361/202243574},
archivePrefix = {arXiv},
       eprint = {2203.10880},
 primaryClass = {astro-ph.CO},
       adsurl = {https://ui.adsabs.harvard.edu/abs/2022A&A...661L...7D}
}

@ARTICLE{astropy.22,
       author = {{Astropy Collaboration} and {Price-Whelan}, Adrian M. and {Lim}, Pey Lian and {Earl}, Nicholas and {Starkman}, Nathaniel and {Bradley}, Larry and {Shupe}, David L. and {Patil}, Aarya A. and {Corrales}, Lia and {Brasseur}, C.~E. and {N{\"o}the}, Maximilian and {Donath}, Axel and {Tollerud}, Erik and {Morris}, Brett M. and {Ginsburg}, Adam and {Vaher}, Eero and {Weaver}, Benjamin A. and {Tocknell}, James and {Jamieson}, William and {van Kerkwijk}, Marten H. and {Robitaille}, Thomas P. and {Merry}, Bruce and {Bachetti}, Matteo and {G{\"u}nther}, H. Moritz and {Aldcroft}, Thomas L. and {Alvarado-Montes}, Jaime A. and {Archibald}, Anne M. and {B{\'o}di}, Attila and {Bapat}, Shreyas and {Barentsen}, Geert and {Baz{\'a}n}, Juanjo and {Biswas}, Manish and {Boquien}, M{\'e}d{\'e}ric and {Burke}, D.~J. and {Cara}, Daria and {Cara}, Mihai and {Conroy}, Kyle E. and {Conseil}, Simon and {Craig}, Matthew W. and {Cross}, Robert M. and {Cruz}, Kelle L. and {D'Eugenio}, Francesco and {Dencheva}, Nadia and {Devillepoix}, Hadrien A.~R. and {Dietrich}, J{\"o}rg P. and {Eigenbrot}, Arthur Davis and {Erben}, Thomas and {Ferreira}, Leonardo and {Foreman-Mackey}, Daniel and {Fox}, Ryan and {Freij}, Nabil and {Garg}, Suyog and {Geda}, Robel and {Glattly}, Lauren and {Gondhalekar}, Yash and {Gordon}, Karl D. and {Grant}, David and {Greenfield}, Perry and {Groener}, Austen M. and {Guest}, Steve and {Gurovich}, Sebastian and {Handberg}, Rasmus and {Hart}, Akeem and {Hatfield-Dodds}, Zac and {Homeier}, Derek and {Hosseinzadeh}, Griffin and {Jenness}, Tim and {Jones}, Craig K. and {Joseph}, Prajwel and {Kalmbach}, J. Bryce and {Karamehmetoglu}, Emir and {Ka{\l}uszy{\'n}ski}, Miko{\l}aj and {Kelley}, Michael S.~P. and {Kern}, Nicholas and {Kerzendorf}, Wolfgang E. and {Koch}, Eric W. and {Kulumani}, Shankar and {Lee}, Antony and {Ly}, Chun and {Ma}, Zhiyuan and {MacBride}, Conor and {Maljaars}, Jakob M. and {Muna}, Demitri and {Murphy}, N.~A. and {Norman}, Henrik and {O'Steen}, Richard and {Oman}, Kyle A. and {Pacifici}, Camilla and {Pascual}, Sergio and {Pascual-Granado}, J. and {Patil}, Rohit R. and {Perren}, Gabriel I. and {Pickering}, Timothy E. and {Rastogi}, Tanuj and {Roulston}, Benjamin R. and {Ryan}, Daniel F. and {Rykoff}, Eli S. and {Sabater}, Jose and {Sakurikar}, Parikshit and {Salgado}, Jes{\'u}s and {Sanghi}, Aniket and {Saunders}, Nicholas and {Savchenko}, Volodymyr and {Schwardt}, Ludwig and {Seifert-Eckert}, Michael and {Shih}, Albert Y. and {Jain}, Anany Shrey and {Shukla}, Gyanendra and {Sick}, Jonathan and {Simpson}, Chris and {Singanamalla}, Sudheesh and {Singer}, Leo P. and {Singhal}, Jaladh and {Sinha}, Manodeep and {Sip{\H{o}}cz}, Brigitta M. and {Spitler}, Lee R. and {Stansby}, David and {Streicher}, Ole and {{\v{S}}umak}, Jani and {Swinbank}, John D. and {Taranu}, Dan S. and {Tewary}, Nikita and {Tremblay}, Grant R. and {de Val-Borro}, Miguel and {Van Kooten}, Samuel J. and {Vasovi{\'c}}, Zlatan and {Verma}, Shresth and {de Miranda Cardoso}, Jos{\'e} Vin{\'\i}cius and {Williams}, Peter K.~G. and {Wilson}, Tom J. and {Winkel}, Benjamin and {Wood-Vasey}, W.~M. and {Xue}, Rui and {Yoachim}, Peter and {Zhang}, Chen and {Zonca}, Andrea and {Astropy Project Contributors}},
        title = "{The Astropy Project: Sustaining and Growing a Community-oriented Open-source Project and the Latest Major Release (v5.0) of the Core Package}",
      journal = {\apj},
         year = 2022,
        month = aug,
       volume = {935},
       number = {2},
          eid = {167},
        pages = {167},
          doi = {10.3847/1538-4357/ac7c74},
archivePrefix = {arXiv},
       eprint = {2206.14220},
 primaryClass = {astro-ph.IM},
       adsurl = {https://ui.adsabs.harvard.edu/abs/2022ApJ...935..167A}
}

@ARTICLE{Gronnow.etal.22,
       author = {{Gr{\o}nnow}, Asger and {Tepper-Garc{\'\i}a}, Thor and {Bland-Hawthorn}, Joss and {Fraternali}, Filippo},
        title = "{The role of the halo magnetic field on accretion through high-velocity clouds}",
      journal = {\mnras},
         year = 2022,
        month = feb,
       volume = {509},
       number = {4},
        pages = {5756-5770},
          doi = {10.1093/mnras/stab3452},
archivePrefix = {arXiv},
       eprint = {2111.12733},
 primaryClass = {astro-ph.GA},
       adsurl = {https://ui.adsabs.harvard.edu/abs/2022MNRAS.509.5756G}
}

@article{Kotecha.etal.22,
  title = {Cosmic Filaments Delay Quenching inside Clusters},
  author = {Kotecha, Sachin and Welker, Charlotte and Zhou, Zihan and Wadsley, James and Kraljic, Katarina and Sorce, Jenny and Rasia, Elena and Roberts, Ian and Gray, Meghan and Yepes, Gustavo and Cui, Weiguang},
  year = 2022,
  month = mar,
  journal = {Monthly Notices of the Royal Astronomical Society},
  volume = {512},
  number = {1},
  pages = {926--944},
  issn = {0035-8711, 1365-2966},
  doi = {10.1093/mnras/stac300}
}

@article{Gronke.etal.22,
  title = {Survival and Mass Growth of Cold Gas in a Turbulent, Multiphase Medium},
  author = {Gronke, Max and Oh, S Peng and Ji, Suoqing and Norman, Colin},
  year = 2022,
  month = feb,
  journal = {Monthly Notices of the Royal Astronomical Society},
  volume = {511},
  number = {1},
  pages = {859--876},
  issn = {0035-8711, 1365-2966},
  doi = {10.1093/mnras/stab3351}
}

@article{Vurm.etal.23,
  title = {Cosmic Gas Highways in {{C-EAGLE}} Simulations},
  author = {Vurm, I. and Nevalainen, J. and Hong, S.E. and Bah{\'e}, Y. M. and Vecchia, C. Dalla and Hein{\"a}m{\"a}ki, P.},
  year = 2023,
  month = may,
  journal = {Astronomy \& Astrophysics},
  volume = {673},
  eprint = {2303.03244},
  primaryclass = {astro-ph.CO},
  pages = {A62},
  issn = {0004-6361, 1432-0746},
  doi = {10.1051/0004-6361/202243904},
  archiveprefix = {arXiv}
}

@article{Pandya.etal.23,
  title = {A {{Unified Model}} for the {{Coevolution}} of {{Galaxies}} and {{Their Circumgalactic Medium}}: {{The Relative Roles}} of {{Turbulence}} and {{Atomic Cooling Physics}}},
  author = {Pandya, Viraj and Fielding, Drummond B. and Bryan, Greg L. and Carr, Christopher and Somerville, Rachel S. and Stern, Jonathan and {Faucher-Gigu{\`e}re}, Claude-Andr{\'e} and Hafen, Zachary and {Angl{\'e}s-Alc{\'a}zar}, Daniel and Forbes, John C.},
  year = 2023,
  month = oct,
  journal = {The Astrophysical Journal},
  volume = {956},
  number = {2},
  pages = {118},
  issn = {0004-637X, 1538-4357},
  doi = {10.3847/1538-4357/acf3ea}
}

@ARTICLE{Abruzzo.etal.23,
       author = {{Abruzzo}, Matthew W. and {Fielding}, Drummond B. and {Bryan}, Greg L.},
        title = "{TuRMoiL of Survival: A Unified Survival Criterion for Cloud-Wind Interactions}",
      journal = {arXiv e-prints},
         year = 2023,
        month = jul,
          eid = {arXiv:2307.03228},
        pages = {arXiv:2307.03228},
          doi = {10.48550/arXiv.2307.03228},
archivePrefix = {arXiv},
       eprint = {2307.03228},
 primaryClass = {astro-ph.GA},
       adsurl = {https://ui.adsabs.harvard.edu/abs/2023arXiv230703228A}
}

@ARTICLE{Faucher-Giguere.Oh.23,
       author = {{Faucher-Gigu{\`e}re}, Claude-Andr{\'e} and {Oh}, S. Peng},
        title = "{Key Physical Processes in the Circumgalactic Medium}",
      journal = {\araa},
         year = 2023,
        month = aug,
       volume = {61},
        pages = {131-195},
          doi = {10.1146/annurev-astro-052920-125203},
archivePrefix = {arXiv},
       eprint = {2301.10253},
 primaryClass = {astro-ph.GA},
       adsurl = {https://ui.adsabs.harvard.edu/abs/2023ARA&A..61..131F}
}

@ARTICLE{Popesso.etal.23,
       author = {{Popesso}, P. and {Concas}, A. and {Cresci}, G. and {Belli}, S. and {Rodighiero}, G. and {Inami}, H. and {Dickinson}, M. and {Ilbert}, O. and {Pannella}, M. and {Elbaz}, D.},
        title = "{The main sequence of star-forming galaxies across cosmic times}",
      journal = {\mnras},
         year = 2023,
        month = feb,
       volume = {519},
       number = {1},
        pages = {1526-1544},
          doi = {10.1093/mnras/stac3214},
archivePrefix = {arXiv},
       eprint = {2203.10487},
 primaryClass = {astro-ph.GA},
       adsurl = {https://ui.adsabs.harvard.edu/abs/2023MNRAS.519.1526P}
}

@article{Ledos.etal.23,
  title = {Stability and {{Ly}} {$\alpha$} Emission of {{Cold Stream}} in the {{Circumgalactic Medium}}: Impact of Magnetic Fields and Thermal Conduction},
  author = {Ledos, Nicolas and Takasao, Shinsuke and Nagamine, Kentaro},
  year = 2023,
  month = dec,
  journal = {Monthly Notices of the Royal Astronomical Society},
  volume = {527},
  number = {4},
  pages = {11304--11326},
  issn = {0035-8711, 1365-2966},
  doi = {10.1093/mnras/stad3814}
}

@ARTICLE{Hidalgo-Pineda.etal.24,
       author = {{Hidalgo-Pineda}, Fernando and {Farber}, Ryan Jeffrey and {Gronke}, Max},
        title = "{Better together: the complex interplay between radiative cooling and magnetic draping}",
      journal = {\mnras},
         year = 2024,
        month = jan,
       volume = {527},
       number = {1},
        pages = {135-149},
          doi = {10.1093/mnras/stad3069},
archivePrefix = {arXiv},
       eprint = {2304.09897},
 primaryClass = {astro-ph.GA},
       adsurl = {https://ui.adsabs.harvard.edu/abs/2024MNRAS.527..135H}
}

@ARTICLE{Das.Gronke.24,
       author = {{Das}, Hitesh Kishore and {Gronke}, Max},
        title = "{Magnetic fields in multiphase turbulence: impact on dynamics and structure}",
      journal = {\mnras},
         year = 2024,
        month = jan,
       volume = {527},
       number = {1},
        pages = {991-1013},
          doi = {10.1093/mnras/stad3125},
archivePrefix = {arXiv},
       eprint = {2307.06411},
 primaryClass = {astro-ph.GA},
       adsurl = {https://ui.adsabs.harvard.edu/abs/2024MNRAS.527..991D}
}

@ARTICLE{Ramesh.etal.24,
       author = {{Ramesh}, Rahul and {Nelson}, Dylan and {Fielding}, Drummond and {Br{\"u}ggen}, Marcus},
        title = "{Zooming in on the circumgalactic medium with GIBLE. The topology and draping of magnetic fields around cold clouds}",
      journal = {\aap},
         year = 2024,
        month = apr,
       volume = {684},
          eid = {L16},
        pages = {L16},
          doi = {10.1051/0004-6361/202348786},
archivePrefix = {arXiv},
       eprint = {2404.01370},
 primaryClass = {astro-ph.GA},
       adsurl = {https://ui.adsabs.harvard.edu/abs/2024A&A...684L..16R}
}

@ARTICLE{Lu.etal.24,
       author = {{Lu}, Yue Samuel and {Mandelker}, Nir and {Oh}, Siang Peng and {Dekel}, Avishai and {van den Bosch}, Frank C. and {Springel}, Volker and {Nagai}, Daisuke and {van de Voort}, Freeke},
        title = "{The structure and dynamics of massive high-z cosmic-web filaments: three radial zones in filament cross-sections}",
      journal = {\mnras},
         year = 2024,
        month = feb,
       volume = {527},
       number = {4},
        pages = {11256-11287},
          doi = {10.1093/mnras/stad3779},
archivePrefix = {arXiv},
       eprint = {2306.03966},
 primaryClass = {astro-ph.CO},
       adsurl = {https://ui.adsabs.harvard.edu/abs/2024MNRAS.52711256L}
}

@ARTICLE{Aung.etal.19,
       author = {{Aung}, Han and {Mandelker}, Nir and {Nagai}, Daisuke and {Dekel}, Avishai and {Birnboim}, Yuval},
        title = "{Kelvin-Helmholtz instability in self-gravitating streams}",
      journal = {\mnras},
         year = 2019,
        month = nov,
       volume = {490},
       number = {1},
        pages = {181-201},
          doi = {10.1093/mnras/stz1964},
archivePrefix = {arXiv},
       eprint = {1903.09666},
 primaryClass = {astro-ph.GA},
       adsurl = {https://ui.adsabs.harvard.edu/abs/2019MNRAS.490..181A}
}

@ARTICLE{Aung.etal.24,
       author = {{Aung}, Han and {Mandelker}, Nir and {Dekel}, Avishai and {Nagai}, Daisuke and {Semenov}, Vadim and {van den Bosch}, Frank C.},
        title = "{Entrainment of hot gas into cold streams: the origin of excessive star formation rates at cosmic noon}",
      journal = {\mnras},
         year = 2024,
        month = aug,
       volume = {532},
       number = {3},
        pages = {2965-2987},
          doi = {10.1093/mnras/stae1673},
archivePrefix = {arXiv},
       eprint = {2403.00912},
 primaryClass = {astro-ph.GA},
       adsurl = {https://ui.adsabs.harvard.edu/abs/2024MNRAS.532.2965A}
}

@ARTICLE{Rost.etal.24,
       author = {{Rost}, Agust{\'\i}n M. and {Nuza}, Sebasti{\'a}n E. and {Stasyszyn}, Federico and {Kuchner}, Ulrike and {Hoeft}, Matthias and {Welker}, Charlotte and {Pearce}, Frazer and {Gray}, Meghan and {Knebe}, Alexander and {Cui}, Weiguang and {Yepes}, Gustavo},
        title = "{The three hundred project: thermodynamical properties, shocks, and gas dynamics in simulated galaxy cluster filaments and their surroundings}",
      journal = {\mnras},
         year = 2024,
        month = jan,
       volume = {527},
       number = {1},
        pages = {1301-1316},
          doi = {10.1093/mnras/stad3208},
archivePrefix = {arXiv},
       eprint = {2310.12245},
 primaryClass = {astro-ph.CO},
       adsurl = {https://ui.adsabs.harvard.edu/abs/2024MNRAS.527.1301R}
}

@article{Stern.etal.24,
  title = {Accretion onto Disc Galaxies via Hot and Rotating {{CGM}} Inflows},
  author = {Stern, Jonathan and Fielding, Drummond and Hafen, Zachary and Su, Kung-Yi and Naor, Nadav and {Faucher-Gigu{\`e}re}, Claude-Andr{\'e} and Quataert, Eliot and Bullock, James},
  year = 2024,
  month = apr,
  journal = {Monthly Notices of the Royal Astronomical Society},
  volume = {530},
  number = {2},
  pages = {1711--1731},
  issn = {0035-8711, 1365-2966},
  doi = {10.1093/mnras/stae824}
}

@article{Voit.etal.24a,
  title = {Equilibrium {{States}} of {{Galactic Atmospheres}}. {{I}}. {{The Flip Side}} of {{Mass Loading}}},
  author = {Voit, G. Mark and Pandya, Viraj and Fielding, Drummond B. and Bryan, Greg L. and Carr, Christopher and Donahue, Megan and Oppenheimer, Benjamin D. and Somerville, Rachel S.},
  year = 2024,
  month = dec,
  journal = {The Astrophysical Journal},
  volume = {976},
  number = {2},
  pages = {150},
  issn = {0004-637X, 1538-4357},
  doi = {10.3847/1538-4357/ad81d6}
}

@article{Voit.etal.24b,
  title = {Equilibrium {{States}} of {{Galactic Atmospheres}}. {{II}}. {{Interpretation}} and {{Implications}}},
  author = {Voit, G. Mark and Carr, Christopher and Fielding, Drummond B. and Pandya, Viraj and Bryan, Greg L. and Donahue, Megan and Oppenheimer, Benjamin D. and Somerville, Rachel S.},
  year = 2024,
  month = dec,
  journal = {The Astrophysical Journal},
  volume = {976},
  number = {2},
  pages = {151},
  issn = {0004-637X, 1538-4357},
  doi = {10.3847/1538-4357/ad81d5}
}

@ARTICLE{Yao.etal.25,
       author = {{Yao}, Zhiyuan and {Mandelker}, Nir and {Oh}, S. Peng and {Aung}, Han and {Dekel}, Avishai},
        title = "{Effects of cloud geometry and metallicity on shattering and coagulation of cold gas, and implications for cold streams penetrating virial shocks}",
      journal = {\mnras},
         year = 2025,
        month = jan,
       volume = {536},
       number = {3},
        pages = {3053-3089},
          doi = {10.1093/mnras/stae2771},
archivePrefix = {arXiv},
       eprint = {2410.12914},
 primaryClass = {astro-ph.GA},
       adsurl = {https://ui.adsabs.harvard.edu/abs/2025MNRAS.536.3053Y}
}

@article{Wang.etal.25,
  title = {{\emph{Eppur }}{{{\emph{Si Muove}}}} : Self-Sustained Streaming Motions in Multiphase {{MHD}}},
  author = {Wang, Chaoran and Oh, S Peng and Jiang, Yan-Fei and Kaul, Ish},
  year = 2025,
  month = nov,
  journal = {Monthly Notices of the Royal Astronomical Society},
  volume = {544},
  number = {4},
  pages = {4119--4145},
  issn = {0035-8711, 1365-2966},
  doi = {10.1093/mnras/staf1968}
}

@article{Kaul.etal.25,
  title = {Tales of Tension: Magnetized Infalling Cold Clouds and Streams in the {{CGM}}},
  author = {Kaul, Ish and Tan, Brent and Oh, S Peng and Mandelker, Nir},
  year = 2025,
  month = may,
  journal = {Monthly Notices of the Royal Astronomical Society},
  volume = {539},
  number = {4},
  pages = {3669--3696},
  issn = {0035-8711, 1365-2966},
  doi = {10.1093/mnras/staf706}
}

@misc{Goldner.etal.25,
  title = {Accretion-{{Driven Turbulence}} in the {{Circumgalactic Medium}}},
  author = {Goldner, Roy and Stern, Jonathan and Fielding, Drummond and {Faucher-Gigu{\`e}re}, Claude-Andr{\'e} and Faerman, Yakov and Kakoly, Aharon},
  year = 2025,
  month = oct,
  number = {arXiv:2510.27678},
  eprint = {2510.27678},
  primaryclass = {astro-ph},
  publisher = {arXiv},
  doi = {10.48550/arXiv.2510.27678},
  archiveprefix = {arXiv}
}

@article{Antipov.etal.25,
  title = {On the Kinematic and Thermodynamic State of Clouds in Complex Wind--Multicloud Environments Using a Friends-of-Friends Analysis},
  author = {Antipov, A and {Banda-Barrag{\'a}n}, W E and Birnboim, Y and Federrath, C and Gnat, O and Br{\"u}ggen, M},
  year = 2025,
  month = jun,
  journal = {Monthly Notices of the Royal Astronomical Society},
  volume = {540},
  number = {4},
  pages = {3798--3817},
  issn = {0035-8711, 1365-2966},
  doi = {10.1093/mnras/staf949}
}

@article{Medlock.etal.26,
  title = {Statistical {{Properties}} of {{Cold Streams}} in {{Massive Star-forming Halos}} in {{TNG50}}},
  author = {Medlock, Isabel and Nagai, Daisuke and Mandelker, Nir and Springel, Volker and {van den Bosch}, Frank C. and Zinger, Elad and Chiang, Barry T.},
  year = 2026,
  month = apr,
  journal = {The Astrophysical Journal},
  volume = {1000},
  number = {2},
  pages = {222},
  issn = {0004-637X, 1538-4357},
  doi = {10.3847/1538-4357/ae4c42}
}

@ARTICLE{Elias.etal.26,
       author = {{Elias}, Guillaume and {Daddi}, Emanuele and {D'Eugenio}, Chiara and {Elbaz}, David and {Franco}, Maximilien and {Gentile}, Fabrizio and {Gobat}, Raphael and {Guo}, Sicen and {Jin}, Shuowen and {Laigle}, Clotilde and {Lu}, Shiying and {Magdis}, Georgios E. and {Magnelli}, Benjamin and {Sillassen}, Nikolaj B. and {Strazzullo}, Veronica and {Tarrasse}, Maxime and {Wang}, Tao and {Zhou}, Luwenjia},
        title = "{Quiescent fractions in high-redshift galaxy groups reflect their hot-or-cold state of gas accretion}",
      journal = {arXiv e-prints},
         year = 2026,
        month = apr,
          eid = {arXiv:2604.22401},
        pages = {arXiv:2604.22401},
archivePrefix = {arXiv},
       eprint = {2604.22401},
 primaryClass = {astro-ph.GA},
       adsurl = {https://ui.adsabs.harvard.edu/abs/2026arXiv260422401E}
}

@article{Paste.etal.26,
  title = {Cosmic Gas Accretion from Filaments onto Galaxy Clusters Using the {{IllustrisTNG}} Simulation},
  author = {Past{\'e}, Jade and Gouin, C{\'e}line and Aghanim, Nabila and Sorce, Jenny G.},
  year = 2026,
  month = apr,
  eprint = {2604.24852},
  journal = {arXiv e-prints},
  primaryclass = {astro-ph.CO},
  publisher = {arXiv},
  doi = {10.48550/arXiv.2604.24852},
  archiveprefix = {arXiv}
}

@article{Gronke.Schneider.26,
  title = {Simulations of Multi-Phase Gas in and around Galaxies},
  author = {Gronke, Max and Schneider, Evan},
  year = 2026,
  month = apr,
  journal = {Living Reviews in Computational Astrophysics},
  volume = {12},
  number = {1},
  pages = {2},
  issn = {2365-0524},
  doi = {10.1007/s41115-026-00025-7}
}

@article{Bustard.Gronke.22,
  author = {Bustard, Chad and Gronke, Max},
  title = {Radiative Turbulent Mixing Layers and the Survival of {Magellanic} Debris},
  journal = {ApJ},
  year = {2022},
  volume = {933},
  number = {2},
  pages = {120},
  doi = {10.3847/1538-4357/ac752b}
}

@ARTICLE{McCourt.etal.12,
       author = {{McCourt}, Michael and {Sharma}, Prateek and {Quataert}, Eliot and {Parrish}, Ian J.},
        title = {Thermal instability in gravitationally stratified plasmas: implications for multiphase structure in clusters and galaxy haloes},
      journal = {\mnras},
         year = {2012},
       volume = {419},
       number = {4},
        pages = {3319--3337},
          doi = {10.1111/j.1365-2966.2011.19972.x},
archivePrefix = {arXiv},
       eprint = {1105.2563},
 primaryClass = {astro-ph.CO}
}

@ARTICLE{Thompson.etal.16,
       author = {{Thompson}, Todd A. and {Quataert}, Eliot and {Zhang}, Dong and {Weinberg}, David H.},
        title = {An origin for multiphase gas in galactic winds and haloes},
      journal = {\mnras},
         year = {2016},
       volume = {455},
       number = {2},
        pages = {1830--1844},
          doi = {10.1093/mnras/stv2428},
archivePrefix = {arXiv},
       eprint = {1507.04362},
 primaryClass = {astro-ph.GA}
}

@ARTICLE{Schneider.etal.20,
       author = {{Schneider}, Evan E. and {Ostriker}, Eve C. and {Robertson}, Brant E. and {Thompson}, Todd A.},
        title = {The Physical Nature of Starburst-driven Galactic Outflows},
      journal = {\apj},
         year = {2020},
       volume = {895},
       number = {1},
        pages = {43},
          doi = {10.3847/1538-4357/ab8ae8},
archivePrefix = {arXiv},
       eprint = {2002.10468},
 primaryClass = {astro-ph.GA}
}

@ARTICLE{Fielding.Bryan.22,
       author = {{Fielding}, Drummond B. and {Bryan}, Greg L.},
        title = {The Structure of Multiphase Galactic Winds},
      journal = {\apj},
         year = {2022},
       volume = {924},
        pages = {82},
          doi = {10.3847/1538-4357/ac2f41},
archivePrefix = {arXiv},
       eprint = {2108.05355},
 primaryClass = {astro-ph.GA}
}

@ARTICLE{Voit.etal.17,
       author = {{Voit}, G. Mark and {Meece}, Greg and {Li}, Yuan and {O'Shea}, Brian W. and {Bryan}, Greg L. and {Donahue}, Megan},
        title = {A Global Model for Circumgalactic and Cluster-core Precipitation},
      journal = {\apj},
         year = {2017},
       volume = {845},
        pages = {80},
          doi = {10.3847/1538-4357/aa7d04},
archivePrefix = {arXiv},
       eprint = {1607.02212},
 primaryClass = {astro-ph.GA}
}



\appendix

\section{Convergence tests}
\label{sec:convergence}
\begin{figure*}
    \centering	
    \includegraphics[width=\textwidth]{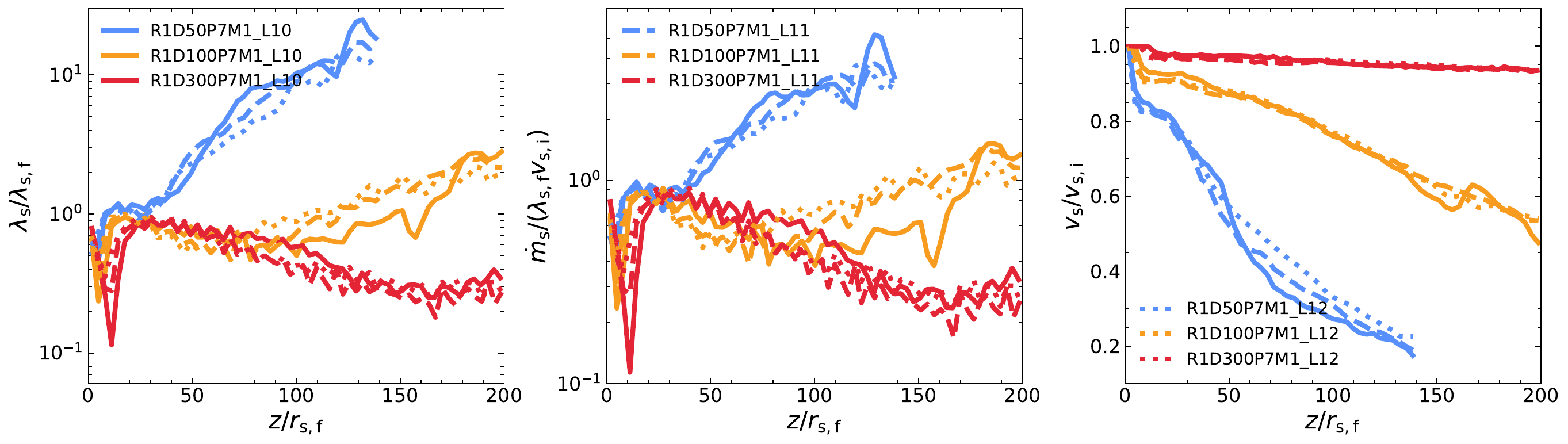}
    \caption{Similar to \cref{fig:profile_mass_size}, but for simulations with different maximum resolutions. Solid, dashed, and dotted curves correspond to highest refinement levels of 10, 11, and 12, respectively.
    }
    \label{fig:profile_mass_res}
\end{figure*}

\begin{figure*}
    \centering	
    \includegraphics[width=\textwidth]{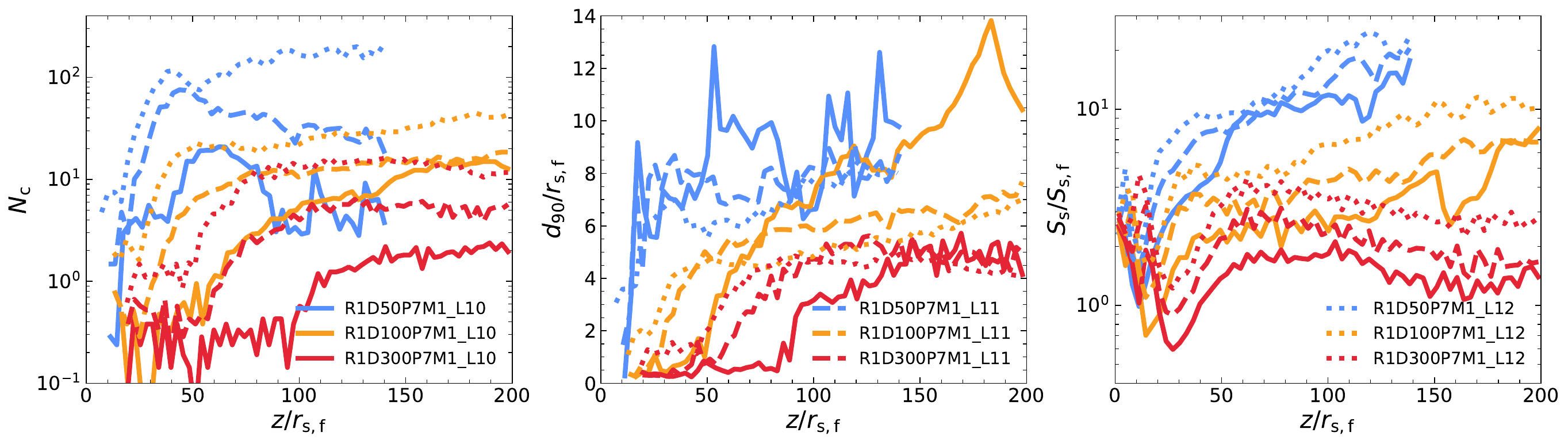}
    \caption{Similar to \cref{fig:profile_morph_size}, but varying the highest refinement levels of simulations.
    }
    \label{fig:profile_morph_res}
\end{figure*}


\begin{figure}
    \centering	
    \includegraphics[width=\columnwidth]{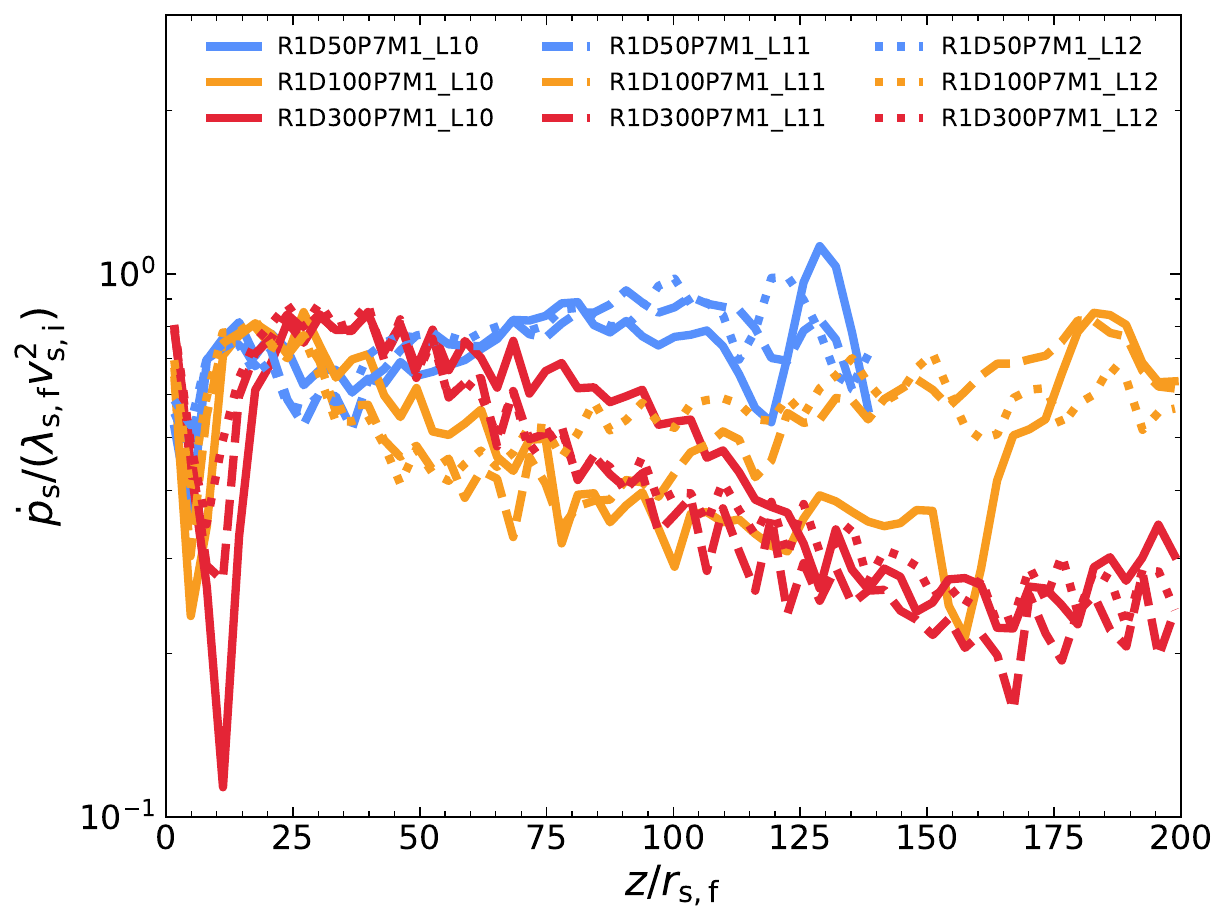}
    \caption{Similar to the central panel of \cref{fig:profile_mass_theory}, but varying the highest refinement levels of simulations.}
    \label{fig:profile_momx_res}
\end{figure}
We test resolution effects by varying the maximum refinement level around the stream region in three fiducial runs. In the level-10 runs, refinement level 11 is removed within \ifm{\rx{cyl}=9.5\,\rx{s,f}}, thereby eliminating the resolution drop at that radius. In contrast, in the level-12 runs, we introduce an additional refinement level within \ifm{\rx{cyl}=14.2\,\rx{s,f}}, ensuring that the refinement level reaches 12 inside \ifm{\rx{cyl}=9.5\,\rx{s,f}}. We present the evolution of cold-gas line density, mass flux, and velocity in \cref{fig:profile_mass_res}; the number of clumps, the 90th-percentile clump distance, and the cold-gas surface area per unit length in \cref{fig:profile_morph_res}; and momentum flux in \cref{fig:profile_momx_res}. The bulk cold-gas profiles retain their qualitative growth or loss trends across all tested resolutions, with only minor differences. Clump counts and interfacial area are more sensitive to resolution. In particular, the downstream decline in \ifm{\Nx{c}} for \run{R1D50P7M1} at the fiducial resolution is not preserved at level 12. Thus, the clump-number profile alone does not establish convergence of its coagulation classification.

\smallskip
Inspection of the movies suggests that the level-12 stream retains an extended cold structure in projection while developing finer interface structure. Continued shear-driven stripping downstream of the initial collapse and rebound may replenish small fragments and offset their removal through coagulation, weakening the downstream decline in \ifm{\Nx{c}}. The visualizations also suggest differences in transverse dispersal: fragments expelled during the initial pressure-driven rebound reach larger distances from the axis, whereas subsequent interface fragments tend to remain closer to the stream and elongate downstream. These differences could affect the cold-gas covering fraction, although their contribution has not been quantified.

\smallskip
Strong resolution dependence of fragmentation has also been reported in radiative cloud--wind simulations by \citet{Sparre.etal.19}. Their cooling-induced fragmentation occurs in a flow that also contains shock compression and velocity shear, so this comparison does not establish that the same mechanism operates in our simulations. Pressure-driven studies without sustained shear found that the qualitative fragmentation or recoagulation outcome was robust across the tested resolutions, despite changes in fragment counts and small-scale morphology \citep{Gronke.Oh.20b,Yao.etal.25}.

\smallskip
The two critical radii shown in \cref{fig:diag_size_chi} are physically motivated estimates based on either velocity shear alone or pressure-driven fragmentation alone. The resolution dependence of shear-driven fragmentation complicates this picture, and we leave a quantitative separation of fragment production, disruption, and coagulation to future work. In contrast to the resolution sensitivity of the clump count, the bulk cold-gas profiles shown in \cref{fig:profile_mass_res} retain their qualitative growth or loss trends across all tested resolutions---supporting the broad survival and disruption trends more robustly than the detailed morphological classification does.

\smallskip
Numerical diffusion broadens the cold--hot interface and produces intermediate-temperature gas that can cool rapidly. However, unresolved thermal interfaces need not imply that numerical diffusion controls the bulk entrainment rate. \citet{Tan.etal.21} find that time-averaged cooling can converge when resolved turbulent transport dominates numerical diffusion, approximately \ifm{D_\mathrm{turb}\sim u'L>D_\mathrm{num}\sim \vx{num}\Delta x}, where \ifm{u'} and \ifm{L} characterize the turbulence and \ifm{\vx{num}} is the velocity scale entering the numerical diffusivity. Our final stream radius spans approximately 10 cells at the fiducial resolution and 20 cells at level 12. Approximating \ifm{L\sim\rx{s,f}} and \ifm{u'\sim\cx{s}}, with the characteristic velocity for numerical diffusivity \ifm{\vx{num}\lesssim\vx{s,i}=\Max{s}\cx{s}\chix{f}^{1/2}}, we obtain \ifm{D_\mathrm{turb}/D_\mathrm{num}\gtrsim10/(\Max{s}\chix{f}^{1/2})} at the fiducial resolution. This gives \ifm{D_\mathrm{turb}/D_\mathrm{num}\gtrsim1} for runs with \ifm{\Max{s}\chix{f}^{1/2}\le10}, i.e. all \ifm{\chix{f}=50} runs, the \ifm{\chix{f}=100} runs with \ifm{\Max{s}\le1}, and \run{R1D300P7M05}, but only \ifm{\gtrsim0.3\text{--}0.6} for the remaining \ifm{\chix{f}=300} runs and the \ifm{\Max{s}=2} runs, where numerical diffusion may play a larger role. The weaker resolution sensitivity of the bulk transport profiles is consistent with turbulence regulating entrainment, while numerical diffusion continues to affect thin interfaces, small fragments, and regions where turbulence has not yet developed.

\section{Sensitivity to interface perturbations}
\label{sec:perturb}
\begin{figure*}
    \centering	
    \includegraphics[width=\textwidth]{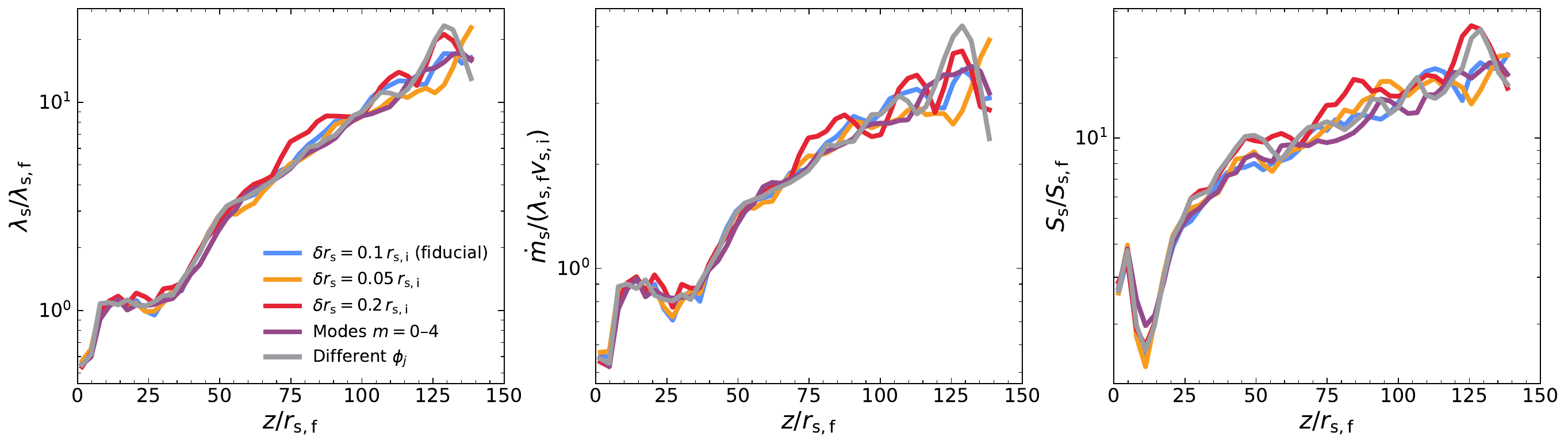}
    \caption{Sensitivity of the representative coagulation run \run{R1D50P7M1} to the imposed interface perturbations. From left to right, the panels show the cold-stream line density \ifm{\lambdax{s}}, the streamwise cold-gas mass flux \ifm{\dotm{s}}, and the cold-stream surface area per unit length \ifm{\Sx{s}}, normalized and time-averaged as in \cref{fig:profile_mass_size,fig:profile_morph_size}. The blue curve is the fiducial run with an rms interface-displacement amplitude of \ifm{\delta\rx{s}=0.1\,\rx{s,i}}. The orange and red curves halve and double this amplitude to \ifm{\delta\rx{s}=0.05\,\rx{s,i}} and \ifm{0.2\,\rx{s,i}}, respectively, while retaining the fiducial phases and mode spectrum. The purple curve extends the azimuthal spectrum from \ifm{m=0\text{--}1} to \ifm{m=0\text{--}4} at fixed longitudinal wavelengths and total rms amplitude. The gray curve adopts a different random realization of the phases \ifm{\phi_j} at the fiducial amplitude and spectrum. All five runs follow nearly identical profiles along the entire penetration length.}
    \label{fig:profile_mass_pert}
\end{figure*}
Our interface perturbations follow the prescription adopted in \citet{Yao.etal.25}, which builds on earlier studies of the Kelvin--Helmholtz instability in cold streams \citep{Mandelker.etal.19,Mandelker.etal.20a}. Those studies examined aspects of the sensitivity of shear-driven evolution to the imposed perturbations. However, they do not establish robustness when pressure-driven fragmentation associated with the RMI is also present. Neither \citet{Yao.etal.25} nor the simulation suite in \cref{tab:sim_table} systematically explores this dependence.

\smallskip
To address this concern, we performed four additional simulations of the representative coagulation case \run{R1D50P7M1}, which undergoes vigorous fragmentation followed by coagulation and shows the strongest cold-mass growth among the fiducial runs (\cref{sec:size}). It is therefore the case in which the outcome might be expected to depend most on the seed perturbations. Two runs vary the rms interface-displacement amplitude to \ifm{\delta\rx{s}=0.05\,\rx{s,i}} and \ifm{0.2\,\rx{s,i}}, compared with the fiducial value of \ifm{0.1\,\rx{s,i}}, while retaining the original phases and mode spectrum. A third adopts a different random realization of the shape-perturbation phases \ifm{\phi_j} while retaining the fiducial amplitude and spectrum. The fourth extends the azimuthal spectrum from \ifm{m=0\text{--}1} to \ifm{m=0\text{--}4}, retaining the original longitudinal wavelengths and total rms displacement amplitude, so that \ifm{N_\mathrm{pert}=5\times49=245}. As in the fiducial setup, the same perturbation parameters enter both the initial conditions and the time-dependent injection aperture in \cref{eq:rinj}.

\smallskip
\cref{fig:profile_mass_pert} compares the cold-gas line density, streamwise mass flux, and interfacial area per unit length along the stream in the five runs. The profiles are nearly indistinguishable over the entire penetration length. The initial compression, the transient drop in interfacial area near the convergence of the cylindrical shock at \ifm{z\sim10\,\rx{s,f}}, the subsequent rebound, and the downstream growth of the line density and mass flux by more than an order of magnitude all occur at the same positions with the same amplitudes. The differences between runs are comparable to the fluctuations within a single run and show no systematic trend with the perturbation amplitude, even though it spans a factor of four. Broadening the azimuthal spectrum to include the fluting modes \ifm{m=2\text{--}4}, which the fiducial spectrum omits, likewise leaves the profiles unchanged.

\smallskip
This insensitivity is consistent with the physical picture developed in the main text. The pressure-driven implosion and rebound are set by the imposed pressure contrast and stream Mach number rather than by the seed perturbations, and they drive the stream into a nonlinear, turbulent state within \ifm{z\lesssim20\,\rx{s,f}}. The subsequent evolution is governed by the competition between shear-driven mixing, radiative cooling, and coagulation, whose rates depend on the bulk stream parameters \ifm{\chix{f}}, \ifm{\rx{s,f}}, \ifm{\Max{s}}, and \ifm{\mathcal{P}} rather than on the details of the initial interface. The perturbations therefore serve mainly to seed the instabilities and to suppress the carbuncle instability at the shock fronts, as discussed in \cref{sec:methods}, and we expect the regime classification in \cref{tab:sim_table} to be robust to the specific perturbation realization.

\section{Steady-state tests}
\label{sec:steadystate}

\begin{figure*}
    \centering
    \includegraphics[width=\textwidth]{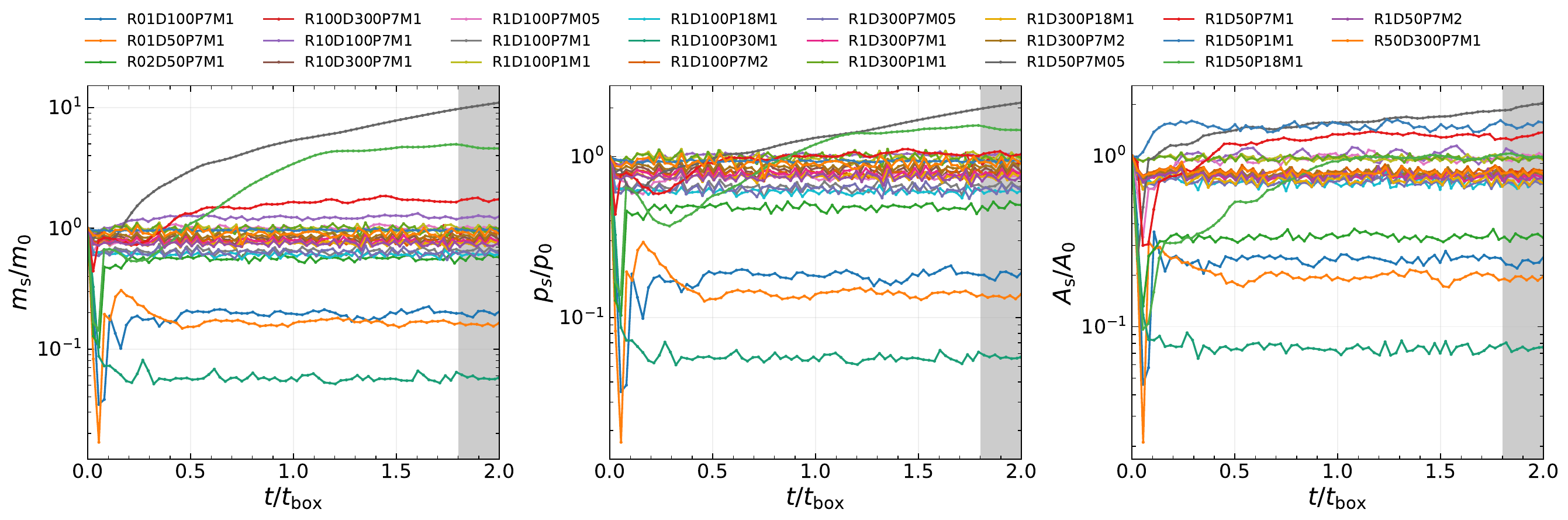}
    \caption{Time evolution of the total cold-gas mass \ifm{\mx{s}} (left), streamwise momentum \ifm{\px{s}} (middle), and cold--hot interfacial area \ifm{\Ax{s}} (right) for the runs listed in the legend. All quantities are evaluated within the selected analysis regions used for the spatial profiles and are normalized by their respective initial values \ifm{m_0}, \ifm{p_0}, and \ifm{A_0}. Cold gas is defined by \ifm{T<\SI{5e4}{K}}, and the interface is measured at \ifm{T=\SI{5e4}{K}}. Time is expressed in units of the box-crossing time, \ifm{\tx{box}=L_\mathrm{box}/\vx{s,i}}. The gray shaded interval, \ifm{1.8\text{--}2\,\tx{box}}, contains the final ten snapshots used to construct the time-averaged profiles. Most runs exhibit approximately stationary late-time fluctuations, while \run{R1D50P7M05} continues to grow.}
    \label{fig:time_evo}
\end{figure*}

To assess the temporal stability of the spatial profiles, \cref{fig:time_evo} shows the evolution of the cold-gas mass, streamwise momentum, and interfacial area within the selected analysis regions. Most runs undergo an initial transient and subsequently fluctuate around nearly constant values, with little systematic evolution during the final averaging interval, \ifm{1.8\text{--}2\,\tx{box}}. This behavior supports the use of the final ten snapshots to characterize their approximately steady bulk properties while averaging over short-term fluctuations.

\smallskip
The approach to steady state is slower in runs with strong cold-gas growth. In \run{R1D50P18M1}, the mass and momentum rise over much of the simulation but approach a plateau toward the final averaging interval. In \run{R1D50P7M05}, however, the mass, momentum, and interfacial area continue to increase through \ifm{2\,\tx{box}}. These measurements already exclude regions dominated by residual pre-existing stream material, so the integrated diagnostics do not establish stationarity for this run. Its time-averaged profiles characterize the late-time flow over the stated interval.

\section{Phase diagram}
\label{sec:phasediag}

\begin{figure*}
    \centering
    \includegraphics[width=\textwidth]{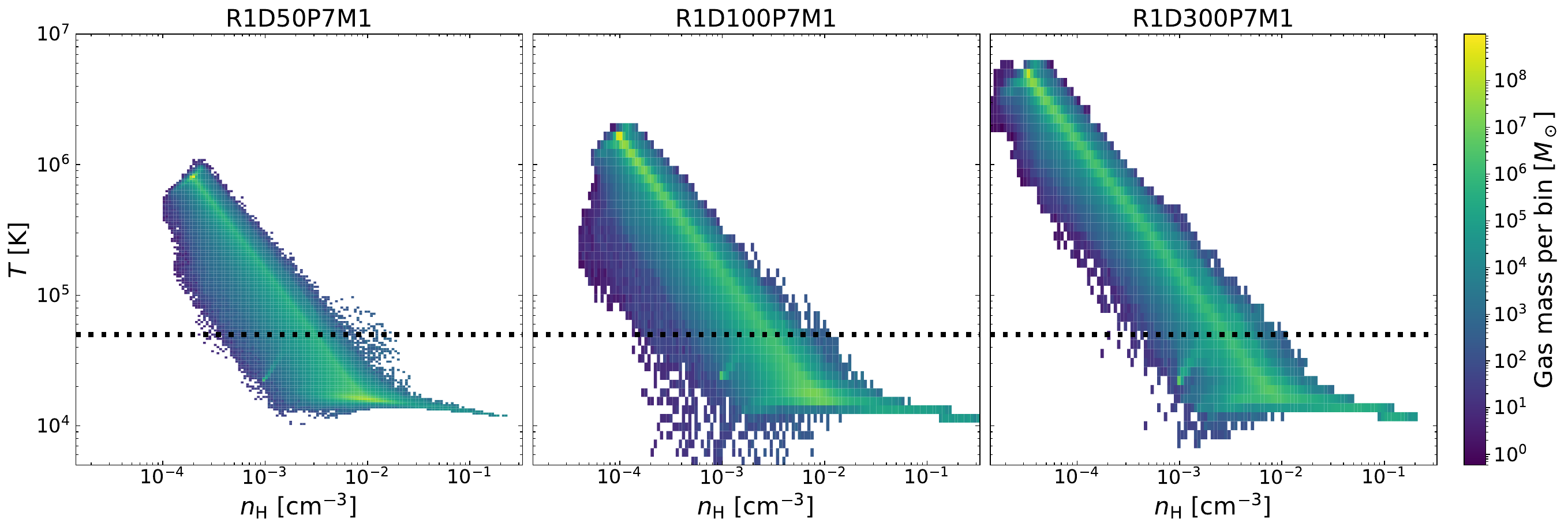}
    \caption{Temperature--density phase diagrams for the three fiducial runs, \run{R1D50P7M1}, \run{R1D100P7M1}, and \run{R1D300P7M1}, from left to right at the final snapshot. These runs have final density contrasts \ifm{\chix{f}=50,100,300}, respectively, at fixed \ifm{\rx{s,f}=\SI{1}{kpc}}, \ifm{\mathcal{P}=7}, and \ifm{\Max{s}=1}. The horizontal axis gives the hydrogen number density, \ifm{\nx{H}}, and the vertical axis gives the gas temperature. Colors indicate the gas mass in each two-dimensional bin, in \si{\Msun}, using a common logarithmic scale. The horizontal dotted line marks the adopted cold-gas threshold, \ifm{T=\SI{5e4}{K}}.}
    \label{fig:phase_diag}
\end{figure*}

The phase diagrams in \cref{fig:phase_diag} illustrate the gas component selected by our cold-gas criterion. All three runs exhibit a cool branch at \ifm{T\sim(1\text{--}2)\times10^4\,\mathrm{K}}, connected to the hot background through an extended, approximately isobaric distribution of intermediate-temperature gas. The cool branch also extends toward higher densities and pressures, consistent with compression of the injected stream. Our threshold, \ifm{T<\SI{5e4}{K}}, includes the cool equilibrium branch and moderate departures from thermal equilibrium while limiting the contribution of warmer transitional gas.

\smallskip
This choice closely matches the definition adopted by \citet{Mandelker.etal.20a} in idealized simulations of radiatively cooling streams, where cold gas is selected using \ifm{T<3\Tx{s}}. Their fiducial stream temperature, \ifm{\Tx{s}=\SI{1.5e4}{K}}, corresponds to a threshold of \ifm{\SI{4.5e4}{K}}. The continuous distribution across our adopted threshold indicates that it is an operational definition of the cold component, rather than a sharp thermodynamic phase boundary. Raising the threshold would include additional intermediate-temperature material and increase the selected gas mass; the cold-gas properties reported here refer specifically to gas below \ifm{\SI{5e4}{K}}.

\bsp	
\label{lastpage}
\end{document}
